\documentclass[a4paper,11pt]{article}
\pdfoutput=1 

\usepackage{jheppub} 

\usepackage[T1]{fontenc}
\usepackage{multirow,bbold,slashed}

\allowdisplaybreaks
 
\definecolor{nicered}{rgb}{0.7,0.1,0.1}
\definecolor{nicegreen}{rgb}{0.1,0.5,0.1}
\definecolor{niceblue}{rgb}{0.0,0.1,0.7}
\hypersetup{colorlinks,citecolor=niceblue,linkcolor=niceblue,urlcolor=niceblue}

\usepackage[normalem]{ulem}

\def \bm#1{\mbox{\boldmath$#1$\unboldmath}}
\def \beq{\begin{equation}}
\def \eeq{\end{equation}}
\def \bea{\begin{eqnarray}}
\def \eea{\end{eqnarray}}

\DeclareMathOperator{\arccot}{arccot}

\title{Precision tests of third-generation four-quark operators: one- and two-loop matching}

\author[a]{Ulrich Haisch}
\author[a,b]{and Luc Schnell}

\affiliation[a]{Max Planck Institute for Physics, \\ Boltzmannstr.~8, 85748 Garching, Germany}
\affiliation[b]{Technische Universit{\"a}t M{\"u}nchen, Physik-Department, \\ James-Franck-Strasse 1, 85748 Garching, Germany}

\emailAdd{haisch@mpp.mpg.de}
\emailAdd{schnell@mpp.mpg.de}

\preprint{MPP-2024-198}

\abstract{We calculate the one- and two-loop matching corrections in the Standard Model effective field theory (SMEFT) that impact electroweak precision measurements and flavour physics observables, focusing on the contributions of third-generation four-quark operators. Our results provide a crucial ingredient for a model-independent analysis of constraints on beyond the Standard Model physics that primarily affects the sector of third-generation four-quark operators. Concise analytic expressions are provided for all considered precision observables, which should facilitate their inclusion into global SMEFT analyses.}

\begin{document} 
\maketitle
\flushbottom

\section{Introduction}
\label{sec:introduction}

Assuming that possible beyond the Standard Model (BSM) contributions arise only from particles with masses far above the electroweak (EW) scale, the Standard Model~(SM) effective field theory~(EFT), also known as SMEFT~\cite{Buchmuller:1985jz,Grzadkowski:2010es,Brivio:2017vri,Isidori:2023pyp}, offers a largely model-independent, systematically improvable quantum-field theoretical framework for analyses of BSM effects in both high- and low-energy processes. In view of not having observed an unambiguous sign of BSM physics at colliders, SMEFT interpretations in form of global fits (for the latest results see~\cite{Brivio:2019ius,Ellis:2020unq,Ethier:2021bye,ATL-PHYS-PUB-2022-037,Celada:2024mcf}) have by now become a well-established and integral part of the LHC physics programme. 

The most sophisticated global SMEFT interpretation performed to date~\cite{Celada:2024mcf} simultaneously considers 50 different dimension-six operators and employs~the full slate of Higgs, diboson, and top-quark production and decay measurements from the~LHC to constrain their Wilson coefficients, partially also incorporating the information provided by the LEP measurements of~EW precision observables~(EWPOs). While several of the Wilson coefficients turn out to be stringently constrained by the combined fit, providing a model-dependent glimpse onto weakly-coupled BSM models with new particle masses in excess of~$1 \, {\rm TeV}$, certain classes of effective interactions remain poorly bounded. One such type of dimension-six operators are four-quark contact interactions that contain only third-generation fields. 

The weakness of the existing constraints on the Wilson coefficients of third-generation four-quark operators is due to the fact that this type of interactions can be probed at tree-level only in~$t \bar t t \bar t$, $t \bar t b \bar b$ or~$b \bar b b \bar b$ production. While based on the full LHC Run~II data set of around~$140 \, {\rm fb}^{-1}$ of integrated luminosity both ATLAS and CMS have found evidence for four-top production~\cite{CMS:2019rvj,ATLAS:2020hpj,ATLAS:2021kqb,CMS:2023zdh,ATLAS:2023ajo}, the resulting limits on the~$t \bar t t \bar t$ production cross section still have sizeable uncertainties of the order of 30\%. LHC measurements of~$t \bar t b \bar b$ production also exist (cf.~\cite{ATLAS:2018fwl,CMS:2019eih} for the latest results of this kind) but are presently limited in their precision as well. It furthermore turns out that meaningful bounds on the Wilson coefficients of the third-generation four-quark operators only arise if terms quadratic in the~$1/\Lambda^2$ expansion are included in the existing global analyses~\cite{Ethier:2021bye,Hartland:2019bjb,Degrande:2024mbg,Celada:2024mcf}, which raises questions about the validity of the SMEFT power counting.

Given the above limitations it seems worthwhile to entertain alternative probes of third-generation four-quark contact interactions. In fact, indirect probes have been discussed in the literature and include~top-quark production processes~\cite{Brivio:2019ius,deBlas:2015aea,Degrande:2020evl,Degrande:2024mbg}, top-quark~\cite{Boughezal:2019xpp} and~$Z$-boson decays~\cite{Hartmann:2016pil,Dawson:2022bxd} as well as Higgs physics~\cite{Gauld:2015lmb,Alasfar:2022zyr,DiNoi:2023ygk,Heinrich:2023rsd}. The main goal of this article is to add further observables to this list, thereby strengthening the power of the indirect constraints on the Wilson coefficients of the four-quark operators involving only heavy fields. To~this~purpose, we perform one- and two-loop matching calculations in the SMEFT, focusing on the contributions of third-generation four-quark operators that impact EWPOs and flavour physics observables. Our calculations include the oblique parameters~$S$,~$T$ and~$U$~\cite{Peskin:1991sw}, the $b \to c \ell \nu$ transition, the~$b \to s\ell^+ \ell^-$ process and the~$B_s$--$\bar B_s$ mixing amplitude. As~a~by-product of our computations, we are also able to cross-check the existing exact results for the heavy four-quark contributions to the decays~$t \to bW$~\cite{Boughezal:2019xpp} and~$Z \to b \bar b$~\cite{Dawson:2022bxd}. Compared to the corresponding SM results, all computations performed in our work are next-to-leading order (NLO) in EW interactions. The analytic~results obtained in this article provide a crucial ingredient for a model-independent analysis of constraints on BSM physics, primarily affecting the subclass of SMEFT dimension-six operators involving four third-generation quark fields. A~comprehensive study of the constraints on BSM models of this kind, derived from fits to $Z$-pole data, oblique corrections and precision flavour observables, will be presented in a companion paper~\cite{inprep}.

This work is structured as follows: in~Section~\ref{sec:theory}, we specify the subset of dimension-six SMEFT operators that are relevant in the context of this article. Section~\ref{sec:calculation} describes the different ingredients of our one- and two-loop matching calculations, while in~Section~\ref{sec:results} we detail the anatomy of our results by deriving compact formulas of the considered observables in the heavy top-quark limit. The numerical impact of the SMEFT corrections to the~$Z$-pole observables, the oblique parameters and the flavour processes of interest are discussed in~Section~\ref{sec:numerics}. Our~main findings are summarised in~Section~\ref{sec:conclusions}. The~full~analytic expressions for the one- and two-loop matching contributions involving third-generation four-quark SMEFT operators that are considered in this article are relegated to~Appendix~\ref{app:exact}. In~Appendix~\ref{app:shifts}, we furthermore give the finite shifts of the relevant two-loop observables associated to renormalising the top-quark mass in the~$\overline{\rm MS}$ scheme instead of on-shell~(OS). Additional technical details concerning the choice of operator basis and the cancellation of anomalous contributions are presented in~Appendix~\ref{app:fierz} and Appendix~\ref{app:anomalies}, respectively. In~Appendix~\ref{app:upFCNCs}, we finally provide a brief discussion of possible up-type flavour-changing neutral current (FCNC) constraints on the Wilson coefficients of third-generation four-quark operators.

\section{Theoretical framework}
\label{sec:theory}

The four-quark dimension-six interactions involving only third-generation fields that we use in the present analysis are constructed in terms of suitable linear combinations of the Wilson coefficients~$c_i$ in the basis of operators introduced by the LHC Top Working Group~(LHCTopWG)~in~\cite{Aguilar-Saavedra:2018ksv}: 
\beq \label{eq:LSMEFT}
{\cal L}_{\rm SMEFT} \supset \sum_i \frac{c_i}{\Lambda^2} \hspace{0.25mm} {\cal O}_i \,.
\eeq
The corresponding operators are defined as
\beq \label{eq:operators}
\begin{split}
{\cal O}^1_{QQ} &= \frac{1}{2} \, (\bar q \gamma_\mu q) (\bar q \gamma^\mu q) \,, \\[2mm]
{\cal O}^8_{QQ} &= \frac{1}{2} \, (\bar q \gamma_\mu T^a q) (\bar q \gamma^\mu T^a q) \,, \\[2mm]
{\cal O}^1_{Qt} &= (\bar q \gamma_\mu q) (\bar t \gamma^\mu t) \,, \\[2mm]
{\cal O}^8_{Qt} &= (\bar q \gamma_\mu T^a q)(\bar t \gamma^\mu T^a t) \,, \\[2mm]
{\cal O}^1_{Qb} &= (\bar q \gamma_\mu q) (\bar b \gamma^\mu b) \,, \\[2mm]
{\cal O}^8_{Qb} &= (\bar q \gamma_\mu T^a q)(\bar b \gamma^\mu T^a b) \,, \\[2mm]
{\cal O}^1_{tt} &= (\bar t \gamma_\mu t)(\bar t \gamma^\mu t) \,, \\[2mm]
{\cal O}^1_{tb} &= (\bar t \gamma_\mu t)(\bar b \gamma^\mu b) \,, \\[2mm]
{\cal O}^8_{tb} &= (\bar t \gamma_\mu T^a t)(\bar b \gamma^\mu T^a b) \,, \\[2mm]
{\cal O}^1_{QtQb} &= (\bar q t) \hspace{0.5mm} \varepsilon \hspace{0.5mm} (\bar q b) \,, \\[2mm]
{\cal O}^8_{QtQb} &= (\bar q T^a t) \hspace{0.5mm} \varepsilon \hspace{0.5mm} (\bar q T^a b) \,.
\end{split}
\eeq
Here~$\Lambda$ is the common new-physics scale that suppresses the dimension-six operators~${\cal O}_i$, $q$~denotes the left-handed third-generation quark~$SU(2)_L$ doublet, while~$t$ and~$b$ are the right-handed top- and bottom-quark~$SU(2)_L$ singlets. Finally,~$T^a = \lambda^a/2$ are the~$SU(3)_C$ generators with~$\lambda^a$ the Gell-Mann matrices and~$\varepsilon = i \sigma^2$ the antisymmetric~$SU(2)$ tensor. Here $\sigma^i$~denotes the usual Pauli matrices. Notice that for~${\cal O}^1_{QtQb}$ and~${\cal O}^8_{QtQb}$ the sum of the hermitian conjugate in~(\ref{eq:LSMEFT}) is understood. Additionally, all Wilson coefficients~$c_i$ are assumed to be real in the following discussion, as we only consider CP-conserving observables in our~article. 

The above basis is also used in the {\tt FeynRules}~\cite{Alloul:2013bka} implementation~called~{\tt dim6top}~\cite{dim6top} which is employed in many SMEFT interpretations of top-quark measurements including the global LHC~Run~II analyses~\cite{Brivio:2019ius,Ellis:2020unq,Ethier:2021bye}. It is important to realise that the choice of operators in~(\ref{eq:operators}) resembles almost exactly the definition of third-generation four-quark operators in the Warsaw operator basis~\cite{Grzadkowski:2010es}. The~only exception is ${\cal O}^8_{QQ}$, which, in the Warsaw operator basis, is expressed as a linear combination of $(\bar q \gamma_\mu q) (\bar q \gamma^\mu q)$ and $(\bar q \gamma_\mu \sigma^i q) (\bar q \gamma^\mu \sigma^i q)$, derived through the use of a Fierz identity~(cf.~Appendix~\ref{app:fierz}). Since Fierz identities generally only hold in $d=4$ dimensions, the results presented below differ from the ones that are obtained in the Warsaw operator basis by finite matrix elements of a Fierz evanescent operator proportional to~$c^8_{QQ}$. This technical detail is important to remember whenever comparing our results to those of existing calculations utilising the Warsaw operator basis.\footnote{In Appendix~\ref{app:fierz}, we provide some details on the change of renormalisation scheme that connects the LHCTopWG to the Warsaw~operator basis. Additionally, we collect the full analytic expressions of all our results both in the LHCTopWG and Warsaw~operator bases digitally at~\cite{analyticresults}.} 

\section{Calculation in a nutshell}
\label{sec:calculation}

All calculations performed below use dimensional regularisation for ultraviolet~(UV) singularities in~$d = 4 - 2 \hspace{0.125mm} \epsilon$ dimensions supplemented by the renormalisation scale~$\mu$ and a~naive anti-commuting~$\gamma_5$~(NDR). We have verified that the use of the NDR scheme leads to consistent results for all considered observables. In particular, working in the LHCTopWG basis of operators~(\ref{eq:operators}) completely avoids the appearances of non-zero traces with $\gamma_5$.\footnote{As discussed in Appendix~\ref{app:anomalies}, in the Warsaw operator basis non-zero traces with $\gamma_5$ appear. These anomalous contributions are irrelevant, however, because they can be removed by local counterterms.} Infrared~(IR) divergences have also been regularised dimensionally. Throughout this article, the full top-quark mass dependence is retained in both the matrix elements and the phase-space integrals, while all other fermion masses and Yukawa couplings are set to zero from the outset in our calculations. The~actual generation and computation of the amplitudes made use of the {\tt Mathematica} packages {\tt FeynRules}, {\tt FeynArts}~\cite{Hahn:2000kx}, {\tt FeynCalc}~\cite{Shtabovenko:2020gxv}, {\tt FormCalc}~\cite{Hahn:2016ebn}, {\tt LiteRed}~\cite{Lee:2013mka} and {\tt Package-X}~\cite{Patel:2015tea}. Our two-loop calculations further rely on the method of regions~\cite{Beneke:1997zp,Smirnov:2002pj,Jantzen:2011nz}, which enables the expansion of loop integrals as series in external momenta with coefficients given by vacuum integrals~\cite{Fleischer:1994ef,Tarasov:1995jf}. We developed a custom code for this expansion, while the analytical expressions for the vacuum integrals were obtained from~\cite{Davydychev:1992mt,Korner:2004rr}.

\begin{figure}[t!]
\begin{center}
\includegraphics[width=0.75\textwidth]{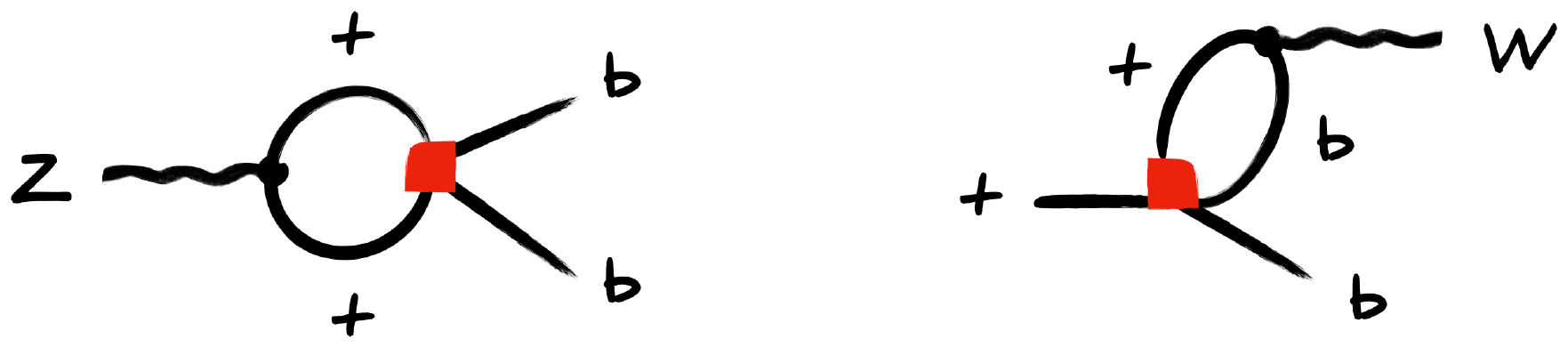}
\end{center}
\vspace{-2mm} 
\caption{\label{fig:decays} Examples of one-loop contributions to the~$Z \to b \bar b$~(left) and~$t \to bW$ (right) decay involving insertions of one of the dimension-six operators (red~squares) introduced in~(\ref{eq:operators}). In~the case of the~$Z \to b \bar b$ transition, instead of top also bottom quarks can be exchanged in the loop. Finally, note that by replacing the external top quark with a charm quark, the right diagram results in a one-loop correction to the $b \to c \ell \nu$ transition.}
\end{figure}

In~Figure~\ref{fig:decays} we display representative Feynman diagrams that give rise to the decays~$Z \to b \bar b$~(left graph) and~$t \to bW$ (right graph) in the presence of the third-generation four-quark dimension-six operators~(\ref{eq:operators}). We calculate the exact on-shell matrix elements, treating both external and internal bottom quarks as massless. The analytic results of the bare one-loop matrix elements of the~$Z \to b \bar b$ and~$t \to b W$ transitions are UV divergent. The corresponding~$1/\epsilon$~poles are removed by performing operator renormalisation in the~$\overline{\rm MS}$ scheme, and we have verified that the divergence structure of our expressions agrees with the direct calculation of the relevant anomalous dimensions presented in the articles~\cite{Jenkins:2013wua,Alonso:2013hga}. The~found agreement constitutes a non-trivial cross-check~of our one-loop SMEFT computations of the~$Z \to b \bar b$ and~$t \to b W$ amplitudes. 

\begin{figure}[t!]
\begin{center}
\includegraphics[width=0.9\textwidth]{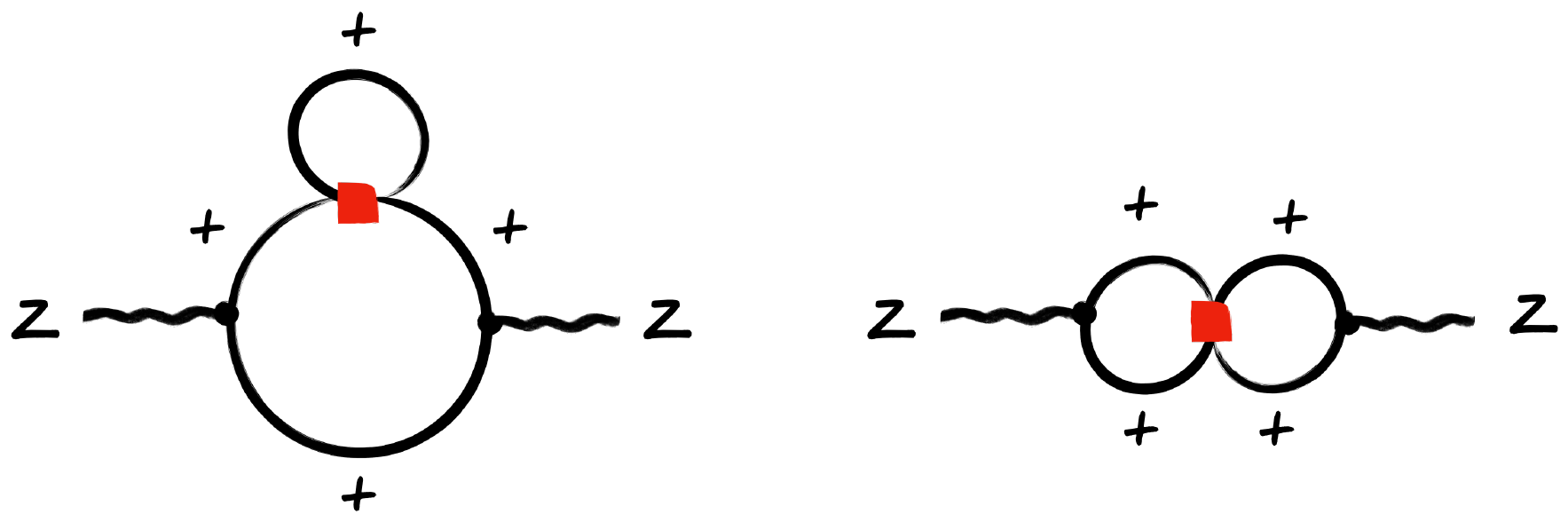}
\end{center}
\vspace{-2mm} 
\caption{\label{fig:peskin} Representative two-loop vacuum polarisation diagrams with insertions of third-generation four-quark dimension-six operators~(red squares) that modify the oblique parameters. Besides the~$ZZ$ vacuum polarisation also the~$Z\gamma$,~$\gamma \gamma$ and~$WW$ self-energies need to be considered and instead of top quarks virtual bottom quarks can be exchanged. }
\end{figure}

The operators~(\ref{eq:operators}) contribute to the oblique parameters first at the two-loop level. Relevant vacuum polarisation diagrams are shown in~Figure~\ref{fig:peskin}. Notice that there are two types of contributions: first, diagrams that involve a correction to the quark propagator~(left~graph) and, second, diagrams where the gauge boson vertices receive corrections~(right~graph). In fact, the subgraph involving the operator insertion in the first type of contributions leads to a renormalisation of the quark masses. To renormalise the bare vacuum polarisation amplitudes one therefore needs to provide counterterms not only for the renormalisation of the operators but also the quark masses. Following~\cite{Gauld:2015lmb,Gauld:2016kuu,Dawson:2018pyl,Cullen:2019nnr,Cullen:2020zof,Alasfar:2022zyr} we adopt a mixed~$\overline{\rm MS\hspace{0.25mm}}$-$\hspace{0.125mm}{\rm OS}$ scheme, in which we renormalise the Wilson coefficients of the dimension-six operators and the quark masses in~$\overline{\rm MS}$ and OS, respectively. Notice~that renormalising the quark masses in the OS scheme has the salient advantage compared to an~$\overline{\rm MS}$ prescription that diagrams like the one displayed on the left-hand side in~Figure~\ref{fig:peskin} do not lead to a contribution to the vacuum polarisations and thus to the oblique parameters.\footnote{For completeness we provide in~Appendix~\ref{app:shifts} analytic expressions for the finite shifts in the oblique parameters that result from changing the renormalisation scheme from OS to~$\overline{\rm MS}$.} After mass renormalisation the bare two-loop vacuum polarisation amplitudes contain both double and single poles of UV origin. As a result of the locality of UV divergences, the structure of the~$1/\epsilon^2$ poles is thereby related to products of one-loop~$1/\epsilon$ subdivergences (see~for~instance~\cite{Chetyrkin:1997fm,Gambino:2003zm}). It can therefore also be derived from the results obtained in~\cite{Jenkins:2013wua,Alonso:2013hga} which serves as a non-trivial cross-check of our two-loop computations. The~$1/\epsilon$ divergences instead determine the two-loop mixing of the operators~(\ref{eq:operators}) into the oblique parameters. Renormalising the Wilson coefficients in the~$\overline{\rm MS}$ scheme then removes all UV poles and one is left with finite vacuum polarisation amplitudes. We add that in our calculation of the two-loop SMEFT corrections to the Peskin-Takeuchi parameters we have set the bottom-quark mass to zero from the very start. In the case of the pure top and the mixed top-bottom quark contributions we then compute the~$S$,~$T$~and~$U$ parameters in terms of the transversal part of the vacuum polarisation tensors and their derivatives evaluated at zero-momentum transfer~$q^2 =0$. The pure bottom-quark contributions to~$S$~and~$U$ are instead expressed through the real part of the transversal part of the vacuum polarisation tensors evaluated on the~$Z$ pole, i.e.~using~$q^2 = M_Z^2$. Notice that in the latter case the finite momentum transfer acts as an IR regulator, which avoids the appearance of~$\ln \left ( m_b^2/\mu^2 \right )$ terms that would appear if a non-zero bottom-quark mass would be used to calculate the derivatives of the transversal part of the vacuum polarisation tensors. 

\begin{figure}[t!]
\begin{center}
\includegraphics[width=0.975\textwidth]{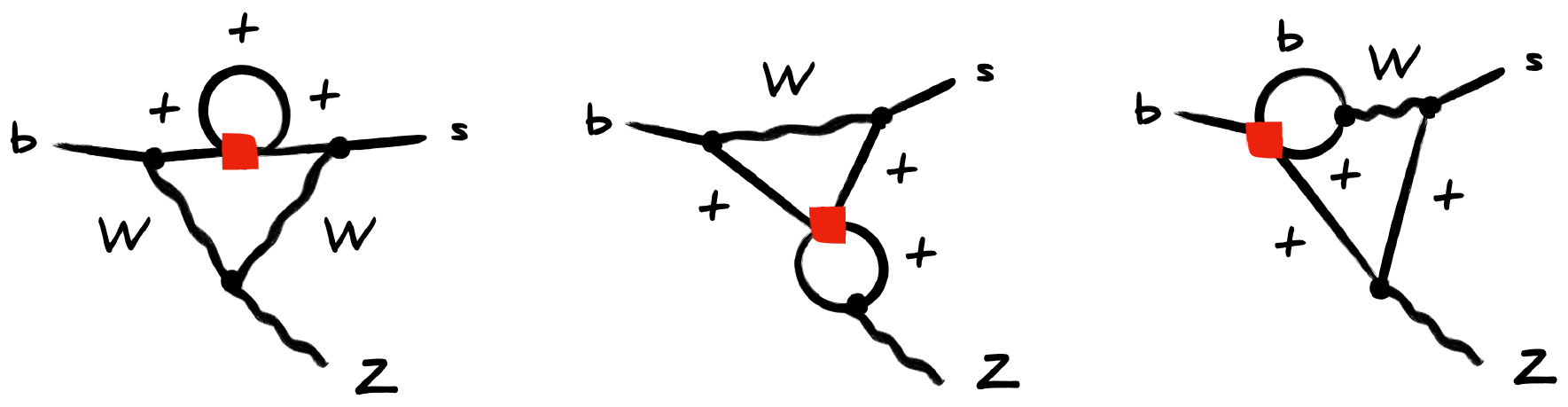}
\end{center}
\vspace{-2mm} 
\caption{\label{fig:penguin} An assortment of two-loop penguin graphs that lead to the~$b \to s Z$ transition with single insertions of the third-generation four-quark dimension-six operators (red squares) appearing in~(\ref{eq:operators}). In~a renormalisable or~$R_\xi$ gauge like the 't~Hooft-Feynman gauge also diagrams with would-be Goldstone bosons need to be considered. Wave function contributions associated to the external quarks must be included as well but are not depicted. }
\end{figure}

\begin{figure}[t!]
\begin{center}
\includegraphics[width=0.633\textwidth]{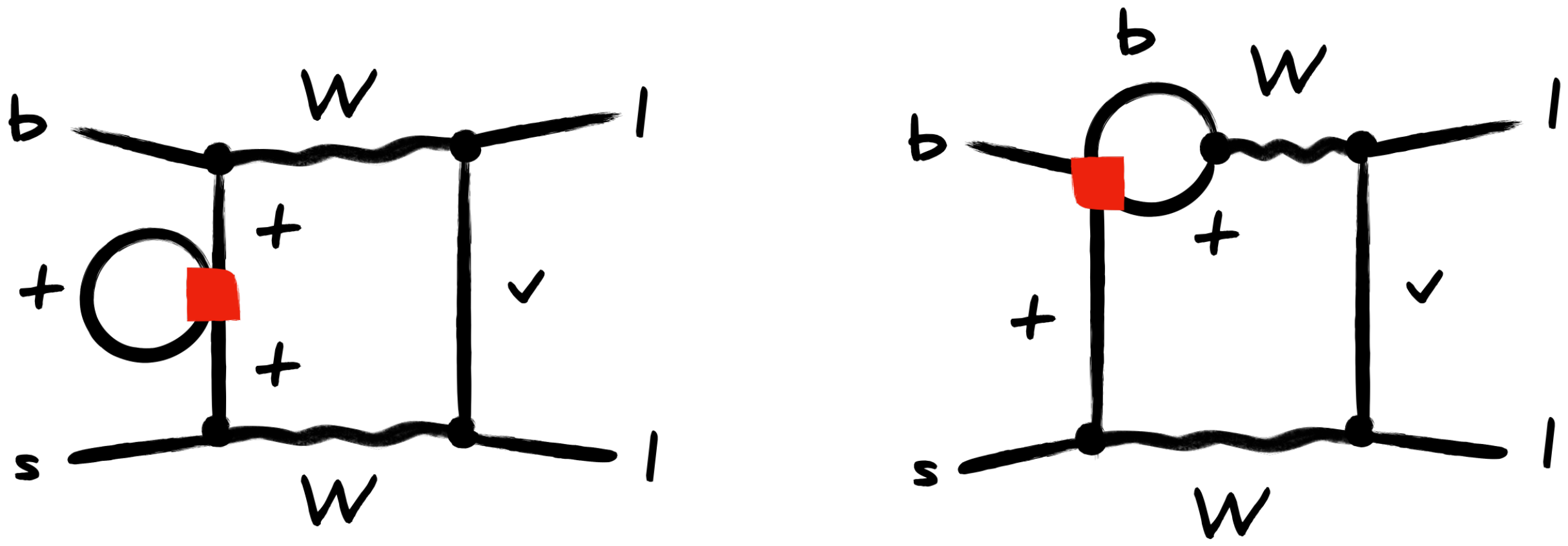}
\end{center}
\vspace{-2mm} 
\caption{\label{fig:sexob} Example two-loop box diagrams with a single insertion of the third-generation four-quark dimension-six operators (red squares) introduced in~(\ref{eq:operators}) that induce the process~$b \to s \ell^+ \ell^-$. Contributions involving would-be Goldstone boson exchange in general also exist but they can be ignored since they only lead to corrections that are power suppressed by small lepton masses. }
\end{figure}

Before discussing our calculation of FCNC processes it is important to recall that the operators introduced in~(\ref{eq:operators}) are written in terms of weak eigenstate fields, while physical observables are described in terms of matrix elements involving mass eigenstates. To make contact with experiment, one therefore has to rotate the weak into mass eigenstates via unitary transformations. After these field redefinitions, there are no FCNC at tree level in the SM, and mixing between different quark generations only occurs in the charged-current interactions being encoded by the Cabibbo-Kobayashi-Maskawa~(CKM) matrix~$V$. When dimension-six four-quark operators such as~(\ref{eq:operators}) are included in the Lagrangian, the mass diagonalisation however leads to tree-level FCNCs either in the down- or up-quark sector or both. Since experimentally the limits from the mixing of neutral mesons,~i.e.~$\Delta F = 2$ processes, are significantly stronger in the down- than in the up-quark sector, choosing the unitary rotations such that tree-level FCNCs associated with~(\ref{eq:operators}) are confined to~$D$--$\bar D$ mixing is preferable from a phenomenological point of view.\footnote{A discussion of the possible constraints on the Wilson coefficients of third-generation four-quark operators arising from up-type FCNCs is provided in Appendix~\ref{app:upFCNCs}.} This option is commonly referred to as down-alignment and corresponds to the choice~$q = (\sum_{\psi=u,c,t} V_{\psi b}^\ast \hspace{0.5mm} \psi_{L}, b_L)^T$ of third-generation~$SU(2)_L$ doublet fields. 

\begin{figure}[t!]
\begin{center}
\includegraphics[width=0.975\textwidth]{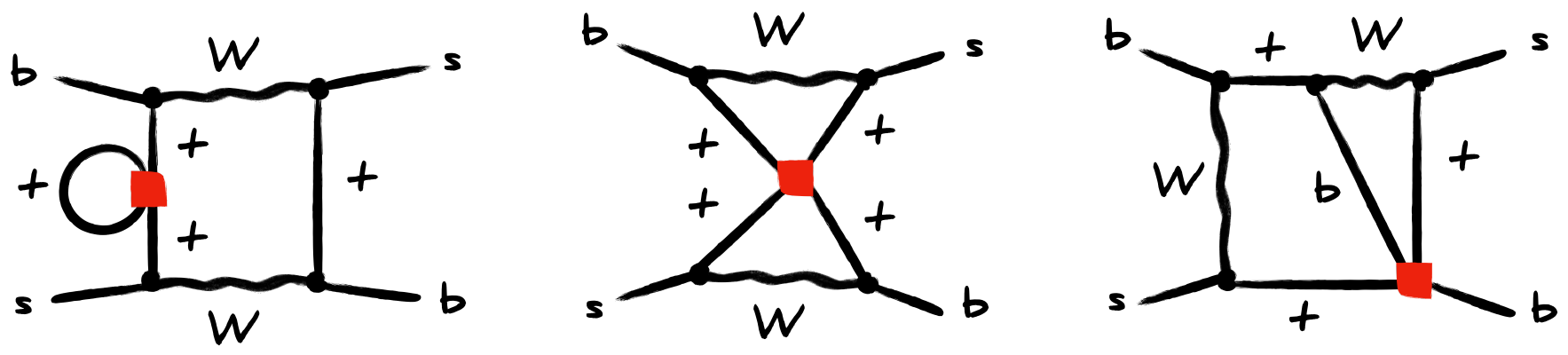}
\end{center}
\vspace{-2mm} 
\caption{\label{fig:boxes} Representative two-loop box diagrams involving insertions of third-generation four-quark operators~(red squares) that contribute to~$B_s$--$\bar B_s$ mixing. Graphs with would-be Goldstone bosons also exist in the 't~Hooft-Feynman gauge but are not explicitly shown. }
\end{figure}

To minimise the FCNC constraints on the Wilson coefficients of the third-generation four-quark contact interactions, we assume down-alignment in the following.\footnote{The orthogonal case of up-alignment has for instance been considered in~\cite{Aguilar-Saavedra:2018ksv,Allwicher:2023shc}.} In such a case insertions of the operators~(\ref{eq:operators}) generate no~$\Delta F = 1$ and~$\Delta F = 2$ transitions in the down-quark sector both at the tree and the one-loop level. At two loops, however, FCNCs involving down-type quarks do necessarily arise from Feynman diagrams involving~$W$-boson exchange. Example contributions to the~$b \to sZ$ transition, the~$b \to s \ell^+ \ell^-$ boxes and the~$B_s$--$\bar B_s$ mixing amplitude are shown in~Figure~\ref{fig:penguin},~\ref{fig:sexob} and~\ref{fig:boxes}, respectively. The diagrams describing the~$b \to s \gamma$ penguin process are obtained by simply replacing the external~$Z$~boson in the diagrams displayed in~Figure~\ref{fig:penguin} by a photon. As in the case of the vacuum polarisations the diagrams that contribute to the relevant FCNC processes in the down-quark sector involve subgraphs that contain UV poles associated to mass and operator renormalisation. Performing the mass renormalisation in the OS scheme, diagrams that involve a correction to the quark propagator do not contribute as they are cancelled by the relevant counterterm.\footnote{The analytic expressions for the finite shifts in the flavour observables that result from a change of the renormalisation scheme from OS to~$\overline{\rm MS}$ can be found in~Appendix~\ref{app:shifts}.} 

In the two-loop calculations of the relevant flavour observables, we utilised the method of regions~\cite{Beneke:1997zp,Smirnov:2002pj,Jantzen:2011nz}, concentrating on the hard regions of the SMEFT diagrams. In these regions, the integration momenta and the masses of the $W$-, $Z$-boson and top quark are significantly larger than the external momenta. This approach enables an integrand-level expansion in the external momenta with coefficients given solely by vacuum integrals, which are much simpler to compute than the full integrals. Moreover, the loop diagrams on the low-energy effective field theory~(LEFT) side vanish because all massive particles have been integrated out (i.e.~the top quark and the heavy EW gauge bosons), leaving only scaleless vacuum integrals. However, it is important to note that divergent contributions may still arise on the LEFT side during hard region matching, stemming from tree-level counterterm diagrams. For example, akin to the SM calculation of the off-shell $b \to s \gamma$ amplitude (cf.~for~example~\cite{Bobeth:1999mk}), obtaining the correct matching corrections to the charm- and up-quark contribution to the photon penguin requires the inclusion of divergent contributions arising from the LEFT. These~LEFT contributions are needed to make the UV-pole structure of the matching corrections local. We carried out the external momentum expansion and reduction using a custom code interfaced to {\tt LiteRed}~\cite{Lee:2013mka}, and validated it by comparing with numerical results from {\tt AMFlow}~\cite{Liu:2022chg}, achieving perfect agreement. The~analytical expressions for the vacuum integrals were taken from~\cite{Davydychev:1992mt,Korner:2004rr}. We add that the calculation of the $b \to s \gamma$ penguin amplitude requires an expansion in external momenta to second order, while in the case of the $b \to s Z$ penguin, the $b \to s \ell^+ \ell^-$ box and the $B_s$--$\bar B_s$ mixing amplitude, the matching calculation can be performed at zeroth order in the external momenta.

Our two-loop calculations involve some further technicalities worth mentioning. First, in the case of the~$b \to sZ$ penguin one has to perform flavour off-diagonal renormalisation of the wave functions of the external quark fields in order to obtain a UV-finite result, while in all other cases operator renormalisation in the $\overline{\rm MS}$ scheme is sufficient to remove all UV poles that do not correspond to a two-loop mixing contribution. Second, in the case of the box contributions to the $b \to s \ell^+ \ell^-$ process and $B_s$--$\bar B_s$ mixing (cf.~Figures~\ref{fig:sexob} and~\ref{fig:boxes}) so-called evanescent operators,~i.e.~operators that vanish algebraically in $d = 4$ dimensions but are non-zero in $d = 4 - 2 \hspace{0.125mm} \epsilon$ dimensions, appear. In the case of the $b \to s \ell^+ \ell^-$ boxes we employ the evansecent operators defined in~\cite{Chetyrkin:1996vx,Chetyrkin:1997gb,Misiak:1999yg} since this is the standard choice in the state-of-the-art calculations of $\Delta F = 1$ processes in the SM~\cite{Gorbahn:2004my}. For the calculation of the $B_s$--$\bar B_s$ mixing amplitudes, we instead use the definition of the evanescent operators as given in~\cite{Buras:1990fn,Herrlich:1996vf}, which are chosen such that the Fierz symmetry of the amplitudes remains manifest in $d = 4 - 2 \hspace{0.125mm} \epsilon$ dimensions. In~Appendix~\ref{app:exact}, we will delve deeper into how the choice of evanescent operators affects our two-loop results of the relevant flavour observables. Third, the $b \to s \gamma$ penguin amplitudes are computed using a background-field version of the 't~Hooft-Feynman gauge for the photon field, as this allows to maintain explicit gauge invariance at the level of off-shell Green's functions~(see for instance~\cite{Cella:1994np,Bobeth:1999mk,Haisch:2002zz} for discussions of this issue in the flavour physics context). Notice that in Figures~\ref{fig:penguin}, \ref{fig:sexob} and~\ref{fig:boxes}, we have only depicted Feynman diagrams involving top quarks. In the case of down-alignment, however, charm- and up-quark contributions also need to be included in the matching calculations. Using the unitarity of the CKM matrix this leads to a Glashow-Iliopoulos-Maiani~(GIM) mechanism in all two-loop FCNC amplitudes calculated in this~article. 

\section{Anatomy of results}
\label{sec:results}

In this section, we provide approximate results for the physical observables discussed above. The~complete analytic expressions are rather lengthy and are therefore provided separately in~Appendix~\ref{app:exact}. We begin our discussion with the~$Z$-boson decay width into bottom quarks~$\Gamma (Z \to b \bar b)$ and the forward-backward asymmetry for bottom quarks~$A_{\rm FB}^{b}$. Other~$Z$-pole observables such as the total~$Z$-boson width~$\Gamma_Z$, the ratio of the~$Z$-boson decay width into bottom quarks, the total hadronic width~$R_b$ and the bottom-quark left-right asymmetry~$A_b$ are straightforwardly derived from the given expressions by realising that the processes~$Z \to \psi \bar \psi$ with~$\psi$ denoting any SM fermion but the bottom quark remain unmodified at the one-loop level in the presence of~(\ref{eq:operators}). For~the~SMEFT corrections to~$\Gamma (Z \to b \bar b)$ and~$A_{\rm FB}^{b}$, we find the following approximations
\beq \label{eq:Zbbapp} 
\begin{split}
\Delta \Gamma (Z \to b \bar b) & \simeq -\frac{\alpha ^2 m_Z}{192 \pi c_w^4 s_w^4} \frac{m_t^2}{m_Z^2}\frac{v^2}{\Lambda^2} \\[1mm]
& \phantom{xx} \times \bigg \{ \Big[ \hspace{0.5mm} 6 \left(2 s_w^2-3\right) (c_{QQ}^1 - c_{Qt}^1) +12 s_w^2 ( c_{Qb}^1 - c_{tb}^1 ) \hspace{0.5mm} \Big] \ln \left ( \frac{m_t^2}{m_Z^2} \right ) \bigg \} \,, \\[2mm]
\Delta A_{\rm FB}^{b} & \simeq \frac{9 \left(8 s_w^4-14 s_w^2+3\right) \alpha}{4 \pi c_w^2 \left(8 s_w^4-12 s_w^2+9\right)^2 \left(8 s_w^4-4 s_w^2+1\right)} \frac{m_t^2}{m_Z^2} \frac{v^2}{\Lambda^2} \\[2mm] 
& \phantom{xx} \times \bigg \{ \Big[ \hspace{0.5mm} 6 s_w^2 (c_{QQ}^1 - c_{Qt}^1) - 3 \left(2 s_w^2-3\right) ( c_{Qb}^1 - c_{tb}^1 ) \hspace{0.5mm} \Big] \ln \left ( \frac{m_t^2}{m_Z^2} \right ) \bigg \} \,.
\end{split} 
\eeq
These results have been obtained by identifying the renormalisation scale~$\mu$ with the~$Z$-boson mass~$m_Z$ and by performing a heavy-top mass expansion~(HTE) of the exact results~(\ref{eq:DGZbb}) to~(\ref{eq:betas}). The HTE is an asymptotic expansion near the limit of infinitely heavy top quark, where $m_t^2 \gg m_Z^2, m_W^2$. Given the choice of renormalisation scale, the Wilson coefficients~$c_i$ in~(\ref{eq:Zbbapp}) are understood to be evaluated at~$m_Z$ as well. Furthermore,~$\alpha$ is the electromagnetic coupling,~$c_w$ and~$s_w$ denote the cosine and sine of the weak mixing and~$v$ is the Higgs vacuum expectation value~(VEV). One notes that the observables~$\Delta \Gamma (Z \to b \bar b)$ and~$\Delta A_{\rm FB}^{b}$ receive logarithmic corrections proportional to~$c_{QQ}^1$,~$c_{Qt}^1$,~$c_{Qb}^1$ and~$c_{tb}^1$ that are enhanced by two powers of the top-quark Yukawa coupling~$y_t = \sqrt{2} \hspace{0.25mm} m_t/v$ with $m_t$ denoting the top-quark mass, which is not the case for~$c_{QQ}^8$ $\big($see (\ref{eq:alphas}) and (\ref{eq:betas})$\big)$. This feature is related to the fact that in the basis~(\ref{eq:operators}), the Wilson coefficient~$c_{QQ}^8$ does not mix into the Wilson coefficients that contribute to the~$Z \to b \bar b$ process at tree level as far as terms proportional to~$y_t^2$ are concerned. This follows from the Yukawa dependence of the renormalisation group~(RG)~equations of~$C_{Hq}^{1 (33)}$,~$C_{Hq}^{3 (33)}$ and~$C_{Hd}^{(33)}$ that have been derived in~\cite{Jenkins:2013wua}. Notice that the results~(\ref{eq:Zbbapp}) only depend on~$c_{QQ}^1$,~$c_{Qt}^1$,~$c_{Qb}^1$ and~$c_{tb}^1$ but not on the remaining Wilson coefficients appearing in~(\ref{eq:operators}). In fact, the contribution proportional to $c^8_{QQ}$ is suppressed by $m_Z^2/m_t^2 \simeq 0.28$, the contributions proportional to~$c^1_{QtQb}$ and~$c^8_{QtQb}$ are suppressed by~$y_b/y_t \simeq 0.02$, while a contribution proportional to~$c_{Qb}^8$ appears at order~$(y_b/y_t)^2 \simeq 4 \cdot 10^{-4}$ relative to~(\ref{eq:Zbbapp}). Since in our calculation the bottom quark is treated as massless, the strongly suppressed terms proportional to $c^1_{QtQb}$, $c^8_{QtQb}$ and $c_{Qb}^8$ have been neglected both in the above approximations and in the exact results~(\ref{eq:DGZbb}) to~(\ref{eq:betas}). We finally remark that subleading terms in the HTE of~$ \Delta \Gamma (Z \to b \bar b)$ and~$\Delta A_{\rm FB}^{b}$ are numerically important and therefore we will use the exact results reported in~Appendix~\ref{app:exact} in our numerical~analysis. 

In the case of the~$t \to bW$ transition we consider three observables, namely the partial decay width~$\Gamma (t \to b W)$ as well as the decay fractions~$F_L$ and~$F_-$ into longitudinal and negative~$W$-boson helicities, respectively. Identifying the renormalisation scale~$\mu$ with~$m_t$, we find\footnote{The positive helicity fraction~$F_+$ does not receive a SMEFT correction from any of the operators introduced in~(\ref{eq:operators}) if the bottom quark is treated as massless.}
\beq \label{eq:tbWapp} 
\begin{split}
& \Delta \Gamma ( t \to b W ) \simeq -\frac{\alpha^2 |V_{tb}|^2 \hspace{0.25mm} m_t^3}{256 \pi s_w^4 \hspace{0.25mm} m_W^2}\frac{m_t^2}{m_W^2} \frac{v^2}{\Lambda^2} \left ( c_{QQ}^1 + \frac{4}{3} \hspace{0.25mm} c_{QQ}^8 \right ) \,, \\[2mm]
& \Delta F_L = -\Delta F_- \simeq \frac{\alpha}{4 \pi s_w^2} \frac{v^2}{\Lambda^2} \left ( c_{Qt}^1 + \frac{4}{3} \hspace{0.25mm} c_{Qt}^8 \right ) \,, 
\end{split} 
\eeq
with~$V_{tb}$ denoting the relevant CKM matrix element. Analogous to~(\ref{eq:Zbbapp}), the above results represent the leading terms in the  HTE of the exact expressions presented in~(\ref{eq:DGtbW}),~(\ref{eq:gammas}) and~(\ref{eq:DFi}). One observes that at relative order~$y_t^2$ only the Wilson coefficients~$c_{QQ}^1$ and~$c_{QQ}^8$ contribute to the partial decay rate~$\Gamma ( t \to b W )$, while in the case of the helicity fractions~$F_L$ and~$F_-$ only~$c_{Qt}^1$ and~$c_{Qt}^8$ render a non-vanishing contribution that is however not enhanced by powers of the top-quark mass. The rate and the helicity fractions of~$t \to b W$ therefore represent complementary probes of~(\ref{eq:operators}) in the sense that they depend to first approximation on two different linear combinations of Wilson coefficients. We add that contributions from the scalar operators~${\cal O}^1_{QtQb}$ and~${\cal O}^8_{QtQb}$ to the~$t \to b W$ observables first arise at relative order~$y_b/y_t$, and thus such effects are not included in our computations.

In the case of the oblique corrections we consider the full set of Peskin-Takeuchi parameters~$S$,~$T$ and~$U$. For the SMEFT corrections induced by single insertions of the third-generation four-quark operators~(\ref{eq:operators}), we obtain the following approximate expressions
\beq \label{eq:STUapp}
\begin{split}
\Delta S & \simeq -\frac{1}{144 \pi^3} \frac{m_t^2}{\Lambda^2} \left ( 24 c_{QQ}^1 + 8 c_{QQ}^8 - 54 c_{Qt}^1 + 96 c_{tt}^1 \right ) \ln^2 \left ( \frac{m_t^2}{m_Z^2} \right ) \,, \\[2mm] 
\Delta T & \simeq -\frac{m_t^2}{256 \pi^3 c_w^2 s_w^2 m_Z^2} \frac{m_t^2}{\Lambda^2} \left ( 42 c_{QQ}^1 + 8 c_{QQ}^8 - 72 c_{Qt}^1 + 96 c_{tt}^1 \right ) \ln^2 \left ( \frac{m_t^2}{m_Z^2} \right ) \,, \\[2mm]
\Delta U & \simeq \frac{1}{48\pi^3} \frac{m_t^2}{\Lambda^2} \left ( 18 c_{QQ}^1 - 18 c_{Qt}^1 \right ) \ln^2 \left ( \frac{m_t^2}{m_Z^2} \right ) \,.
\end{split}
\eeq
In order to arrive at these results, we have identified in the exact results presented in~Appendix~\ref{app:exact} the renormalisation scale~$\mu$ with~$m_Z$ and performed a HTE, keeping only the leading-logarithmic terms. One observes that while all Peskin-Takeuchi parameters~$S$, $T$ and $U$ receive double-logarithmic corrections, the contributions to~$\Delta T$ are parametrically enhanced by two powers of~$y_t$ relative to the ones arising in the case of~$\Delta S$ and $\Delta U$. Notice that the exact same feature is observed in BSM models with new gauge bosons that couple dominantly to the third generation of quarks~\cite{Burdman:1999us,Haisch:2011up}. Recent studies~\cite{Allwicher:2023aql,Garosi:2023yxg,Allwicher:2023shc,Stefanek:2024kds} have also identified the emergence of double-logarithmic corrections in the Peskin-Takeuchi parameters due to operator mixing in the~SMEFT. Numerically, it turns out that the formulas~(\ref{eq:STUapp}) are not very good approximations, so that in our phenomenological analysis we will employ the complete two-loop expressions for~$\Delta S$,~$\Delta T$ and~$\Delta U$ reported in~(\ref{eq:Sexact}) to~(\ref{eq:thetas}). We finally note that the two-loop contributions to the~Peskin-Takeuchi parameters proportional to the Wilson coefficients~$c^1_{QtQb}$ and~$c^8_{QtQb}$ are suppressed by~$y_b/y_t \simeq 0.02$ compared to the corrections included in~(\ref{eq:STUapp}). As~a~result, we will not take into account effects in the $S$, $T$ and $U$ parameters associated to insertions of the scalar operators ${\cal O}^1_{QtQb}$ and~${\cal O}^8_{QtQb}$ in our~work. 

Like the $t \to b W$ process, the $b \to c \ell \nu$ transition also receives one-loop corrections from some of the third-generation four-quark operators introduced in~(\ref{eq:operators}). Identifying the renormalisation scale $\mu$ with the bottom-quark mass, we find the following expression for the SMEFT correction to the relevant partial decay rate
\beq \label{eq:DeltaGammaSL}
\Delta \Gamma ( b \to c \ell \nu ) = -\frac{\alpha}{24 \pi s_w^2} \frac{m_t^2}{m_W^2} \frac{v^2}{\Lambda^2} \left [ \ln \left ( \frac{m_t^2}{m_b^2} \right ) + \frac{1}{2} \right ] \left ( c_{QQ}^1 + \frac{4}{3} \hspace{0.25mm} c_{QQ}^8 \right ) \Gamma ( b \to c \ell \nu )_{\rm SM} \,, 
\eeq
with 
\beq \label{eq:GammaslSM}
\Gamma ( b \to c \ell \nu )_{\rm SM} = \frac{G_F^2 \hspace{0.25mm} |V_{cb}|^2 \hspace{0.25mm} m_b^5}{192 \pi^2} \left ( 1 - 8 \rho - 12 \rho^2 \ln \rho + 8 \rho^3 - \rho^4 \right ) \,, 
\eeq
the leading-order total semileptonic decay width in the SM~\cite{Manohar:2000dt}. Here $G_F$ denotes the Fermi constant as extracted from muon decay and $\rho = m_c^2/m_b^2$ with~$m_b$ and~$m_c$ the bottom- and charm-quark mass, respectively. One sees that relative to the SM the Wilson coefficients~$c_{QQ}^1$ and~$c_{QQ}^8$ contribute to the partial decay rate~$\Gamma ( b \to c \ell \nu)$ at order $y_t^2$ and that this contribution is logarithmically enhanced. The latter feature is well-known from the computation of the short-distance EW logarithmic corrections to the total semileptonic decay width in the SM~\cite{Sirlin:1981ie,Brod:2008ss,Gorbahn:2022rgl,Bigi:2023cbv}. Additionally, we note that contributions from the scalar operators~${\cal O}^1_{QtQb}$ and~${\cal O}^8_{QtQb}$ to $ \Gamma ( b \to c \ell \nu )$ first appear at a relative order of~$y_b/y_t$. Consequently, these effects are not included in our calculations. With this exception,~(\ref{eq:DeltaGammaSL}) represents an exact result.

The modification of the $b \to s Z$ penguin amplitude that arises from single insertions of the third-generation four-quark operators~(\ref{eq:operators}), is given by the following approximate formula: 
\beq \label{eq:DeltaCapp} 
\begin{split}
\Delta C & \simeq -\frac{\alpha}{1536 \pi s_w^2} \frac{m_t^4}{m_W^4} \frac{v^2}{\Lambda^2} \left [ \frac{63 - 8 \pi^2}{2} \left ( c_{QQ}^1 + \frac{4}{3} \hspace{0.25mm} c_{QQ}^8 \right ) \right . \\[2mm]
& \hspace{3.75cm} \left . - \frac{57 - 4 \pi^2}{2} \left ( c_{Qt}^1 + \frac{4}{3} \hspace{0.25mm} c_{Qt}^8 \right ) - 12 c_{tt}^1 \right ] \,.
\end{split}
\eeq
This result is derived by expanding~(\ref{eq:Cexact}) and~(\ref{eq:kappas}) up to leading power in the HTE. The renormalisation scale $\mu$ is thereby fixed to $m_t$. This choice eliminates logarithmic corrections of the form $\ln \left (m_t^2/m_W^2 \right )$ to leading order in the HTE. Notice that the SMEFT corrections included in~(\ref{eq:DeltaCapp}) are all parametrically enhanced by four powers of~$y_t$. This~scaling is anticipated based on the behaviour of the one-loop $b \to s Z$ contribution within the~SM, which is proportional to $y_t^2$ in the limit of an infinitely large top-quark mass. The~contributions to $\Delta C$ proportional to the Wilson coefficients $c_{Qb}^1$, $c_{Qb}^8$, $c_{QtQb}^1$ and $c_{QtQb}^8$ are all suppressed by powers of $y_b/y_t \simeq 0.02$ compared to~(\ref{eq:DeltaCapp}), making them phenomenologically insignificant. We~therefore will not include such effects in our article. Given the numerical significance of higher-order terms in the HTE, we will once more utilise the exact expression for~$\Delta C$ in our numerical studies. 

The SMEFT corrections to the $b \to s \ell^+ \ell^-$ box contribution that arise from single insertions of one of the third-generation four-quark operators~(\ref{eq:operators}), is approximated by 
\beq \label{eq:DeltaBapp}
\Delta B \simeq \frac{\alpha}{128 \pi s_w^2} \frac{m_t^2}{m_W^2} \frac{v^2}{\Lambda^2} \left ( c_{QQ}^1 + \frac{4}{3} \hspace{0.25mm} c_{QQ}^8 \right ) \,.
\eeq
This approximation has been derived from the exact result as given in~(\ref{eq:Bexact}) and~(\ref{eq:xis}) by identifying the renormalisation scale~$\mu$ with~$m_t$ and performing a HTE, keeping only the leading terms. Notice that to relative order $y_t^2$, only the Wilson coefficients $c_{QQ}^1$ and $c_{QQ}^8$ appear in~(\ref{eq:DeltaBapp}). This is an expected feature that follows from the structure of the relevant $\Delta F = 1$ one-loop box contribution in the SM, which scales as a constant in the limit of infinite top-quark mass, and the leading heavy-top behaviour of $\Delta \Gamma (t \to bW)$ as given in~(\ref{eq:tbWapp}). The contributions to $\Delta B$ proportional to $c_{Qb}^1$, $c_{Qb}^8$, $c_{QtQb}^1$ and $c_{QtQb}^8$ are relatively suppressed by at least a single power of $y_b/y_t \simeq 0.02$, and are neither included in~(\ref{eq:DeltaBapp}) nor in our numerics. Since higher-order terms in the HTE are numerically important, we will again employ the exact result for~$\Delta B$ in our phenomenological analysis. 

In the case of the off-shell $b \to s \gamma$ penguin amplitude, we find the following approximate expression 
\beq \label{eq:DeltaDapp}
\begin{split}
 \Delta D & \simeq -\frac{\alpha}{1296 \pi s_w^2} \frac{m_t^2}{m_W^2} \frac{v^2}{\Lambda^2} \left \{ \left [ \frac{45}{4} \hspace{0.125mm} \ln \left ( \frac{m_t^2}{m_W^2} \right ) - \frac{279 - 21 \pi^2}{2} \right ] \left ( c_{QQ}^1 + \frac{4}{3} \hspace{0.25mm} c_{QQ}^8 \right ) \right . \\[2mm]
 & \hspace{3.75cm} \left . + \left ( 14 + 3 \pi^2 \right ) \left ( c_{Qt}^1 + \frac{4}{3} \hspace{0.25mm} c_{Qt}^8 \right ) - 18 c_{tt}^1 \right \} \,,
\end{split} 
\eeq
for the SMEFT contribution that is due to single insertions of the third-generation four-quark operators~(\ref{eq:operators}). This approximation has been obtained from the exact result provided in~(\ref{eq:Dexact}) and~(\ref{eq:rhos}) by setting the renormalisation scale~$\mu$ to~$m_t$ and performing a HTE, retaining only the leading terms. We observe that, given our choice of renormalisation scale, only the operators ${\cal O}_{QQ}^1$ and ${\cal O}_{QQ}^8$ receive logarithmically enhanced corrections. However, these corrections are numerically subdominant, indicating the poor convergence of the HTE for the off-shell $b \to s \gamma$ penguin amplitude. Additionally, it is noteworthy that the overall correction $\Delta D$ scales as $y_t^2$. This expected behaviour arises from the structure of the relevant $\Delta F = 1$ photon penguin contribution in the SM, which scales as a constant in the limit of infinite top-quark mass. Contributions to the coefficient $\Delta D$ that are proportional to $c_{Qb}^1$, $c_{Qb}^8$, $c_{QtQb}^1$ and $c_{QtQb}^8$ are relatively suppressed by at least a factor of $y_b/y_t \simeq 0.02$, and are therefore neither included in~(\ref{eq:DeltaDapp}) nor our numerical~analysis.

The corrections to the $\Delta F = 2$ amplitude that arise from single insertions of the third-generation four-quark operators~(\ref{eq:operators}), take the following approximate form 
\beq \label{eq:DeltaFapp} 
\begin{split}
\Delta F & \simeq -\frac{\alpha}{192 \pi s_w^2} \frac{m_t^4}{m_W^4} \frac{v^2}{\Lambda^2} \left [ \left ( 9 - 2 \pi^2 \right ) c_{QQ}^1 + \frac{27 - 2 \pi^2}{3} \hspace{0.5mm} c_{QQ}^8 \right . \\[2mm] 
& \hspace{3.5cm} \left. - \frac{135 - 12 \pi^2}{4} \hspace{0.5mm} c_{Qt}^1 - \frac{105 - 12 \pi^2}{4} \hspace{0.5mm} c_{Qt}^8 - 3 c_{tt}^1 \right ] \,. 
\end{split} 
\eeq 
This result has been obtained from the exact expressions~(\ref{eq:Fexact}) and~(\ref{eq:omegas}), with the renormalisation scale $\mu$ set to $m_t$, and performing a HTE that keeps only the leading terms. Our choice of renormalisation scale effectively removes $\ln \left ( m_t^2/m_W^2 \right )$ terms in~(\ref{eq:DeltaFapp}). It is noteworthy that, akin to the case of~(\ref{eq:DeltaCapp}), the dominant SMEFT corrections to the $\Delta F = 2$ amplitude scale as $y_t^4$. Once again, this is an expected feature, originating from the $y_t^2$ dependence of the one-loop SM $\Delta F = 2$ amplitude. The contributions of the operators ${\cal O}_{Qb}^1$, ${\cal O}_{Qb}^8$, ${\cal O}_{Qt}^1$, ${\cal O}_{Qt}^8$, ${\cal O}_{QtQb}^1$ and ${\cal O}_{QtQb}^8$ to~(\ref{eq:DeltaFapp}) are suppressed relative to the terms included above by at least one power of $y_b/y_t \simeq 0.02$. They are therefore neglected in the subsequent analysis. Since the HTE of~$\Delta F$ shows poor convergence, we use the exact results in our numerical studies.

\section{Phenomenological analysis}
\label{sec:numerics}

As input for our numerical analysis we use~$G_F = 1.166379 \cdot 10^{-5} \, {\rm GeV}^{-2}$,~$m_W = 80.379 \, {\rm GeV}$ and~$m_Z = 91.1876 \, {\rm GeV}$~\cite{ParticleDataGroup:2022pth}. The Higgs VEV is calculated from the Fermi constant using~$v = 2^{-1/4} \hspace{0.25mm} G_F^{-1/2} = 246.22 \, {\rm GeV}$, while the value of the electromagnetic coupling and the weak mixing angle is derived as~$\alpha =\sqrt{2} \hspace{0.25mm} G_F \hspace{0.25mm} m_W^2 s_w^2/\pi = 1/132.184$ and~$s_w^2 = 1-{m_W^2}/{m_Z^2} = 0.2230$. For the mass of the top quark we employ the OS value~$m_t = 172.5 \, {\rm GeV}$, while we use a Higgs mass of~$m_h = 125 \, {\rm GeV}$ and~$|V_{tb}| = 0.999142$. 

Using these input parameters, we find the following numerical results in the case of the~$Z$-pole observables: 
\beq \label{eq:Zpolenumerical} 
\begin{split}
\frac{\Gamma_Z}{\Gamma_Z^{\rm SM}} & = 1 + \left ( 3.83\hspace{0.25mm} c_{QQ}^1 - 0.09\hspace{0.25mm} c_{QQ}^8 - 4.58\hspace{0.25mm} c_{Qt}^1 - 0.59\hspace{0.25mm} c_{Qb}^1 + 0.80\hspace{0.25mm} c_{tb}^1 \right ) \cdot 10^{-4} \,, \\[2mm]
\frac{R_\ell}{R_\ell^{\rm SM}} & = 1 + \left ( 5.49\hspace{0.25mm} c_{QQ}^1 - 0.13\hspace{0.25mm} c_{QQ}^8 - 6.57\hspace{0.25mm} c_{Qt}^1 - 0.85\hspace{0.25mm} c_{Qb}^1 + 1.15\hspace{0.25mm} c_{tb}^1 \right ) \cdot 10^{-4} \,, \\[2mm]
\frac{R_b}{R_b^{\rm SM}} & = 1 + \left ( 25.43\hspace{0.25mm} c_{QQ}^1 - 0.60\hspace{0.25mm} c_{QQ}^8 - 30.42\hspace{0.25mm} c_{Qt}^1 - 3.93\hspace{0.25mm} c_{Qb}^1 + 5.31\hspace{0.25mm} c_{tb}^1 \right ) \cdot 10^{-4} \,, \\[2mm]
\frac{R_c}{R_c^{\rm SM}} & = 1 - \left ( 5.49\hspace{0.25mm} c_{QQ}^1 - 0.13\hspace{0.25mm} c_{QQ}^8 - 6.57\hspace{0.25mm} c_{Qt}^1 - 0.85\hspace{0.25mm} c_{Qb}^1 + 1.15\hspace{0.25mm} c_{tb}^1 \right ) \cdot 10^{-4} \,, \\[2mm]
\frac{A_{\rm FB}^{b}}{A_{\rm FB}^{b, {\rm SM}}} & = 1 + \left ( 2.37\hspace{0.25mm} c_{QQ}^1 - 0.06\hspace{0.25mm} c_{QQ}^8 - 2.83\hspace{0.25mm} c_{Qt}^1 + 13.85\hspace{0.25mm} c_{Qb}^1 - 16.22\hspace{0.25mm} c_{tb}^1 \right ) \cdot 10^{-4} \,, \\[2mm]
\frac{A_b}{A_b^{\rm SM}} & = 1 + \left ( 1.63\hspace{0.25mm} c_{QQ}^1 - 0.04\hspace{0.25mm} c_{QQ}^8 - 1.95\hspace{0.25mm} c_{Qt}^1 + 9.53\hspace{0.25mm} c_{Qb}^1 - 11.15\hspace{0.25mm} c_{tb}^1 \right ) \cdot 10^{-4} \,, 
\end{split} 
\eeq
where~$\ell = e, \mu, \tau$. Here, we have set the renormalisation scale~$\mu$ to~$m_Z$ and fixed the new-physics scale to~$\Lambda = 1 \, {\rm TeV}$. The SM values of the observables entering~(\ref{eq:Zpolenumerical}) are given by~$\Gamma_Z^{\rm SM} = 2.49411 \, {\rm GeV}$,~$R_e^{\rm SM} = 20.736$,~$R_\mu^{\rm SM} = 20.736$,~$R_\tau^{\rm SM} = 20.781$,~$R_b^{\rm SM} = 0.21582$,~$R_c^{\rm SM} = 0.17221$,~$A_{\rm FB}^{b, {\rm SM}} = 0.1029$ and~$A_b^{\rm SM} = 0.9347$~\cite{ALEPH:2005ab,Workman:2022ynf}. Two features of the above formulas are worth to note. First, the observables~$R_b$,~$A_{\rm FB}^{b}$ and~$A_b$ are more sensitive to the Wilson coefficients of the third-generation four-quark operators~(\ref{eq:operators}) than~$\Gamma_Z$,~$R_\ell$ and~$R_c$. Second, the ratio of the~$Z$-boson decay width into bottom quarks and the total hadronic width~$R_b$ probe mostly~$c_{QQ}^1$ and~$c_{Qt}^1$, while the bottom-quark asymmetries~$A_{\rm FB}^{b}$ and~$A_b$ have a pronounced sensitivity to~$c_{Qb}^1$ and~$c_{tb}^1$. This feature is reminiscent of the SM where~$R_b$ ($A_{\rm FB}^{b}$ and~$A_b$) has (have) a strong sensitivity to left-handed~(right-handed) currents --- see for instance~\cite{Field:1997gz,Chanowitz:2001bv}. Also notice that the numerical coefficients in~(\ref{eq:Zpolenumerical}) depend in a non-negligible way on the choice of the top-quark mass and the weak mixing angle. In~particular, choosing an~$\overline{\rm MS}$ mass instead of an OS mass for~$m_t$ would generically lead to coefficients smaller in magnitude in the above predictions. Given that the leading terms in~(\ref{eq:Zbbapp}) scale as~$m_t^2/m_Z^2 \hspace{0.25mm} \ln \left ( m_t^2/m_Z^2 \right )$ or~$m_t^2/m_Z^2$, this is an expected feature. We add that the predictions in~(\ref{eq:Zpolenumerical}) agree well with the numerical results for the SMEFT expressions for the relevant~$Z$-pole observables published in~\cite{Dawson:2022bxd} and provided in machine readable format at~\cite{SDPPG}. This agreement serves as an important cross-check of our computations. In~this context, it is also worth recalling that the results in~\cite{Dawson:2022bxd} were obtained in the Warsaw operator basis, whereas our results~(\ref{eq:Zpolenumerical}) correspond to the choice~(\ref{eq:operators}) of third-generation four-quark operators. Details on the change of renormalisation scheme that connects the LHCTopWG to the Warsaw~operator basis can be found in Appendix~\ref{app:fierz}.

Setting the renormalisation scale~$\mu$ to~$m_t$ and fixing the new-physics scale to~$\Lambda = 1 \, {\rm TeV}$, we obtain for the three observables relevant in the case of the~$t \to b W$ decay the expressions
\beq \label{eq:tbWnumerical} 
\begin{split}
\frac{\Gamma_t}{\Gamma_t^{\rm SM}} & = 1 - \left ( 1.75\hspace{0.25mm} c_{QQ}^1 + 2.34\hspace{0.25mm} c_{QQ}^8 + 1.91\hspace{0.25mm} c_{Qt}^1 + 2.55\hspace{0.25mm} c_{Qt}^8 \right ) \cdot 10^{-4} \,, \\[2mm]
\frac{F_L}{F_L^{\rm SM}} & = 1 + \left ( 0.90 \hspace{0.25mm} c_{Qt}^1 + 1.21 \hspace{0.25mm} c_{Qt}^8 \right ) \cdot 10^{-4}\,, \\[2mm]
\frac{F_-}{F_-^{\rm SM}} & = 1 - \left ( 2.02 \hspace{0.25mm} c_{Qt}^1+ 2.69 \hspace{0.25mm} c_{Qt}^8 \right ) \cdot 10^{-4} \,.
\end{split} 
\eeq
The SM values of the total decay width of the top quark and the helicity fractions of the~$W$ boson are given by~$\Gamma_t^{\rm SM} = 1.3216 \, {\rm GeV}$,~$F_L^{\rm SM} = 0.689$ and~$F_-^{\rm SM} = 0.309$~\cite{Czarnecki:2010gb,Gao:2012ja}. We~emphasise that for the same input the formulas~(\ref{eq:tbWnumerical}) agree with the results obtained in~\cite{Boughezal:2019xpp} after fixing obvious typographical mistakes in (38), (39)~and~(40) of the arXiv~version~2 of that paper.\footnote{In (38) an overall factor of $g/\pi$ is missing and the overall signs in (39) and (40) should both be reversed. Here~$g$~denotes the $SU(2)_L$ gauge coupling.} One also has to bear in mind that~\cite{Boughezal:2019xpp} employs the Warsaw operator basis, while the numerical results~(\ref{eq:tbWnumerical}) correspond to the choice~(\ref{eq:operators}) of third-generation four-quark operators. The comparison represents an independent cross-check~of our calculations.

For~$\mu = m_Z$ and~$\Lambda = 1 \, {\rm TeV}$ our numerical results for the oblique parameters read 
\bea \label{eq:STUnumerical} 
\begin{split}
\Delta S & = \left ( -12.4 \hspace{0.25mm} c_{QQ}^1 -4.0 \hspace{0.25mm} c_{QQ}^8 + 19.4 \hspace{0.25mm} c_{Qt}^1 + 5.1 \hspace{0.25mm} c_{Qt}^8 + 2.5 \hspace{0.25mm} c_{Qb}^1 -21.8 \hspace{0.25mm}c_{tt}^1 -2.5 \hspace{0.25mm}c_{tb}^1 \right ) \cdot 10^{-4} \,, \\[2mm] 
\Delta T & = \left ( -74.1 \hspace{0.25mm} c_{QQ}^1 -38.4 \hspace{0.25mm} c_{QQ}^8 +90.6\hspace{0.25mm} c_{Qt}^1 -168.2 \hspace{0.25mm}c_{tt}^1 \right ) \cdot 10^{-4} \,, \\[2mm]
\Delta U & = \left ( 19.2 \hspace{0.25mm} c_{QQ}^1 +4.6 \hspace{0.25mm} c_{QQ}^8 -17.2 \hspace{0.25mm} c_{Qt}^1 + 1.2 \hspace{0.25mm} c_{Qt}^8 +7.3 \hspace{0.25mm} c_{tt}^1 \right ) \cdot 10^{-4} \,.
\end{split}
\eea
The size of the numerical coefficients imply that~$\Delta T$ is the most sensitive oblique parameter to probe BSM physics of the form~(\ref{eq:operators}) with~$\Delta S$ generically providing weaker and~$\Delta U$ providing the least important constraints. Notice that in contrast to~(\ref{eq:Zpolenumerical}) and~(\ref{eq:tbWnumerical}),~$\Delta S$, $\Delta T$ and $\Delta U$ allow to test the purely right-handed four-top quark operator proportional to the Wilson coefficient~$c_{tt}^1$. We furthermore note that the oblique parameters depend on~$c_{QQ}^1$,~$c_{QQ}^8$,~$c_{Qt}^1$,~$c_{Qb}^1$ and~$c_{tb}^1$ in a different way than the~$Z$-pole observables~(\ref{eq:Zpolenumerical}). We~expect that this feature will help resolve certain flat directions in global SMEFT analyses, which would otherwise remain untested if only one set of observables were included in the Wilson coefficient fit.

In order to constrain the Wilson coefficients $c_{QQ}^1$ and $c_{QQ}^8$ from the $b \to c \ell \nu$ decay, one can consider the ratio 
\beq \label{eq:Rlnu}
R_{ \ell \nu} = \frac{{\rm Br} \left ( B \to X_c \ell \nu \right )}{{\rm Br} \left ( B \to X_c \ell \nu \right )_{\rm SM}} = 1 - \left ( 11.7 \hspace{0.25mm} c_{QQ}^1 + 15.6 \hspace{0.25mm} c_{QQ}^8 \right ) \cdot 10^{-4} \ \,.
\eeq
This expression has been obtained from~(\ref{eq:DeltaGammaSL}) and~(\ref{eq:GammaslSM}) using the state-of-the-art prediction~${\rm Br} \left ( B \to X_c \ell \nu \right )_{\rm SM} \simeq 10.66\%$~\cite{Fael:2020tow,Bordone:2021oof} for the inclusive semileptonic branching ratio in the SM, fixing the new-physics scale to $\Lambda = 1 \, {\rm TeV}$. Notice that~(\ref{eq:Rlnu}) depends on the Wilson coefficients of the third-generation four-quark vector operators~(\ref{eq:operators}) only through the combination $c_{QQ}^1 + 4/3 \hspace{0.5mm} c_{QQ}^8$. The corresponding flat direction therefore cannot be resolved by precision measurements of the inclusive semileptonic branching ratio or the moments of the kinematic distributions in~$B \to X_c \ell \nu$. 

To parameterise the two-loop SMEFT contributions to the $b \to s \ell^+ \ell^-$ processes, we employ the LEFT Hamiltonian 
\beq \label{eq:Heffbsll}
{\cal H}_{\rm eff}^{\Delta B = 1} = -\frac{4 G_F}{\sqrt{2}} \hspace{0.25mm} V_{ts}^\ast V_{tb} \hspace{0.25mm} \big ( C_9 \hspace{0.25mm} Q_9 + C_{10} \hspace{0.25mm} Q_{10} \big ) \,, 
\eeq
where
\beq \label{eq:Q9Q10} 
Q_9 = \frac{e^2}{16 \hspace{0.125mm} \pi^2} \left ( \bar s \hspace{0.25mm} \gamma_\mu P_L \hspace{0.25mm} b \right ) \left ( \bar \ell \hspace{0.25mm} \gamma^\mu \hspace{0.25mm} \ell \right ) \,, \qquad 
Q_{10} = \frac{e^2}{16 \hspace{0.125mm} \pi^2} \left ( \bar s \hspace{0.25mm} \gamma_\mu P_L \hspace{0.25mm} b \right ) \left ( \bar \ell \hspace{0.25mm} \gamma^\mu \gamma_5 \hspace{0.25mm} \ell \right ) \,,
\eeq
are the vector and axial-vector semi-leptonic operator, respectively, $P_L$ projects onto left-handed fermionic fields and $e$ is the QED coupling constant. The SM values of the Wilson coefficients are $C_9^{\rm SM} \simeq 4.3$ and $C_{10}^{\rm SM} \simeq -4.2$~\cite{Bobeth:2003at,Huber:2005ig,Bobeth:2013uxa,Hermann:2013kca,Bobeth:2013tba,Beneke:2017vpq}. In terms of the $Z$-penguin, box and photon-penguin contributions $\Delta C$, $\Delta B$ and $\Delta D$, the two-loop SMEFT corrections to the Wilson coefficients of the operators~(\ref{eq:Q9Q10}) are given by 
\beq \label{eq:DC9DC10analytic}
\Delta C_9 = \frac{1}{s_w^2} \Big[ \left ( 1-4s_w^2 \right ) \Delta C - \Delta B - s_w^2 \Delta D \Big ] \,, \qquad 
\Delta C_{10} = -\frac{1}{s_w^2} \left ( \Delta C - \Delta B \right ) \,.
\eeq
Notice that in contrast to $\Delta C$, $\Delta B$ and $\Delta D$, the linear combinations $\Delta C_9$ and $\Delta C_{10}$ are gauge independent. Setting the renormalisation scale $\mu$ equal to $m_t$ and fixing the new-physics scale to $\Lambda = 1 \, {\rm TeV}$, we obtain the following numerical results 
\beq \label{eq:DC9DC10numeric} 
\begin{split}
\Delta C_{9} & = \left ( 0.16 \hspace{0.25mm} c_{QQ}^1 + 0.21 \hspace{0.25mm} c_{QQ}^8 + 1.49 \hspace{0.25mm} c_{Qt}^1 + 1.99 \hspace{0.25mm} c_{Qt}^8 + 0.21 \hspace{0.25mm} c_{tt}^1 \right ) \cdot 10^{-4} \,, \\[2mm]
\Delta C_{10} & = -\left ( 4.77 \hspace{0.25mm} c_{QQ}^1 + 6.35 \hspace{0.25mm} c_{QQ}^8 + 6.40 \hspace{0.25mm} c_{Qt}^1 + 8.53 \hspace{0.25mm} c_{Qt}^8 + 9.53 \hspace{0.25mm} c_{tt}^1 \right ) \cdot 10^{-4} \,.
\end{split}
\eeq
One observes that the numerical coefficients present in $\Delta C_9$ are approximately one order of magnitude smaller compared to those in $\Delta C_{10}$. This feature can be understood from~(\ref{eq:DeltaCapp}),~(\ref{eq:DeltaBapp}) and~(\ref{eq:DeltaDapp}), which shows that $\Delta C$ is parametrically enhanced by a factor of $m_t^4/m_W^4 \simeq 20$ with respect to $\Delta B$ and $\Delta D$. However, the numerically dominant $Z$-penguin contribution enters $\Delta C_9$ with a relative factor of $1 - 4 s_w^2 \simeq 0.1$ compared to~$\Delta C_{10}$. Considering only $\Delta C$, one therefore expects $\Delta C_{10}/\Delta C_{9} \simeq 10$, which approximately matches the ratio of the numerical results~(\ref{eq:DC9DC10numeric}). 

A clean probe of the Wilson coefficient $C_{10}$ is provided by the purely leptonic decay~$B_s \to \mu^+ \mu^-$. To linear order in the correction $\Delta C_{10}$, one can write 
\beq \label{eq:Rmumu}
R_{\mu^+\mu^-} = \frac{{\rm Br} \left ( B_s \to \mu^+ \mu^ - \right )}{{\rm Br} \left ( B_s \to \mu^+ \mu^ - \right )_{\rm SM}} = 1 - 0.24 \hspace{0.25mm} \Delta C_{10} \,, 
\eeq
where we have used the SM value ${\rm Br} \left ( B_s \to \mu^+ \mu^ - \right )_{\rm SM} \simeq 3.66 \cdot 10^{-9}$~\cite{Beneke:2019slt} as normalisation. Constraints on both $\Delta C_{9}$ and $\Delta C_{10}$ can also be derived from global analyses of $b \to s \ell^+ \ell^-$ data~---~see~\cite{Geng:2021nhg,Hurth:2021nsi,Alguero:2021anc,Ciuchini:2021smi,Gubernari:2022hxn,Greljo:2022jac} for recent examples of such global fits. Since the SM predictions for most $b \to s \ell^+ \ell^-$ branching ratios and angular observables can potentially be affected by large hadronic uncertainties, the constraints on the shifts $\Delta C_{9}$ and $\Delta C_{10}$ derived from global $b \to s \ell^+ \ell^-$ analyses are less straightforward to interpret. Consequently, the limits on the Wilson coefficients $c_{QQ}^1$, $c_{QQ}^8$, $c_{Qt}^1$, $c_{Qt}^8$ and $c_{tt}^1$ that follow from the various $b \to s \ell^+ \ell^-$ observables are theoretically less robust than the bounds that derive from a combination of~(\ref{eq:DC9DC10numeric}) and~(\ref{eq:Rmumu}).

In the case of $B_s$--$\bar B_s$ mixing, we employ the following LEFT Hamiltonian
\beq \label{eq:Heffmix}
{\cal H}_{\rm eff}^{\Delta B = 2} = \frac{G_F^2 \hspace{0.25mm} m_W^2}{4 \pi^2} \hspace{0.25mm} \left ( V_{ts}^\ast V_{tb} \right )^2 \hspace{0.25mm} C_Q \hspace{0.25mm} Q + {\rm h.c.} \,, 
\eeq
with 
\beq \label{eq:Qmix}
Q = \left ( \bar s \gamma_\mu P_L b \right ) \left ( \bar s \gamma^\mu P_L b \right ) \,.
\eeq
In the SM, the Wilson coefficient $C_Q$ is given at the scale $\mu = m_t$ by $C_Q^{\rm SM} = F_{\rm SM} \simeq 2.3$~\cite{Buras:1990fn}. At the same scale and using $\Lambda = 1 \, {\rm TeV}$, the two-loop SMEFT corrections to the Wilson coefficient of the operator~(\ref{eq:Qmix}) are given by 
\beq \label{eq:DF}
\Delta F = \left ( 26.3 \hspace{0.25mm} c_{QQ}^1 - 0.3 \hspace{0.25mm} c_{QQ}^8 + 2.4 \hspace{0.25mm} c_{Qt}^1 - 3.6 \hspace{0.25mm} c_{Qt}^8 + 4.3 \hspace{0.25mm} c_{tt}^1 \right ) \cdot 10^{-4} \,. 
\eeq
Notice that the numerical coefficients present in $\Delta F$ and $\Delta C_{10}$ are roughly of the same size, while they are smaller than those present in $\Delta T$. A observable that is directly sensitive to~(\ref{eq:DF}) is the ratio: 
\beq \label{eq:Delta}
R_{\Delta} = \frac{\Delta M_{s}}{\Delta M_{s}^{\rm SM}} = 1 + 0.43 \hspace{0.25mm} \Delta F \,.
\eeq
Above we have retained only the terms linear in $\Delta F$. The central value of the latest SM prediction for the mass difference in $B_s$--$\bar{B}_s$ mixing is $\Delta M_{s}^{\rm SM} \simeq 18.23 \, {\rm ps}^{-1}$~\cite{Albrecht:2024oyn}.

\section{Conclusions}
\label{sec:conclusions}

In this article, we have initiated a comprehensive study of indirect constraints on third-generation four-quark operators that arise in the SMEFT at the level of dimension six. Our~work is motivated by the observation that such operators are only very weakly constrained by the existing~$t \bar t t \bar t$ and~$t \bar b t \bar b$ production measurements performed during Run~II of the~LHC~\cite{Brivio:2019ius,Ellis:2020unq,Ethier:2021bye}. We have first calculated the matching corrections to the decay processes~$t \to bW$ and~$Z \to b \bar b$, which receive contributions from four-quark operators involving only heavy fields at the one-loop level. Our results agree with those obtained in~\cite{Boughezal:2019xpp} and~\cite{Dawson:2022bxd}, respectively. We~have~then extended our analysis to the two-loop level and computed the dominant matching corrections to the Peskin-Takeuchi parameters~$S$,~$T$ and~$U$ as well as the $b \to s\ell^+ \ell^-$ transition and the~$B_s$--$\bar B_s$ mixing amplitude. As a by-product, we have also found the one-loop matching corrections to the $b \to c \ell \nu$ process. All computations are formally at NLO in EW interactions. Regarding non-Higgs observables, the processes examined in our paper encompass all topologies with at least a quadratic sensitivity to the top-quark Yukawa coupling~$y_t$. See~for~example~\cite{UH,Giudice:2015toa}.

To leading power in the bottom-quark mass or Yukawa coupling, the considered observables are sensitive to~$c_{QQ}^1$,~$c_{QQ}^8$,~$c_{Qt}^1$ ,~$c_{Qt}^8$ ,~$c^1_{Qb}$,~$c_{tb}^1$ and~$c_{tt}^1$, while contributions proportional to the Wilson coefficients~$c^8_{Qb}$,~$c^8_{tb}$,~$c^1_{QtQb}$ and~$c^8_{QtQb}$ enter at subleading order in~$y_b$. This~feature renders the latter effects irrelevant in practice and we have therefore neglected power-suppressed terms throughout our work. The sensitivity to the individual Wilson coefficients is however observable-dependent. While the ratio of the~$Z$-boson decay width into bottom quarks and the total hadronic width~$R_b$ test mostly~$c_{QQ}^1$ and~$c_{Qt}^1$, the bottom-quark asymmetries~$A_{\rm FB}^b$ and~$A_b$ depend mainly on~$c_{Qb}^1$ and~$c_{tb}^1$. The oblique parameters~$S$ and~$T$ instead represent sensitive probes of the Wilson coefficients~$c_{QQ}^1$,~$c_{QQ}^8$~$c_{Qt}^1$ and~$c_{tt}^1$ but to a lesser extent also allow to constrain~$c_{Qt}^8$,~$c_{Qb}^1$ and~$c_{tb}^1$. It furthermore turns out that the $U$~parameter and the three considered~$t \to bW$ observables,~i.e.~the total top decay width~$\Gamma_t$ and the~$W$-boson helicity fractions~$F_L$ and~$F_-$, only have a limited constraining power in practice. In the case of flavour physics, we found that for down-alignment only the Wilson coefficients $c_{QQ}^1$ and $c_{QQ}^8$ can be probed at the one-loop level through precision measurements of $b \to c \ell \nu$ processes. However, the two-loop matching corrections to both the $b \to s \ell^+ \ell^-$ and the $B_s$--$\bar{B}_s$ mixing amplitude depend on $c_{QQ}^1$,~$c_{QQ}^8$,~$c_{Qt}^1$,~$c_{Qt}^8$ and~$c_{tt}^1$. Specifically, the $Z$-penguin and the $B_s$--$\bar{B}_s$ mixing amplitude receive $y_t^4$-enhanced corrections, which can be probed in a theoretically clean manner through measurements of the $B_s \to \mu^+ \mu^-$ branching ratio and the mass difference $\Delta M_s$ in neutral $B_s$-meson mixing, respectively. The scalar Wilson coefficients~$c^1_{QtQb}$ and~$c^8_{QtQb}$ remain unconstrained by the observables studied in this article but can be tested in Higgs physics~\cite{Gauld:2015lmb,Alasfar:2022zyr,DiNoi:2023ygk,Heinrich:2023rsd}. 

In order to obtain the most stringent bounds on the full set of third-generation four-quark operators~(\ref{eq:operators}), one should combine top-quark and Higgs data, along with information from EWPOs and flavour physics. The~analytic results provided in the publications~\cite{Gauld:2015lmb,Alasfar:2022zyr,DiNoi:2023ygk,Heinrich:2023rsd} for Higgs production and decay, as well as those presented here for the $Z \to b \bar b$ and $t \to b W$ observables (see also~\cite{Boughezal:2019xpp,Dawson:2022bxd}), and the oblique parameters and relevant FCNC processes should facilitate such comprehensive SMEFT analyses. Two points merit further mention. First, the studied indirect constraints arise at first order in the $1/\Lambda^2$ expansion and are associated with momentum scales below $m_t$. In contrast, the limits from relevant top-quark processes typically stem from contributions of order $1/\Lambda^4$~\cite{Ethier:2021bye,Hartland:2019bjb,Degrande:2024mbg,Celada:2024mcf} and involve a momentum transfer of at least $2 \hspace{0.125mm} m_t$. From the perspective of EFT applicability, indirect constraints therefore seem more robust than the direct bounds obtained from top-quark production processes. Second, many of the observables calculated in this article exhibit logarithmically enhanced corrections that depend on the new-physics~scale~$\Lambda$. These~contributions are linked to the RG flow in the SMEFT, thereby increasing the sensitivity to TeV-scale BSM physics (see also~\cite{Allwicher:2023aql,Garosi:2023yxg,Allwicher:2023shc,Stefanek:2024kds} for related discussions). 

Given that the collider constraints from top-quark production can now be calculated automatically at the one-loop level in QCD~\cite{Degrande:2020evl,Degrande:2024mbg} and in view of the remarkable progress in SMEFT fits (cf.~\cite{Brivio:2019ius,Ellis:2020unq,Ethier:2021bye,ATL-PHYS-PUB-2022-037} for the latest results of this~kind), obtaining better and more robust determinations of the Wilson coefficients of third-generation four-quark operators just seems to be a question of time and effort. First steps in this direction have been undertaken for instance in the works~\cite{Bruggisser:2022rhb,Garosi:2023yxg,Allwicher:2023shc,Bartocci:2023nvp,Stefanek:2024kds}. Our own comprehensive analysis of the constraints on third-generation four-quark operators, derived from fits to $Z$-pole data, oblique corrections and precision flavour observables, will be presented elsewhere~\cite{inprep}.

\acknowledgments We thank Paolo Gambino for collaboration in the initial stage of this project. This work has been triggered by the talk of Eleni Vryonidou at the Workshop on Tools for High Precision LHC Simulations at Ringberg Castle~\cite{EV}. UH and LS thank the organisers for an invitation to this exciting workshop at this beautiful location. We also acknowledge helpful and prompt communications with Radja~Boughezal and Ramona Gr{\"o}ber concerning their works~\cite{Boughezal:2019xpp} and~\cite{Alasfar:2022zyr}, respectively. A big thank you also goes to Ben~Stefanek for interesting email exchanges and making us aware of~\cite{Allwicher:2023aql,Garosi:2023yxg,Allwicher:2023shc,Stefanek:2024kds} and to Peter~Stangl for encouraging us to work harder. We are also thankful to an anonymous referee for raising thoughtful and pertinent questions that enabled us to improve our original manuscript. LS thanks the International Max Planck Research School~(IMPRS) on ``Elementary Particle Physics'' as well as the Collaborative~Research~Center SFB1258 for~support.

\appendix

\newpage 

\section{Exact formulas}
\label{app:exact}

In this~appendix, we present the exact one- and two-loop results for all the observables considered in this article. The given expressions correspond to the choice~(\ref{eq:operators}) of third-generation four-quark operators. For the convenience of the readers, these materials are also available in electronic format at~\cite{analyticresults}. This webpage additionally contains the analytic expressions for the exact one- and two-loop results in the Warsaw operator basis. 

For the SMEFT correction to the~$Z \to b \bar b$ decay rate we obtain the one-loop result 
\beq \label{eq:DGZbb}
\begin{split}
\Delta \Gamma (Z \to b \bar b) = \frac{\alpha^2 m_Z}{288 \pi c_w^4 s_w^4} \frac{v^2}{\Lambda^2} \, \sum_{i=\phantom{}_{QQ}^1, \phantom{}_{QQ}^8, \phantom{}_{Qt}^1, \phantom{}_{Qb}^1, \phantom{}_{tb}^1} \alpha_i \hspace{0.25mm} c_i \,, 
\end{split} 
\eeq
with 
\bea \label{eq:alphas} 
\begin{split}
\alpha_{QQ}^{1} & = -\frac{1}{9} \left(2 s_w^2-3\right) \bigg \{ -36 \left(8 s_w^2+3\right) x_{t/Z} -98 s_w^2+48 + 9 \left(4 s_w^2+9 x_{t/Z}-3\right) \ln x_{t/Z} \\[1mm] 
& \phantom{xxxx} \hspace{0cm} + 36 \left[ s_w^2 \left (8 x_{t/Z}+4 \right )+3 \left (x_{t/Z}-1 \right )\right] f_1(x_{t/Z}) -3 \left(4 s_w^2+27 x_{t/Z}+3\right) \ln x_{\mu/Z} \bigg \} \,, \\[2mm]
\alpha_{QQ}^{8} & = -\frac{8}{27} \left(2 s_w^2 - 3 \right)^2 \left ( 1 + \frac{3}{2} \ln x_{\mu/Z} \right ) \,, \\[2mm] 
\alpha_{Qt}^{1} & = -\frac{1}{3} \left(2 s_w^2-3\right) \bigg \{ -4 s_w^2 \left (24 x_{t/Z} +11 \right )+108 x_{t/Z} +3 \left(4 s_w^2-9 x_{t/Z}\right) \ln x_{t/Z} \\[1mm] 
& \phantom{xxxx} + 12 \left[ s_w^2 (8 x_{t/Z} +4)-9 x_{t/Z}\right] f_1(x_{t/Z}) - 3 \left(4 s_w^2 - 9 x_{t/Z} \right) \ln x_{\mu/Z} \bigg \} \,, \\[2mm] 
\alpha_{Qb}^{1} & = 2\hspace{0.25mm} s_w^2 \hspace{0.5mm}\bigg \{ 8 s_w^2 (4 x_{t/Z} +1) +12 x_{t/Z} -1 - \left( 4 s_w^2+ 9 x_{t/Z}-3\right) \ln x_{t/Z} \\[1mm] 
& \phantom{xxxx} - 4 \left[ s_w^2 \left (8 x_{t/Z}+4 \right )+3 \left (x_{t/Z}-1 \right )\right] f_1(x_{t/Z}) + 3 \left( 3 x_{t/Z} +1\right) \ln x_{\mu/Z} \bigg \} \,, \\[2mm]
\alpha_{tb}^{1} & = \frac{2}{3} \hspace{0.25mm} s_w^2 \hspace{0.5mm} \bigg \{ s_w^2 \left (96 x_{t/Z} +44 \right ) -108 x_{t/Z} - 3 \left(4 s_w^2 - 9 x_{t/Z}\right) \ln x_{t/Z} \\[1mm] 
& \phantom{xxxx} -12 \left[ s_w^2 \left (8 x_{t/Z} +4 \right )-9 x_{t/Z} \right] f_1(x_{t/Z}) + 3 \left(4 s_w^2-9 x_{t/Z} \right) \ln x_{\mu/Z} \bigg \} \,. 
\end{split}
\eea
In the case of the forward-backward asymmetry for bottom quarks we instead find 
\beq \label{eq:DAFBbb}
\begin{split}
\Delta A_{\rm FB}^ {b} = \frac{9 \left(8 s_w^4-14 s_w^2+3\right) \alpha}{16 \pi c_w^2 \left(8 s_w^4-12 s_w^2+9\right)^2 \left(8 s_w^4-4 s_w^2+1\right)} \frac{v^2}{\Lambda^2} \, \sum_{i=\phantom{}_{QQ}^1, \phantom{}_{QQ}^8, \phantom{}_{Qt}^1, \phantom{}_{Qb}^1, \phantom{}_{tb}^1} \beta_i \hspace{0.25mm} c_i \,, 
\end{split} 
\eeq
with 
\bea \label{eq:betas} 
\begin{split}
\beta_{QQ}^{1} & = \frac{8}{27} \hspace{0.25mm} s_w^2 \hspace{0.5mm} \bigg \{ -36 \left(8 s_w^2+3\right) x_{t/Z} -98 s_w^2+48 + 9 \left(4 s_w^2+9 x_{t/Z}-3\right) \ln x_{t/Z} \\[1mm] 
& \phantom{xxxxx} \hspace{0cm} + 36 \left[ s_w^2 \left (8 x_{t/Z}+4 \right )+3 \left (x_{t/Z}-1 \right )\right] f_1(x_{t/Z}) \\[1mm]
& \phantom{xxxxx} -3 \left(4 s_w^2+27 x_{t/Z} +3\right) \ln x_{\mu/Z} \bigg \} \,, \\[2mm]
\beta_{QQ}^{8} & = \frac{64}{81} \hspace{0.25mm} s_w^2 \left(2 s_w^2-3 \right) \left ( 1 + \frac{3}{2} \ln x_{\mu/Z} \right ) \,, \\[2mm] 
\beta_{Qt}^{1} & = \frac{8}{9} \hspace{0.25mm} s_w^2 \hspace{0.5mm} \bigg \{ -4 s_w^2 \left (24 x_{t/Z} +11 \right )+108 x_{t/Z} +3 \left(4 s_w^2-9 x_{t/Z}\right) \ln x_{t/Z} \\[1mm] 
& \phantom{xxxxx} \hspace{0cm} + 12 \left[ s_w^2 (8 x_{t/Z} +4)-9 x_{t/Z}\right] f_1(x_{t/Z}) - 3 \left(4 s_w^2 - 9 x_{t/Z} \right) \ln x_{\mu/Z} \bigg \} \,, \\[2mm] 
\beta_{Qb}^{1} & = -\frac{4}{9} \hspace{0.25mm} \bigg \{ -8 s_w^4 \left (24 x_{t/Z} +11 \right )+6 s_w^2 \left (36 x_{t/Z} +23 \right )+108 x_{t/Z}-54 \\[1mm] 
& \phantom{xxxxx} \hspace{0cm} + 3 \left(2 s_w^2-3\right) \left(4 s_w^2+9 x_{t/Z}-3\right) \ln x_{t/Z} \\[1mm]
& \phantom{xxxxx} \hspace{0cm} + 12 \left(2 s_w^2-3\right) \left[ s_w^2 (8 x_{t/Z}+4)+3 \left (x_{t/Z}-1 \right ) \right ] f_1 (x_{t/Z}) \\[1mm]
& \phantom{xxxxx} \hspace{0cm} -3 \left[ 8 s_w^4 + 6 s_w^2 \left (3 x_{t/Z} -1 \right )-27 x_{t/Z} \right] \ln x_{\mu/Z} \bigg \} \,, \\[2mm]
\beta_{tb}^{1} & = -\frac{4}{9} \left ( 2 s_w^2-3 \right ) \hspace{0.5mm} \bigg \{ -4 s_w^2 \left (24 x_{t/Z} + 11 \right ) +108 x_{t/Z} +3 \left(4 s_w^2 - 9 x_{t/Z} \right) \ln x_{t/Z} \\[1mm] 
& \phantom{xxxxx} \hspace{0cm} +12 \left[ s_w^2 \left (8 x_{t/Z} +4 \right )-9 x_{t/Z} \right] f_1(x_{t/Z}) - 3 \left(4 s_w^2 - 9 x_{t/Z} \right) \ln x_{\mu/Z} \bigg \} \,. 
\end{split}
\eea
Here, we have introduced the abbreviations~$x_{t/Z} = m_t^2/m_Z^2$ and~$x_{\mu/Z} = \mu^2/m_Z^2$. The~function $f_1(x)$ appearing in~(\ref{eq:alphas}) and~(\ref{eq:betas}) takes the form 
\beq \label{eq:f1}
f_1(x) = \frac{1}{2} \hspace{0.25mm} \bigg [ 1 + \sqrt{4x-1} \arccot \left ( \sqrt{4x-1} \right ) \bigg ] \,. 
\eeq
Notice that this function is normalised such that~$f_1 (\infty) = 1$ and that an expansion of~$f_1 (x)$ around~$x \to \infty$ does not generate logarithms. 

In the case of~$t \to bW$ we find for the decay rate the one-loop result 
\beq \label{eq:DGtbW}
\begin{split}
\Delta \Gamma ( t \to b W )= \frac{\alpha^2 |V_{tb}|^2 \hspace{0.125mm} m_t}{64 \pi s_w^4}\frac{x_{t/W}^3-3 x_{t/W}+2}{x_{t/W}^2} \frac{v^2}{\Lambda^2} \, \sum_{i=\phantom{}_{QQ}^1, \phantom{}_{QQ}^8, \phantom{}_{Qt}^1, \phantom{}_{Qt}^8} \gamma_i \hspace{0.25mm} c_i \,, 
\end{split} 
\eeq
with 
\bea \label{eq:gammas} 
\begin{split}
\gamma_{QQ}^{1} & =\frac{1}{18} \bigg [ -3 x_{t/W}^2-9 x_{t/W}+8 + 3 \left(x_{t/W}^2+x_{t/W}-2\right) f_2(x_{t/W}) \\[1mm] 
& \phantom{xxxxx}+ 3 \left (3 x_{t/W} -2 \right ) \ln x_{\mu/t} \bigg ] \,, \\[2mm] 
\gamma_{QQ}^{8} & = \frac{4}{3} \hspace{0.5mm} \gamma_{QQ}^{1} \,, \\[2mm] 
\gamma_{Qt}^{1} & = -\frac{3 x_{t/W}}{2 (x_{t/W}+2)} \,, \\[2mm] 
\gamma_{Qt}^{8} & = \frac{4}{3} \hspace{0.5mm} \gamma_{Qt}^{1} \,.
\end{split}
\eea
Above, we have defined~$x_{t/W} = m_t^2/m_W^2$ and~$x_{\mu/t} = \mu^2/m_t^2$. The function $f_2(x)$ entering~(\ref{eq:gammas}) reads 
\beq \label{eq:f2}
f_2(x) = 2- (x-1) \ln \left(\frac{x}{x-1}\right) \,,
\eeq
and satisfies~$f_2( \infty) = 1$. Like~(\ref{eq:f1}) also the expression~(\ref{eq:f2}) does not develop logarithms in the limit~$x \to \infty$. The relevant helicity fractions of the~$W$ boson in the~$t\to bW$ decay can instead be written as 
\beq \label{eq:DFi}
\begin{split}
\Delta F_L = -\Delta F_- = \frac{\alpha}{4 \pi s_w^2} \frac{x_{t/W} \left ( x_{t/W} - 1 \right ) }{\left ( x_{t/W} + 2 \right )^2} \frac{v^2}{\Lambda^2} \left ( c_{Qt}^1 + \frac{4}{3} \hspace{0.25mm} c_{Qt}^8 \right ) \,, 
\end{split}
\eeq
Notice that since~$F_+ = 1 - F_L - F_-$, it follows that~$\Delta F_+ = 0$, meaning that the set of third-generation four-quark operators~(\ref{eq:operators}) does not induce a correction to the positive helicity fraction of the~$W$ boson when bottom quarks are treated as massless. We add that our results~(\ref{eq:DGtbW}),~(\ref{eq:gammas}) and~(\ref{eq:DFi}) agree with the expressions given in the arXiv version 2 of~\cite{Boughezal:2019xpp} after correcting typographical errors in~(38),~(39)~and~(40) of the latter publication. To be precise, in (38) an overall factor of $g/\pi$ is missing and the overall signs in (39)~and~(40) should both be reversed. One also has to take into account that the results in~\cite{Boughezal:2019xpp} were obtained in the Warsaw operator basis, although the~choice of operator basis only affects the result for $\gamma_{QQ}^{8}$.

In the case of the Peskin-Takeuchi parameters~$S$ we obtain at the two-loop level the SMEFT correction 
\beq \label{eq:Sexact}
\Delta S = -\frac{1}{144 \pi^3} \frac{m_t^2}{\Lambda^2} \sum_{i=\phantom{}_{QQ}^1, \phantom{}_{QQ}^8, \phantom{}_{Qt}^1, \phantom{}_{Qt}^8,\phantom{}_{Qb}^1, \phantom{}_{tt}^1, \phantom{}_{tb}^1} \zeta_i \hspace{0.25mm} c_i \,, \\[2mm] 
\eeq
with 
\beq \label{eq:zetas} 
\begin{split}
\zeta_{QQ}^1 &= 24 \ln^2 x_{\mu/t} - 36 \ln x_{\mu/t} \ln x_{\mu/Z} - 108 \ln x_{\mu/t} + 9 \,, \\[2mm]
\zeta_{QQ}^8 &= 8 \ln^2 x_{\mu/t} - 28 \ln x_{\mu/t} + 12 \,, \\[2mm]
\zeta_{Qt}^1 &= -54 \ln^2 x_{\mu/t} + 36 \ln x_{\mu/t} \ln x_{\mu/Z} + 159 \ln x_{\mu/t} \,, \\[2mm]
\zeta_{Qt}^8 &= 60 \ln x_{\mu/t} \,, \\[2mm]
\zeta_{Qb}^1 &= 18 \ln x_{\mu/t} \ln x_{\mu/Z} + 30 \ln x_{\mu/t} \,, \\[2mm]
\zeta_{tt}^1 &= 96 \ln^2 x_{\mu/t} - 120 \ln x_{\mu/t} + 18\,, \\[2mm]
\zeta_{tb}^1 &= -18 \ln x_{\mu/t} \ln x_{\mu/Z} - 30 \ln x_{\mu/t} \,.
\end{split}
\eeq
The corresponding corrections to the~$T$ parameter are given by 
\beq \label{eq:Texact}
\Delta T = -\frac{m_t^2}{256 \pi^3 c_w^2 s_w^2 m_Z^2} \frac{m_t^2}{\Lambda^2} \sum_{i=\phantom{}_{QQ}^1, \phantom{}_{QQ}^8, \phantom{}_{Qt}^1, \phantom{}_{tt}^1} \eta_i \hspace{0.25mm} c_i \,, \\[2mm] 
\eeq
with 
\beq \label{eq:etas} 
\begin{split}
\eta_{QQ}^1 &= 42 \ln^2 x_{\mu/t} - 18 \ln x_{\mu/t} + \frac{9}{2} \,, \\[2mm]
\eta_{QQ}^8 &= 8 \ln^2 x_{\mu/t} - 24 \ln x_{\mu/t} + 6 \,, \\[2mm]
\eta_{Qt}^1 &= -72 \ln^2 x_{\mu/t} \,, \\[2mm]
\eta_{tt}^1 &= 96 \ln^2 x_{\mu/t} - 48 \ln x_{\mu/t} \,. \\[2mm]
\end{split}
\eeq
For the parameter~$U$ we finally find that the SMEFT corrections due to~(\ref{eq:operators}) can be written as follows 
\beq \label{eq:Uexact}
\Delta U = \frac{1}{48 \pi^3} \frac{m_t^2}{\Lambda^2} \sum_{i=\phantom{}_{QQ}^1, \phantom{}_{QQ}^8, \phantom{}_{Qt}^1, \phantom{}_{Qt}^8, \phantom{}_{tt}^1} \theta_i \hspace{0.25mm} c_i \,, \\[2mm] 
\eeq
where 
\beq \label{eq:thetas} 
\begin{split}
\theta_{QQ}^1 &= 18 \ln^2 x_{\mu/t} -18 \ln x_{\mu/t} \ln x_{\mu/Z} - 47 \ln x_{\mu/t} + 7 \,, \\[2mm]
\theta_{QQ}^8 &= -\frac{32}{3} \ln x_{\mu/t} + \frac{28}{3} \,, \\[2mm]
\theta_{Qt}^1 &= -18 \ln^2 x_{\mu/t} + 18 \ln x_{\mu/t} \ln x_{\mu/Z} + 48 \ln x_{\mu/t} + \frac{9}{2} \,, \\[2mm]
\theta_{Qt}^8 &= 6 \,, \\[2mm]
\theta_{tt}^1 &= -24 \ln x_{\mu/t} + 6 \,.
\end{split}
\eeq

Using the unitarity of the CKM matrix, the two-loop SMEFT corrections to the $b \to sZ$ penguin amplitude can be written in the following form 
\beq \label{eq:Cexact}
\Delta C = \frac{\alpha}{1536 \pi s_w^2} \frac{v^2}{\Lambda^2} \sum_{i=\phantom{}_{QQ}^1, \phantom{}_{QQ}^8, \phantom{}_{Qt}^1, \phantom{}_{Qt}^8,\phantom{}_{tt}^1} \kappa_i \hspace{0.25mm} c_i \,, \\[2mm] 
\eeq
where
\begin{align} \label{eq:kappas}
\kappa_{QQ}^1 &= -\frac{x_{t/W} \left ( 15 x_{t/W}^2 - 29 x_{t/W} - 18 \right ) }{2 \left (x_{t/W}-1 \right )} \nonumber \\[2mm]
& \phantom{=.} + \frac{2 x_{t/W} \left ( 12 x_{t/W}^3 - 63 x_{t/W}^2 + 49 x_{t/W} - 6 \right )}{\left (x_{t/W}-1 \right )^2} \hspace{0.25mm} \ln x_{t/W} \nonumber \\[2mm]
& \phantom{=.} - 12 x_{t/W}^2 \left ( x_{t/W} - 3 \right ) \hspace{0.25mm} \ln^2 x_{t/W} - 24 x_{t/W}^2 \left ( x_{t/W} - 3 \right ) \hspace{0.5mm} {\rm Li}_2 \left (1-x_{t/W} \right ) \nonumber \\[2mm]
& \phantom{=.} - 4 \pi^2 \hspace{0.25mm} x_{t/W}^2 \left ( x_{t/W} - 4 \right ) + f_3(x_{t/W}) \ln x_{\mu/t} - 12 x_{t/W}^2 \ln ^2 x_{\mu/t} \,, \nonumber \\[4mm]
\kappa_{QQ}^8 &= -\frac{2 x_{t/W} \left ( 15 x_{t/W}^2 - 29 x_{t/W} - 18 \right ) }{3 \left (x_{t/W}-1 \right )} \nonumber \\[2mm]
& \phantom{=.} + \frac{8 x_{t/W} \left ( 12 x_{t/W}^3 - 63 x_{t/W}^2 + 49 x_{t/W} - 6 \right )}{3 \left (x_{t/W}-1 \right )^2} \hspace{0.25mm} \ln x_{t/W} \nonumber \\[2mm]
& \phantom{=.} - 16 x_{t/W}^2 \left ( x_{t/W} - 3 \right ) \hspace{0.25mm} \ln^2 x_{t/W} - 32 x_{t/W}^2 \left ( x_{t/W} - 3 \right ) \hspace{0.5mm} {\rm Li}_2 \left (1-x_{t/W} \right ) \\[2mm]
& \phantom{=.} - \frac{16}{3} \hspace{0.25mm} \pi^2 \hspace{0.25mm} x_{t/W}^2 \left ( x_{t/W} - 4 \right ) + f_4(x_{t/W}) \ln x_{\mu/t} - 16 x_{t/W}^2 \ln ^2 x_{\mu/t} \,, \nonumber \\[4mm]
\kappa_{Qt}^1 & = \frac{3 x_{t/W} \left ( 19 x_{t/W}^2 + 7 x_{t/W} + 14 \right )}{2 \left (x_{t/W}-1 \right )} - \frac{12 x_{t/W} \left ( x_{t/W}^2 + 3 x_{t/W} + 1 \right )}{\left (x_{t/W}-1 \right )^2} \hspace{0.25mm} \ln x_{t/W} \nonumber \\[2mm]
& \phantom{=.} + 6 x_{t/W}^2 \hspace{0.25mm} \ln^2 x_{t/W} + 12 x_{t/W}^2 \hspace{0.5mm} {\rm Li}_2 \left (1-x_{t/W} \right ) + f_5(x_{t/W}) \ln x_{\mu/t} + 54 x_{t/W}^2 \ln ^2 x_{\mu/t} \,, \nonumber \\[4mm]
\kappa_{Qt}^8 & = \frac{2 x_{t/W} \left ( 19 x_{t/W}^2 + 7 x_{t/W} + 14 \right ) }{x_{t/W}-1} - \frac{16 x_{t/W} \left ( x_{t/W}^2+ 3 x_{t/W} + 1 \right )}{\left (x_{t/W}-1 \right )^2} \hspace{0.25mm} \ln x_{t/W} \nonumber \\[2mm]
& \phantom{=.} + 8 x_{t/W}^2 \hspace{0.25mm} \ln^2 x_{t/W} + 16 x_{t/W}^2 \hspace{0.5mm} {\rm Li}_2 \left (1-x_{t/W} \right ) + f_6(x_{t/W}) \ln x_{\mu/t} + 24 x_{t/W}^2 \ln ^2 x_{\mu/t} \,, \nonumber \\[4mm]
\kappa_{tt}^1 &= \frac{12 x_{t/W}^2 \left (x_{t/W}+5 \right )}{x_{t/W}-1} -\frac{72 x_{t/W}^2}{\left (x_{t/W}-1 \right )^2} \ln x_{t/W} + f_7(x_{t/W}) \ln x_{\mu/t} -96 x_{t/W}^2 \ln ^2 x_{\mu/t} \,. \nonumber
\end{align}
Here ${\rm Li}_2 (z)$ denotes the usual dilogarithm and the functions $f_3(x)$, $f_4 (x)$, $f_5 (x)$, $f_6 (x)$ and~$f_7(x)$ appearing in~(\ref{eq:kappas}) take the~form 
\beq 
\begin{split} 
f_3 (x) & = \frac{6 x \left( 17 x^2 - 25 x + 2 \right)}{x - 1} - \frac{36 x^2 \left ( x - 2 \right )}{\left ( x - 1 \right )^2} \hspace{0.25mm} \ln x \,, \\[4mm]
f_4 (x) & = \frac{8 x \left( 5 x^2 - 13 x + 2 \right)}{x - 1} + \frac{48 x^3}{\left ( x - 1 \right )^2} \hspace{0.25mm} \ln x \,, \\[4mm] 
f_5 (x) & = -\frac{12 x \left( 6 x^2 - 18 x - 1 \right)}{x - 1} + \frac{12 x^2 \left ( 6 x - 19 \right )}{\left ( x - 1 \right )^2} \hspace{0.25mm} \ln x \\[4mm]
f_6 (x) & = \frac{8 x \left( 3 x^2 + 3 x + 2 \right)}{x - 1} - \frac{64 x^2}{\left ( x - 1 \right )^2} \hspace{0.25mm} \ln x \\[4mm]
f_7 (x) & = \frac{72 x^2 \left (x-5 \right )}{x-1} + \frac{288 x^2}{\left (x-1 \right )^2} \hspace{0.25mm} \ln x \,.
\end{split}
\eeq
The correction~(\ref{eq:Cexact}) is gauge-dependent, with the expressions~(\ref{eq:kappas}) corresponding to the~'t~Hooft-Feynman gauge. Note that when $x_{\mu/t} = 1$, the coefficients in~(\ref{eq:kappas}) adhere to the relationship $\kappa_{QQ}^8 = 4/3 \hspace{0.25mm} \kappa_{QQ}^1$ and $\kappa_{Qt}^8 = 4/3 \hspace{0.25mm} \kappa_{Qt}^1$, which is a straightforward consequence of the $SU(3)_C$ colour algebra. We also observe that our final results~(\ref{eq:kappas}) do not depend on~$s_w^2$, although such a dependence appears in the intermediate steps of the calculation. This is a non-trivial consequence of the EW gauge structure and its breaking,~i.e.~${SU(2)_L \times U(1)_Y \to U(1)_Q}$, which implies that the $Z$-penguin amplitude arises through ${SU(2)_L \text{-breaking}}$ terms and involves only left-handed quarks. As in the SM, the top-quark Yukawa coupling provides the dominant $SU(2)_L$ breaking in the SMEFT as~well.

The $b \to s \ell^+ \ell^-$ box correction~$\Delta B$ due to single insertions of the third-generation four-quark dimension-six operators~(\ref{eq:operators}) takes the following form 
\beq \label{eq:Bexact}
\Delta B = -\frac{\alpha}{384 \pi s_w^2} \frac{v^2}{\Lambda^2} \sum_{i=\phantom{}_{QQ}^1, \phantom{}_{QQ}^8, \phantom{}_{Qt}^1, \phantom{}_{Qt}^8} \xi_i \hspace{0.25mm} c_i \,. \\[2mm] 
\eeq
After the GIM mechanism, we obtain 
\bea \label{eq:xis}
\begin{split}
\xi_{QQ}^1 &= -12 x_{t/W}^2 - 12 x_{t/W}^2 \hspace{0.25mm} \ln x_{t/W} + 3 x_{t/W}^2 \left ( 2 x_{t/W} - 1\right ) \ln ^2 x_{t/W} \\[2mm] 
& \phantom{xx} + 6 x_{t/W}^2 \left ( 2 x_{t/W} - 1\right ) {\rm Li}_2 \left (1-x_{t/W} \right ) + \pi ^2 \hspace{0.25mm} x_{t/W}^2 \left ( 2 x_{t/W} - 1\right ) + 6 f_8 (x_{t/W}) \ln x_{\mu/t} \,, \\[2mm]
\xi_{QQ}^8 & = \frac{4}{3} \hspace{0.5mm} \xi_{QQ}^1 \,, \\[2mm] 
\xi_{Qt}^1 &= \frac{6 x_{t/W} \hspace{0.5mm} \big ( 2 x_{t/W} - 3 \big) }{x_{t/W}-1} + \frac{6 x_{t/W} \hspace{0.5mm} \big ( 2 x_{t/W}^2 - 3 x_{t/W}+2 \big) }{ \left (x_{t/W}-1 \right )^2} \hspace{0.25mm} \ln x_{t/W} \\[2mm]
& \phantom{xx} -6 x_{t/W}^2 \hspace{0.25mm} \ln ^2 x_{t/W} -12 x_{t/W}^2 \hspace{0.25mm} {\rm Li}_2 \left (1-x_{t/W} \right ) - 2 \pi ^2 x_{t/W}^2 - 12 f_9 (x_{t/W}) \ln x_{\mu/t} \,, \hspace{6mm} \\[2mm]
 \xi_{Qt}^8 & = \frac{4}{3} \hspace{0.5mm} \xi_{Qt}^1 \,.
\end{split}
\eea
The functions $f_8(x)$ and $f_9(x)$ entering~(\ref{eq:xis}) read
\beq \label{eq:f3}
\begin{split}
f_8(x) & = \frac{x^2}{x-1} - \frac{x^2}{\left (x-1 \right )^2} \hspace{0.25mm}\ln x \,, \\[2mm]
f_9(x) & = \frac{x}{x-1} - \frac{x^2}{\left (x-1 \right )^2} \hspace{0.25mm}\ln x \,. 
\end{split}
\eeq
We emphasise that the correction~(\ref{eq:Bexact}) is gauge-dependent. The results given in~(\ref{eq:xis}) correspond to the 't~Hooft-Feynman gauge. Moreover, it is worth mentioning that due to the GIM mechanism, the two-loop $b \to s \ell^+ \ell^-$ box correction~(\ref{eq:Bexact}) remains unaltered by the particular definition of evanescent operators within the $\Delta F = 1$ sector. 

The off-shell photon penguin contribution $\Delta D$, which arises at the two-loop level due to single insertions of the third-generation four-quark dimension-six operators~(\ref{eq:operators}), can be expressed as
\beq \label{eq:Dexact}
\Delta D = \frac{\alpha}{1296 \pi s_w^2} \frac{v^2}{\Lambda^2} \sum_{i=\phantom{}_{QQ}^1, \phantom{}_{QQ}^8, \phantom{}_{Qt}^1, \phantom{}_{Qt}^8, \phantom{}_{tt}^1} \rho_i \hspace{0.25mm} c_i \,, \\[2mm] 
\eeq
where 
\begin{align} \label{eq:rhos}
\rho_{QQ}^1 &= -\frac{x_{t/W} \left(12 x_{t/W}^4-204 x_{t/W}^3+337 x_{t/W}^2-145 x_{t/W}+12\right)}{2 \left (x_{t/W}-1 \right)^3} \nonumber \\[2mm]
& \phantom{xx} -\frac{x_{t/W} \left(24 x_{t/W}^5+177 x_{t/W}^4-1094 x_{t/W}^3+2315 x_{t/W}^2-2118 x_{t/W}+672\right)}{4 \left (x_{t/W}-1 \right )^4} \hspace{0.25mm} \ln x_{t/W} \nonumber \\[2mm] 
& \phantom{xx} + \frac{3 x_{t/W} \left(2 x_{t/W}^4+14 x_{t/W}^3-67 x_{t/W}^2+114 x_{t/W}-54\right)}{2 \left (x_{t/W}-1 \right )^2} \hspace{0.25mm} \ln^2 x_{t/W} \nonumber \\[2mm]
& \phantom{xx} + \frac{3 x_{t/W} \left(2 x_{t/W}^4+14 x_{t/W}^3-67 x_{t/W}^2+114 x_{t/W}-54\right)}{\left (x_{t/W}-1 \right )^2} \hspace{0.5mm} {\rm Li}_2 \left (1-x_{t/W} \right ) \nonumber \\[2mm]
& \phantom{xx} + \pi ^2 x_{t/W} \left(x_{t/W}^2+9 x_{t/W}-27\right) + f_{10}(x_{t/W}) \ln x_{\mu/t} + 36 x_{t/W} \ln^2 x_{\mu/t} \,, \nonumber \\[4mm]
\rho_{QQ}^8 &= -\frac{2 x_{t/W} \left(12 x_{t/W}^4-204 x_{t/W}^3+337 x_{t/W}^2-145 x_{t/W}+12\right)}{3 \left (x_{t/W}-1 \right )^3} \nonumber \\[2mm]
& \phantom{xx} -\frac{x_{t/W} \left(24 x_{t/W}^5+177 x_{t/W}^4-1094 x_{t/W}^3+2315 x_{t/W}^2-2118 x_{t/W}+672\right)}{3 \left (x_{t/W}-1 \right)^4} \hspace{0.25mm} \ln x_{t/W} \nonumber \\[2mm] 
& \phantom{xx} + \frac{2 x_{t/W} \left(2 x_{t/W}^4+14 x_{t/W}^3-67 x_{t/W}^2+114 x_{t/W}-54\right)}{\left (x_{t/W}-1 \right )^2}\hspace{0.25mm} \ln^2 x_{t/W} \nonumber \\[2mm]
& \phantom{xx} + \frac{4 x_{t/W} \left(2 x_{t/W}^4+14 x_{t/W}^3-67 x_{t/W}^2+114 x_{t/W}-54\right)}{\left (x_{t/W}-1 \right )^2} \hspace{0.5mm} {\rm Li}_2 \left (1-x_{t/W} \right ) \nonumber \\[2mm]
& \phantom{xx} + \frac{4}{3} \hspace{0.25mm} \pi ^2 x_{t/W} \left(x_{t/W}^2+9 x_{t/W}-27\right) + f_{11}(x_{t/W}) \ln x_{\mu/t} + 48 x_{t/W} \ln^2 x_{\mu/t} \,, \nonumber \\[4mm]
\rho_{Qt}^1 &= -\frac{x_{t/W} \left(56 x_{t/W}^3+93 x_{t/W}^2-3 x_{t/W}-92\right)}{4 \left (x_{t/W}-1 \right )^3} \nonumber \\[-2mm] \\[-2mm] 
& \phantom{xx} + \frac{9 x_{t/W} \left(11 x_{t/W}^3+9 x_{t/W}^2-30 x_{t/W}+13\right)}{2 \left (x_{t/W}-1 \right )^4} \hspace{0.25mm} \ln x_{t/W} \nonumber \\[2mm] 
& \phantom{xx} + \frac{9 x_{t/W} \left(2 x_{t/W}^2+10 x_{t/W}-3\right)}{2 \left (x_{t/W}-1 \right )^2} \hspace{0.25mm} \ln^2 x_{t/W} \nonumber \\[2mm]
& \phantom{xx} + \frac{9 x_{t/W} \left(2 x_{t/W}^2+10 x_{t/W}-3\right)}{\left (x_{t/W}-1 \right)^2} \hspace{0.5mm} {\rm Li}_2 \left (1-x_{t/W} \right ) \nonumber \\[2mm]
& \phantom{xx} + f_{12}(x_{t/W}) \ln x_{\mu/t} -\frac{63}{2} \hspace{0.25mm} x_{t/W} \ln^2 x_{\mu/t} \,, \nonumber \\[4mm]
\rho_{Qt}^8 &= -\frac{x_{t/W} \left(56 x_{t/W}^3+93 x_{t/W}^2-3 x_{t/W}-92\right)}{3 \left (x_{t/W}-1 \right)^3} \nonumber \\[2mm]
& \phantom{xx} +\frac{6 x_{t/W} \left(11 x_{t/W}^3+9 x_{t/W}^2-30 x_{t/W}+13\right)}{\left (x_{t/W}-1 \right)^4} \hspace{0.25mm} \ln x_{t/W} \nonumber \\[2mm] 
& \phantom{xx} + \frac{6 x_{t/W} \left(2 x_{t/W}^2+10 x_{t/W}-3\right)}{\left (x_{t/W}-1 \right)^2} \hspace{0.25mm} \ln^2 x_{t/W} \nonumber \\[2mm]
& \phantom{xx} + \frac{12 x_{t/W} \left(2 x_{t/W}^2+10 x_{t/W}-3\right)}{\left (x_{t/W}-1 \right)^2} \hspace{0.5mm} {\rm Li}_2 \left (1-x_{t/W} \right ) \nonumber \\[2mm]
& \phantom{xx} + f_{13}(x_{t/W}) \ln x_{\mu/t} + 30 x_{t/W} \ln^2 x_{\mu/t} \,, \nonumber \\[4mm]
\rho_{tt}^1 &= \frac{18 x_{t/W} \left (x_{t/W}+5 \right)}{x_{t/W}-1} -\frac{108 x_{t/W}}{\left (x_{t/W}-1 \right )^2} \hspace{0.25mm} \ln x_{t/W} \nonumber \\[2mm] 
& \phantom{=.} + f_{14}(x_{t/W}) \ln x_{\mu/t} -144 x_{t/W} \ln^2 x_{\mu/t} \,, \nonumber
\end{align}
with
\beq
\begin{split} 
f_{10} (x) & = -\frac{3 x \left( 281 x^3 - 857 x^2 + 866 x - 296 \right)}{4 \left ( x - 1 \right )^3} \\[1mm]
& \phantom{xx} + \frac{3 x \left( 15 x^4 + 56 x^3 - 266 x^2 + 300 x -108 \right)}{2 \left ( x - 1 \right )^4} \hspace{0.25mm} \ln x \,, \\[2mm]
f_{11} (x) & = -\frac{x \left( 137 x^3 - 425 x^2 + 434 x - 152 \right)}{\left ( x -1 \right )^3} \\[1mm]
& \phantom{xx} + \frac{2 x \left( 15 x^4 - 16 x^3 - 50 x^2 + 84 x -36 \right)}{\left ( x - 1 \right )^4} \hspace{0.25mm} \ln x \,, \\[2mm]
f_{12} (x) & = -\frac{3 x \left ( 43 x + 191 \right )}{2 \left ( x - 1 \right )} + \frac{27 x \left (10 x + 3 \right )}{\left ( x - 1 \right )^2} \hspace{0.25mm}\ln x \,, \\[2mm]
f_{13} (x) & = \frac{2 x \left ( 11 x -137 \right ) }{x - 1} + \frac{36 x \left ( 6 x + 1 \right )}{\left ( x - 1 \right )^2} \hspace{0.25mm} \ln x \,, \\[2mm]
f_{14} (x) & = \frac{108 x \left ( x - 5 \right )}{x - 1}+ \frac{432 x}{\left ( x - 1 \right )^2} \hspace{0.25mm} \ln x \,.
\end{split}
\eeq
To obtain the coefficients in~(\ref{eq:rhos}), we have again employed the GIM mechanism. Like~$\Delta C$ and~$\Delta B$, $\Delta D$ is also gauge-dependent, and the expressions provided above correspond to the~'t Hooft-Feynman gauge. Additionally, note that for $x_{\mu/t} = 1$, the coefficients in~(\ref{eq:rhos}) satisfy $\rho_{QQ}^8 = 4/3 \hspace{0.25mm} \rho_{QQ}^1$ and $\rho_{Qt}^8 = 4/3 \hspace{0.25mm} \rho_{Qt}^1$, as a consequence of the $SU(3)_C$ colour algebra.

The box correction~$\Delta F$ describes the modification of the $B_s$--$\bar B_s$ mixing amplitude due to single insertions of the third-generation four-quark dimension-six operators~(\ref{eq:operators}). After~making use of the unitarity of the CKM matrix, we~obtain 
\beq \label{eq:Fexact}
\Delta F = \frac{\alpha}{192 \pi s_w^2} \frac{v^2}{\Lambda^2} \sum_{i=\phantom{}_{QQ}^1, \phantom{}_{QQ}^8, \phantom{}_{Qt}^1, \phantom{}_{Qt}^8, \phantom{}_{tt}^1} \omega_i \hspace{0.25mm} c_i \,, \\[2mm] 
\eeq
where 
\begin{align} \label{eq:omegas}
\omega_{QQ}^1 &= -\frac{x_{t/W}^2 \left ( 33 x_{t/W}^2 - 68 x_{t/W} + 19 \right )}{\left ( x_{t/W} - 1 \right)^2} \nonumber \\[2mm]
& \phantom{=.} - \frac{2 x_{t/W}^2 \left ( 12 x_{t/W}^3 - 15 x_{t/W}^2 - x_{t/W} + 12 \right )}{\left ( x_{t/W} - 1 \right)^3} \hspace{0.25mm} \ln x_{t/W} \nonumber \\[2mm] 
& \phantom{=.} + \frac{6 x_{t/W}^2 \left ( 2 x_{t/W}^3 - x_{t/W}^2 - 10 x_{t/W} + 4 \right )}{\left (x_{t/W} - 1 \right)^2} \hspace{0.25mm} \ln^2 x_{t/W} \nonumber \\[2mm]
& \phantom{=.} + \frac{12 x_{t/W}^2 \left( 2 x_{t/W}^2 + x_{t/W} - 9 \right)}{x_{t/W} - 1} \hspace{0.5mm} {\rm Li}_2 \left (1-x_{t/W} \right ) \nonumber \\[2mm]
& \phantom{=.} + 4 \pi^2 x_{t/W}^2 \left ( x_{t/W} + 2 \right ) + f_{15} (x_{t/W}) \ln x_{\mu/t} \,, \nonumber \\[4mm]
\omega_{QQ}^8 &= -\frac{x_{t/W}^2 \left ( 123 x_{t/W}^2 - 254 x_{t/W} + 67 \right )}{3 \left ( x_{t/W} - 1 \right)^2} \nonumber \\[2mm]
& \phantom{=.} - \frac{4 x_{t/W}^2 \left ( 24 x_{t/W}^3 - 39 x_{t/W}^2 + 16 x_{t/W} + 15 \right )}{3 \left ( x_{t/W} - 1 \right)^3} \hspace{0.25mm} \ln x_{t/W} \nonumber \\[2mm] 
& \phantom{=.} + \frac{2 x_{t/W}^2 \left ( 8 x_{t/W}^3 - 13 x_{t/W}^2 - 4 x_{t/W} + 4 \right )}{\left (x_{t/W} - 1 \right)^2} \hspace{0.25mm} \ln^2 x_{t/W} \nonumber \\[2mm]
& \phantom{=.} + \frac{4 x_{t/W}^2 \left( 8 x_{t/W}^2 - 5 x_{t/W} - 9 \right)}{x_{t/W} - 1} \hspace{0.5mm} {\rm Li}_2 \left (1-x_{t/W} \right ) \nonumber \\[2mm]
& \phantom{=.} + \frac{8}{3} \hspace{0.5mm} \pi^2 x_{t/W}^2 \left ( 2 x_{t/W} + 1 \right ) + f_{16} (x_{t/W}) \ln x_{\mu/t} \,, \nonumber \\[4mm]
\omega_{Qt}^1 &= \frac{3 x_{t/W} \left ( 45 x_{t/W}^3 - 66 x_{t/W}^2 + 101 x_{t/W} - 32 \right )}{4 \left ( x_{t/W} - 1 \right)^2} \nonumber \\[-2mm] \\[-2mm] 
& \phantom{=.} + \frac{6 x_{t/W} \left ( 4 x_{t/W}^3 - 17 x_{t/W}^2 + 11 x_{t/W} - 4 \right )}{\left ( x_{t/W} - 1 \right)^3} \hspace{0.25mm} \ln x_{t/W} \nonumber \\[2mm] 
& \phantom{=.} - \frac{3 x_{t/W}^2 \left ( 5 x_{t/W}^3 - 17 x_{t/W}^2 + 19 x_{t/W} + 3 \right )}{\left (x_{t/W} - 1 \right)^3} \hspace{0.25mm} \ln^2 x_{t/W} \nonumber \\[2mm]
& \phantom{=.} - \frac{30 x_{t/W}^3 - 42 x_{t/W}^2 - 24}{x_{t/W} - 1} \hspace{0.5mm} {\rm Li}_2 \left (1-x_{t/W} \right ) \nonumber \\[2mm]
& \phantom{=.} -4 \pi^2 \left ( 2 x_{t/W}^2 - x_{t/W} - 1 \right ) + f_{17} (x_{t/W}) \ln x_{\mu/t} + 9 x_{t/W}^2 \ln^2 x_{\mu/t} \,, \nonumber \\[4mm]
\omega_{Qt}^8 &= \frac{x_{t/W} \left ( 105 x_{t/W}^3 - 42 x_{t/W}^2 + 161 x_{t/W} - 32 \right )}{4 \left ( x_{t/W} - 1 \right)^2} \nonumber \\[2mm]
& \phantom{=.} + \frac{4 x_{t/W} \left ( 11 x_{t/W}^3 - 40 x_{t/W}^2 + 19 x_{t/W} - 2 \right )}{\left ( x_{t/W} - 1 \right)^3} \hspace{0.25mm} \ln x_{t/W} \nonumber \\[2mm] 
& \phantom{=.} - \frac{x_{t/W}^2 \left ( 23 x_{t/W}^3 - 71 x_{t/W}^2 + 73 x_{t/W} -15 \right )}{\left (x_{t/W} - 1 \right)^3} \hspace{0.25mm} \ln^2 x_{t/W} \nonumber \\[2mm]
& \phantom{=.} - \frac{46 x_{t/W}^3 - 50 x_{t/W}^2 - 8}{x_{t/W} - 1} \hspace{0.5mm} {\rm Li}_2 \left (1-x_{t/W} \right ) \nonumber \\[2mm]
& \phantom{=.} -\frac{4}{3} \hspace{0.5mm} \pi^2 \left ( 8 x_{t/W}^2 - x_{t/W} - 1 \right ) + f_{18} (x_{t/W}) \ln x_{\mu/t} + 9 x_{t/W}^2 \ln^2 x_{\mu/t} \,, \nonumber \\[4mm]
\omega_{tt}^1 &= \frac{3 x_{t/W}^2 \left (x_{t/W}^2+ 10 x_{t/W} -47 \right )}{\left(x_{t/W}-1\right)^2}-\frac{36 x_{t/W}^2 \left (x_{t/W}-7 \right )}{\left (x_{t/W}-1 \right )^3} \hspace{0.25mm} \ln x_{t/W} \nonumber \\[2mm] 
& \phantom{=.} -\frac{108 x_{t/W}^2}{\left(x_{t/W}-1 \right)^4} \hspace{0.25mm} \ln^2 x_{t/W} + f_{19} (x_{t/W}) \hspace{0.25mm} \ln x_{\mu/t} -12 x_{t/W}^2 \ln^2 x_{\mu/t} \,. \nonumber 
\end{align} 		
The functions $f_{15} (x)$, $f_{16} (x)$, $f_{17} (x)$, $f_{18} (x)$ and $f_{19} (x)$ appearing in~(\ref{eq:omegas}) take the~form 
\beq
\begin{split}
f_{15} (x) & = \frac{6 x^2 \left (3 x^2 - 13 x + 8 \right )}{\left ( x - 1 \right )^2} + \frac{12 x^3 \left ( 3 x - 2 \right ) }{\left ( x - 1 \right )^3}\hspace{0.25mm} \ln x \,, \\[2mm]
f_{16} (x) & = \frac{4 x^2 \left (3 x^2 - 20 x + 13 \right )}{\left ( x - 1 \right)^2} + \frac{16 x^3 \left ( 3 x - 2 \right)}{\left ( x - 1 \right )^3} \hspace{0.25mm} \ln x \,, \\[2mm]
f_{17} (x) & = \frac{6 x^2 \left ( x + 8 \right )}{x - 1} + \frac{6 x^2 \left( 2 x - 11 \right)}{\left ( x - 1 \right )^2} \hspace{0.25mm} \ln x \,, \\[2mm]
f_{18} (x) & = \frac{2 x^2 \left ( 4 x + 23 \right )}{x - 1} + \frac{2 x^2 \left( 2 x - 29 \right)}{\left ( x - 1 \right )^2} \hspace{0.25mm} \ln x \,, \\[2mm]
f_{19} (x) & = \frac{12 x^2 \left ( x - 7 \right )}{x - 1} +\frac{72 x^2}{\left ( x - 1 \right )^2} \hspace{0.25mm} \ln x \,.
\end{split}
\eeq
We emphasise that the rational terms of $\omega_{QQ}^1$, $\omega_{QQ}^8$, $\omega_{Qt}^1$ and $\omega_{Qt}^8$ depend on the definition of the one-loop evanescent $\Delta F = 2$ operator. In our two-loop calculation, we have employed~\cite{Buras:1990fn,Herrlich:1996vf} 
\beq \label{eq:evaQmix}
E_Q = \big ( \bar s \gamma_\mu \gamma_\nu \gamma_\lambda P_L b \big ) \big ( \bar s \gamma^\mu \gamma^\nu \gamma^\lambda P_L b \big ) - \left ( 16 - 4 \epsilon \right ) Q \,, 
\eeq
with the operator $Q$ given in~(\ref{eq:Qmix}). Conversely, the remaining terms in~(\ref{eq:omegas}) are independent of the specific choice made for~(\ref{eq:evaQmix}). Additionally, we note that, owing to the GIM mechanism, the two-loop result in~(\ref{eq:Fexact}) remains unaffected by the specific definition of evanescent operators within the $\Delta F = 1$ sector, which play a crucial role in the renormalisation of subdivergences.

\section{$\bm{\overline{\rm MS}}$ shifts}
\label{app:shifts}

In this~appendix, we present the analytic expressions for the finite shifts of the two-loop results that follow from changing the renormalisation scheme of the top-quark mass from~OS to $\overline{\rm MS}$. In the case of the Peskin-Takeuchi parameters we find 
\bea \label{eq:PTshifts}
\begin{split}
\delta S & = \frac{1}{24 \pi^3} \frac{m_t^2}{\Lambda^2} \left (c_{Qt}^1 + \frac{4}{3} \hspace{0.25mm}c_{Qt}^8 \right ) \left ( 1 + 2 \ln x_{\mu/t} \right ) \,, \\[2mm]
\delta T & = -\frac{3 \hspace{0.25mm} m_t^2}{64\pi^3 c_w^2 s_w^2 m_Z^2} \frac{m_t^2}{\Lambda^2} \left ( c_{Qt}^1 + \frac{4}{3} \hspace{0.25mm}c_{Qt}^8 \right ) \left ( 1 + 2 \ln x_{\mu/t} \right ) \,, \\[2mm]
\delta U & = -\frac{1}{8 \pi^3} \frac{m_t^2}{\Lambda^2} \left (c_{Qt}^1 + \frac{4}{3} \hspace{0.25mm}c_{Qt}^8 \right ) \left ( 1 + 2 \ln x_{\mu/t} \right ) \,. 
\end{split}
\eea
Notice that these expressions all involve the combination~$c_{Qt}^1 + 4/3 \hspace{0.5mm} c_{Qt}^8$ of Wilson coefficients, which is precisely the same factor that also appears in the counterterms of the heavy-quark masses~\cite{Gauld:2015lmb,Alasfar:2022zyr,Jenkins:2013zja}. 

In the case of the flavour observables we obtain the following results for the shifts between the OS and~$\overline{\rm MS}$ schemes
\beq \label{eq:flavourshifts} 
\begin{split}
\delta C & = -\frac{1}{8 \pi^2} \frac{m_t^2}{\Lambda^2} \left ( c_{Qt}^1 + \frac{4}{3} \hspace{0.25mm} c_{Qt}^8 \right ) g_1(x_{t/W}) \left (1 + 2 \ln x_{\mu/t} \right ) \,, \\[2mm]
\delta B & = -\frac{1}{8 \pi^2} \frac{m_t^2}{\Lambda^2} \left ( c_{Qt}^1 + \frac{4}{3} \hspace{0.25mm} c_{Qt}^8 \right ) g_2 (x_{t/W}) \left (1 + 2 \ln x_{\mu/t} \right ) \,, \\[2mm]
\delta D & = -\frac{1}{8 \pi^2} \frac{m_t^2}{\Lambda^2} \left ( c_{Qt}^1 + \frac{4}{3} \hspace{0.25mm} c_{Qt}^8 \right ) g_3 (x_{t/W}) \left (1 + 2 \ln x_{\mu/t} \right ) \,, \\[2mm]
\delta F & = -\frac{1}{8 \pi^2} \frac{m_t^2}{\Lambda^2} \left ( c_{Qt}^1 + \frac{4}{3} \hspace{0.25mm} c_{Qt}^8 \right ) g_4 (x_{t/W}) \left (1 + 2 \ln x_{\mu/t} \right )\,, 
\end{split}
\eeq
with
\bea \label{eq:gis} 
\begin{split} 
g_1 (x) & = \frac{x \left ( x^2 + x + 8 \right )}{4 \left ( x - 1 \right )^2} - \frac{x \left( 4 x + 1 \right)}{2 \left ( x - 1 \right )^3} \hspace{0.25mm}\ln x \,, \\[2mm]
g_2 (x) & = \frac{x}{\left( x - 1 \right)^2}-\frac{x \left( x + 1 \right)}{2 \left( x - 1 \right)^3}\hspace{0.25mm} \ln x \,, \\[2mm]
g_3 (x) & = -\frac{3 x^4 - 46 x^3 + 79 x^2 - 32 x + 8}{9 \left ( x - 1 \right )^4} - \frac{2 x^2 \left( 3 x^2 - 3 x - 2 \right)}{3 \left ( x - 1 \right )^5} \hspace{0.25mm}\ln x \,, \\[2mm]
g_4 (x) & = \frac{x \left( x^3 + 3 x^2 + 18 x - 4 \right)}{8 \left ( x - 1 \right )^3} - \frac{9 x^3}{4 \left ( x - 1 \right)^4}\hspace{0.25mm} \ln x \,.
\end{split}
\eea
Notice that all expressions in~(\ref{eq:flavourshifts}) are again proportional to the combination~$c_{Qt}^1 + 4/3 \hspace{0.5mm} c_{Qt}^8$ of Wilson coefficients, which is expected because these corrections are entirely due to the renormalisation of the top-quark mass. It is also important to realise that the results~(\ref{eq:flavourshifts}) receive contributions from both $W$- and would-be Goldstone-boson exchange. Diagrams with a would-be Goldstone boson involve the top-quark Yukawa coupling, which receives the same correction in the SMEFT proportional to $c_{Qt}^1 + 4/3 \hspace{0.5mm} c_{Qt}^8$ as the mass. Throughout our work, we renormalise the top-quark mass and Yukawa coupling in the same scheme. In the case of the results~(\ref{eq:kappas}),~(\ref{eq:rhos}) and~(\ref{eq:omegas}), this implies that one has to subtract a finite Yukawa counterterm in diagrams with a would-be Goldstone boson to obtain the quoted on-shell scheme expressions.

\section{Scheme change}
\label{app:fierz}

In this appendix, we provide some details on the change of renormalisation scheme that connects the LHCTopWG to the Warsaw operator basis. In the Warsaw operator basis, the LHCTopWG operators ${\cal O}^1_{QQ}$ and ${\cal O}^8_{QQ}$ can be written in terms of suitable linear combinations of 
\beq \label{eq:WarsawKuKu}
Q^{(1)}_{qq} = (\bar q \gamma_\mu q) (\bar q \gamma^\mu q) \,, \qquad 
Q^{(3)}_{qq} = (\bar q \gamma_\mu \sigma^i q) (\bar q \gamma^\mu \sigma^i q) \,.
\eeq
Using the Dirac Fierz identity 
\beq \label{eq:diracfierz}
(\gamma_\mu P_L)_{ij} \otimes (\gamma^\mu P_L)_{kl} = - (\gamma_\mu P_L)_{il} \otimes (\gamma^\mu P_L)_{kj} \,, 
\eeq
as well as the $SU(3)_C$ colour Fierz identity 
\beq \label{eq:colourfierz}
(T^a)_{ij} \otimes (T^a)_{kl} = \frac{1}{2} \left ( \mathbb{1}_{il} \otimes \mathbb{1}_{kj} - \frac{1}{3} \hspace{0.25mm} \mathbb{1}_{ij} \otimes \mathbb{1}_{kl} \right ) \,, 
\eeq
where $\mathbb{1}$ denotes the $3 \times 3$ unit matrix, it can be shown that the operators ${\cal O}^1_{QQ}$ and ${\cal O}^8_{QQ}$ are related to the operators introduced in~(\ref{eq:WarsawKuKu}) as follows:
\beq \label{eq:OPTopToWarsaw}
{\cal O}^1_{QQ} = \frac{1}{2} \hspace{0.25mm} Q^{(1)}_{qq} \,, \qquad 
{\cal O}^8_{QQ} = \frac{1}{24} \hspace{0.25mm}Q^{(1)}_{qq} + \frac{1}{8} \hspace{0.25mm} Q^{(3)}_{qq} + E_{QQ} \,.
\eeq
Here $E_{QQ}$ represents a Fierz evanescent operator that arises because~(\ref{eq:diracfierz}) only holds in $d = 4$ but not in $d = 4 - 2 \hspace{0.125mm} \epsilon$ dimensions. 

To better illustrate the impact of the Fierz evanescent operator $E_{QQ}$, let us consider a simple yet instructive example~\cite{ThesisLuc}. Since the one-loop corrections to the $t \to b W$ decay amplitude due to~(\ref{eq:WarsawKuKu}) have been computed analytically in~\cite{Boughezal:2019xpp}, we will in the following focus on the case of the top-quark decay. For the part proportional to the tree-level $tbW$ vertex of the bare one-loop scattering amplitude corresponding to the right diagram in~Figure~\ref{fig:decays} with single insertions of ${\cal O}^1_{QQ}$ and ${\cal O}^8_{QQ}$, we obtain the following~result
\beq \label{eq:AtbWLHCTopWG}
{\cal A} \left ( t \to b W \right )_{\rm LHCTopWG} = \frac{g}{16 \sqrt{2} \pi^2} \frac{m_t^2}{\Lambda^2} \left ( \frac{\mu^2}{m_t^2} \right )^\epsilon \left ( \frac{1}{\epsilon} - \frac{1}{2} \right ) \left ( c_{QQ}^1 + \frac{4}{3} \hspace{0.125mm} c_{QQ}^8 \right ) \,, 
\eeq
where we have suppressed the spinor chain $\bar t \hspace{0.5mm} \slashed{\varepsilon}_W P_L \hspace{0.25mm} b$ with $\varepsilon_W$~denoting the polarisation vector of the external $W$ boson. The~analogous expression resulting from insertions of the operators introduced in~(\ref{eq:WarsawKuKu}) is found to be
\beq \label{eq:AtbWarsaw}
{\cal A} \left ( t \to b W \right )_{\rm Warsaw} = \frac{g}{16 \sqrt{2} \pi^2} \frac{m_t^2}{\Lambda^2} \left ( \frac{\mu^2}{m_t^2} \right )^\epsilon \left [ \left ( \frac{1}{\epsilon} - \frac{1}{2} \right ) \left ( 2 \hspace{0.25mm} C_{qq}^{(1)} + 10 \hspace{0.25mm} C_{qq}^{(3)} \right ) + 12 \hspace{0.25mm} C_{qq}^{(3)} \right ] \,,
\eeq
with $C^{(1)}_{qq}$ and $C^{(3)}_{qq}$ denoting the Wilson coefficients associated to $Q^{(1)}_{qq}$ and $Q^{(3)}_{qq}$. 

Before continuing our discussion on the impact of the Fierz evanescent operator $E_{QQ}$, the following comment is in order. The UV pole in~(\ref{eq:AtbWarsaw}) encodes the mixing of the operators given in~(\ref{eq:WarsawKuKu}) into the operator
\beq \label{eq:QHq3}
Q^{(3)}_{H q} = (H^\dagger \hspace{0.05mm} i \overset{\leftrightarrow}{D} \ \!\!\!^{\; i}_\mu \hspace{0.25mm} H) (\bar q \gamma^\mu \sigma^i q) \,. 
\eeq
Here $H^\dagger \hspace{0.125mm} i \overset{\leftrightarrow}{D} \ \!\!\!^{\; i}_\mu \hspace{0.25mm} H = i H^\dagger \big (\sigma^i D_\mu - \overset{\leftarrow}{D}_\mu \sigma^i \big ) H$ with $H$ the usual Higgs doublet. The contributions proportional to $y_t^2$ to the corresponding anomalous dimensions can be deduced from~(\ref{eq:AtbWarsaw}). In agreement with~\cite{Jenkins:2013wua}, we find
\beq \label{eq:ADMs}
\gamma_{\,_{Hq}^{(3)},_{qq}^{(1)}} = -\frac{2 \hspace{0.25mm} y_t^2}{(4 \pi)^2} \,, \qquad \gamma_{\,_{Hq}^{(3)},_{qq}^{(3)}} = -\frac{10 \hspace{0.25mm} y_t^2}{(4 \pi)^2} \,.
\eeq
These anomalous dimensions appear in the RG equation that describes the renormalisation scale dependence of the Wilson coefficient of~(\ref{eq:QHq3}) in the following way 
\beq \label{eq:RGflow}
\mu \hspace{0.5mm} \frac{d C_{Hq}^{(3)}}{d \mu} = \sum_{i=1,3} \gamma_{\,_{Hq}^{(3)},_{qq}^{(i)}} \hspace{0.5mm} C_{qq}^{(i)} \,. 
\eeq

To determine whether the results in~(\ref{eq:AtbWLHCTopWG}) and (\ref{eq:AtbWarsaw}) are actually equivalent, we need to apply a rotation that transforms the LHCTopWG into the Warsaw operator basis. Ignoring a possible contribution from $E_{QQ}$, the relevant transformation reads
\beq \label{eq:WCWarsawToLHCTopWG}
C^{(1)}_{qq} = \frac{1}{2} \hspace{0.25mm} c^1_{QQ} + \frac{1}{24} \hspace{0.25mm} c^8_{QQ} \,, \qquad 
C^{(3)}_{qq} = \frac{1}{8} \hspace{0.25mm} c^8_{QQ} \,, 
\eeq
in terms of the Wilson coefficients. By inserting~(\ref{eq:WCWarsawToLHCTopWG}) into~(\ref{eq:AtbWarsaw}) and then taking the difference between the transformed result and~(\ref{eq:AtbWLHCTopWG}), we obtain
\beq \label{eq:evadiff}
{\cal A} \left ( t \to b W \right )_{\rm Warsaw} - {\cal A} \left ( t \to b W \right )_{\rm LHCTopWG} = \frac{3 g}{32 \sqrt{2} \pi^2} \frac{m_t^2}{\Lambda^2} \hspace{0.5mm} c_{QQ}^8 \,. 
\eeq
It can be observed that while this difference is UV finite, it is not zero. The UV finiteness is, of course, a straightforward consequence of the scheme-independence of the full one-loop SMEFT anomalous dimension matrix. The constant remainder in~(\ref{eq:evadiff}) can be attributed to the fact that (\ref{eq:WCWarsawToLHCTopWG}) only hold in $d = 4$ dimensions. In $d = 4 - 2 \hspace{0.125mm} \epsilon$ dimensions, the Fierz evanescent operator $E_{QQ}$ from~(\ref{eq:OPTopToWarsaw}) must be included, as it corresponds to a finite rational contribution to ${\cal A} \left ( t \to b W \right )$. In fact, since we have not included $E_{QQ}$ in the transformation~(\ref{eq:WCWarsawToLHCTopWG}), the difference~(\ref{eq:evadiff}), which is proportional to $c_{QQ}^8$, is exactly the contribution due to the Fierz evanescent operator. 

Applying the above reasoning to the calculation of $\Delta \Gamma (t \to b W)$, $\Delta F_L$ and $\Delta F_-$, it~is clear that compared to the LHCTopWG results presented in~(\ref{eq:DGtbW}),~(\ref{eq:gammas}) and~(\ref{eq:DFi}), only the result for $\gamma_{QQ}^8$ is modified if the computation is performed using the Warsaw operator basis from the start. In fact, taking into account the contribution of the Fierz evanescent operator~$E_{QQ}$, we find agreement with the results obtained in~\cite{Boughezal:2019xpp}, except for obvious typographical mistakes in (38), (39) and (40) of the arXiv version 2 of that article. We~finally remark that the inclusion of the Fierz evanescent operator $E_{QQ}$ modifies most of the exact results presented in Appendix~\ref{app:exact}. However, when calculations are performed using the Warsaw operator basis from the beginning, only the dependence on $c_{QQ}^8$ of the one- and two-loop matching corrections are affected. 

\section{Anomalous contributions}
\label{app:anomalies}

In this appendix, we demonstrate that non-zero traces involving $\gamma_5$ emerge during the intermediate steps of the two-loop matching calculations outlined in~Section~\ref{sec:calculation} when using the Warsaw operator basis. It turns out that these anomalous contributions are all proportional to the Wilson coefficient $C_{qq}^{(3)}$ of the operator $Q_{qq}^{(3)}$ introduced in~(\ref{eq:WarsawKuKu}). Since the operator~$Q_{qq}^{(3)}$ is not present in the LHCTopWG operator basis~(\ref{eq:operators}) used in the main body of our article, the occurrence of non-zero traces with $\gamma_5$ is avoided entirely when this alternative set of operators is employed. 

If one insists on using the Warsaw operator basis, more care is required to handle contributions involving anomalous fermionic traces that arise in intermediate steps of our computations. The appearance of these terms might lead one to mistakenly conclude that arbitrary choices of the SMEFT Wilson coefficients could generally cause gauge anomalies. However, one can show~\cite{Bonnefoy:2020tyv,Feruglio:2020kfq,Cornella:2022hkc,Cohen:2023hmq,Cohen:2023gap} that the anomalous contributions depending on the Wilson coefficients are “irrelevant”, in the sense that they can always be removed by adding local counterterms,~i.e.~Wess-Zumino~(WZ) terms~\cite{Wess:1971yu}, to the SMEFT Lagrangian. As~a~result, the condition for the cancellation of “relevant” gauge anomalies in the SMEFT is the same as in the SM and only dependent on the gauge quantum numbers of the fermionic sector. These~findings align with the naive expectations from an EFT perspective: since the SMEFT encompasses scenarios involving heavy BSM physics where the full SM gauge symmetry is maintained, a matching computation within the same framework will always yield the necessary anomaly cancellation.

\begin{figure}[t!]
\begin{center}
\includegraphics[width=0.633\textwidth]{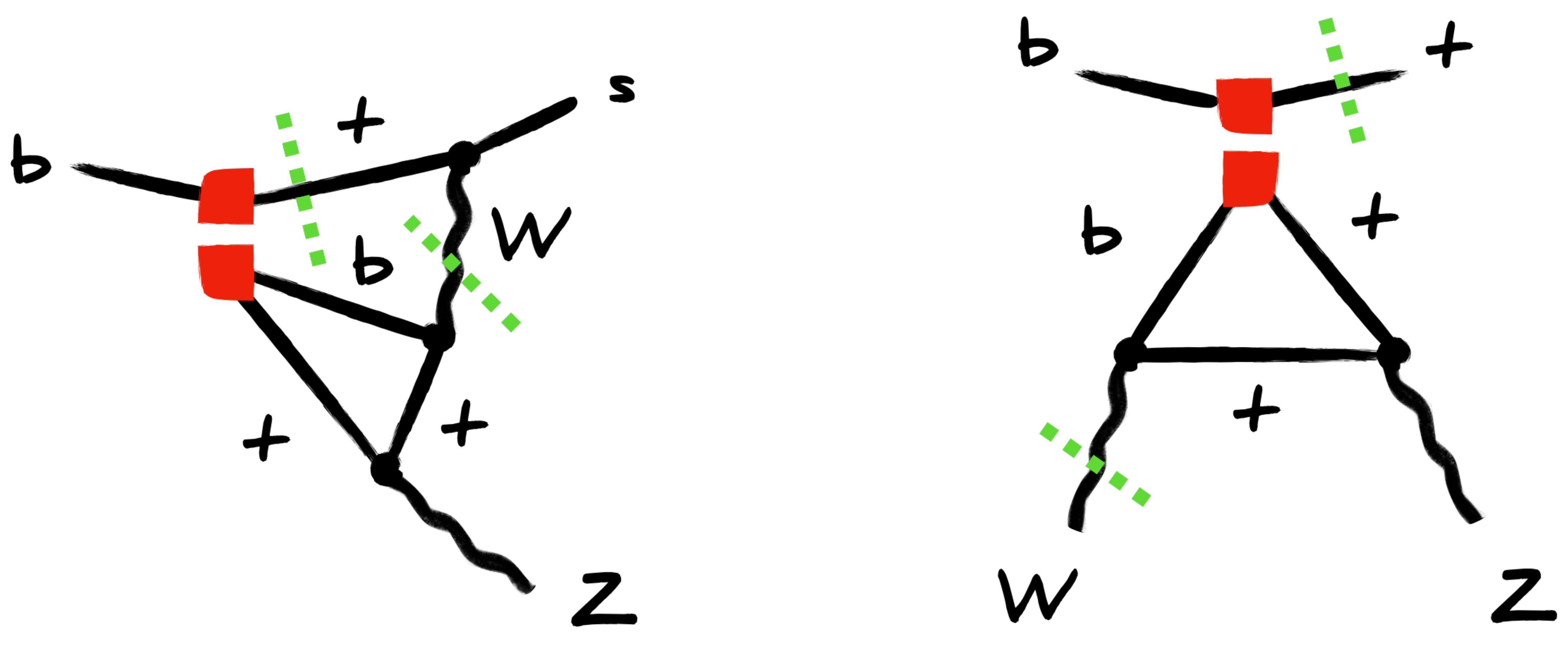}
\end{center}
\vspace{-2mm} 
\caption{\label{fig:anomaly} An example of an anomalous two-loop $b \to sZ$ penguin graph (left), arising from the insertion of the operator $Q_{qq}^{(3)}$ introduced in~(\ref{eq:WarsawKuKu}). Notice that the operator insertion (red squares) is constructed by joining two quarks belonging to the same quark bilinear. The anomalous one-loop subdiagram (right) therefore involves a fermionic trace. Note that the closed fermionic loop contains not only top, but also charm and up quarks, weighted by the relevant CKM factor. Similar graphs also have to be considered in the case of the off-shell $b \to s \gamma$ penguin.}
\end{figure}

After this general discussion, we will determine the set of local counterterms needed to cancel the spurious anomalous contributions that arise in our two-loop matching calculations. Similar WZ terms have been previously examined in \cite{Durieux:2018ggn,Degrande:2020evl}. Figure~\ref{fig:anomaly} shows representative anomalous Feynman diagrams that appear in the calculation of the $b \to sZ$ penguin amplitude in the Warsaw operator basis (\ref{eq:WarsawKuKu}). Similar diagrams also need to be considered for the off-shell $b \to s \gamma$ penguin. We denote the four-point correlator corresponding to the right graph in~Figure~\ref{fig:anomaly} as~$\Gamma_{\mu \nu} ^V (p_1,p_2)$, where $V = \gamma, Z$, and the Lorentz indices $\mu$ and $\nu$ are associated with the incoming momenta $p_1$ and $p_2$ of the~$V$~and the $W$ boson, respectively. The four-point correlator obeys the following anomalous Ward identities:
\beq \label{eq:WIs} 
p_1^\mu \hspace{0.5mm} \Gamma_{\mu \nu}^V (p_1,p_2) = i \hspace{0.25mm} a_V {\cal A}_{\nu} (p_1, p_2) \,, \qquad 
p_2^\nu \hspace{0.5mm} \Gamma_{\mu \nu}^V (p_1,p_2) = -i \hspace{0.25mm} a_V {\cal A}_{\mu} (p_1, p_2)
\eeq
with
\beq \label{eq:calA}
{\cal A}_{\mu} (p_1, p_2) = \frac{1}{2 \pi^2} \, \epsilon_{\mu \alpha \beta \gamma} \left (\bar t \hspace{0.5mm} \gamma^\alpha P_L \hspace{0.25mm} b \right ) p_1^\beta \hspace{0.25mm} p_2^\gamma \,, 
\eeq
where $\epsilon_{\mu \nu \alpha \beta}$ is the four-dimensional Levi-Civita tensor with $\epsilon_{0123} = 1$. To derive the results given in~(\ref{eq:WIs}), we have used the Breitenlohner-Maison-'t~Hooft-Veltman scheme~\cite{tHooft:1972tcz,Breitenlohner:1977hr,Breitenlohner:1975hg,Breitenlohner:1976te}. This scheme is algebraically consistent and introduces Dirac and Lorentz tensors living in~$d$,~$4$~and $2 \hspace{0.125mm} \epsilon$ dimensions, while $\gamma_5$ is a purely four-dimensional object. 

The coefficient $a_V$ encodes the dependence of the anomalous Ward identities~(\ref{eq:WIs}) on the type of involved neutral gauge boson. In the case of an external photon, we obtain 
\beq \label{eq:agamma}
a_\gamma = \frac{\sqrt{2} \hspace{0.25mm} \pi \alpha}{s_w} \left ( Q_d + Q_u \right ) V_{tb} \hspace{0.25mm} \frac{C_{qq}^{(3)}}{\Lambda^2} \, , 
\eeq
where $Q_d = -1/3$ and $Q_u = 2/3$ denote the electric charge of the down- and up-type quarks, respectively. Notice that in order to arrive at~(\ref{eq:agamma}) we made use of the unitarity of the CKM matrix,~i.e.~$\sum_{\psi = u,c,t} |V_{\psi b}|^2 = 1$. Since the photon couples vectorially, the expression~(\ref{eq:agamma}) receives only contributions from diagrams involving a single axial-vector current. For an external $Z$ boson, we instead find the following coefficient 
\beq \label{eq:aZ}
a_Z = -\frac{\sqrt{2} \hspace{0.25mm} \pi \alpha}{c_w} \left ( Q_d + Q_u \right ) V_{tb} \hspace{0.25mm} \frac{C_{qq}^{(3)}}{\Lambda^2} \, .
\eeq
Given that the third components of the weak isospin of the left-handed quarks satisfy the identity $T_d^3 = -T_u^3 = -1/2$, the expression~(\ref{eq:aZ}) does not receive any contribution from the axial-vector coupling of the $Z$ boson, which is proportional to $T_\psi^3$. As a result, the coefficients~(\ref{eq:agamma}) and (\ref{eq:aZ}) satifsy $a_Z = -s_w/c_w \hspace{0.5mm} a_\gamma$. 

The anomalous contributions~(\ref{eq:WIs}) proportional to $C_{qq}^{(3)}$ can now be cancelled by adding a suitable local counterterm to the SMEFT Lagrangian. The WZ term that cancels the triangle anomaly in the two-loop $b \to s \gamma$ and $b \to sZ$ penguin amplitudes, has the following~form: 
\beq \label{eq:LWZ} 
\begin{split}
{\cal L}_{\text{WZ}} & = -i \hspace{0.5mm} \frac{a_\gamma}{2 \pi^2} \, \epsilon_{\alpha \beta \gamma \delta} \left (\bar t \hspace{0.5mm} \gamma^\alpha P_L \hspace{0.25mm} b \right ) \left ( A^\beta \hspace{0.25mm} \partial^\delta W^{+ \gamma } + W^{+ \beta} \hspace{0.25mm} \partial^\delta \hspace{-0.75mm} A^\gamma \right ) \\[2mm]
& \phantom{xx} - i \hspace{0.5mm} \frac{a_Z}{2 \pi^2} \, \epsilon_{\alpha \beta \gamma \delta} \left (\bar t \hspace{0.5mm} \gamma^\alpha P_L \hspace{0.25mm} b \right ) \left ( Z^\beta \hspace{0.25mm} \partial^\delta W^{+ \gamma } + W^{+ \beta} \hspace{0.25mm} \partial^\delta \hspace{-0.5mm} Z^\gamma \right ) + {\rm h.c.}
 \end{split} 
\eeq
Here $A_\mu$, $Z_\mu$ and $W_\mu^+$ are the photon, $Z$- and positively charged $W$-boson field, respectively, and the expressions for the coefficients~$a_\gamma$ and $a_Z$ have already been given in~(\ref{eq:agamma}) and~(\ref{eq:aZ}). The~WZ~term~(\ref{eq:LWZ}) exactly cancels the anomalous contributions to the one-loop $\bar t b \to W \gamma$ and $\bar t b \to W Z$ subdiagrams~---~cf.~the right graph in Figure~\ref{fig:anomaly}. As~a~result, our final two-loop expressions for the $b \to s \gamma$ and $b \to sZ$ penguin amplitudes do not receive any anomalous contribution proportional to the Wilson coefficient $C_{qq}^{(3)}$, as they should not. We finally note that if the external current in~(\ref{eq:calA}) is $\bar c \hspace{0.5mm} \gamma^\alpha P_L \hspace{0.25mm} b$, the CKM factor $V_{cb}$ appears instead of~$V_{tb}$ in~(\ref{eq:agamma}),~(\ref{eq:aZ}) and~(\ref{eq:LWZ}). The~remaining arguments and formulas stay the same.
 
\section{Up-type FCNCs}
\label{app:upFCNCs}

In this appendix, we provide a concise discussion of the possible constraints on the Wilson coefficients of the operators introduced in~(\ref{eq:operators}) arising from up-type FCNCs. As in the main part of our work, we assume down-alignment. We begin our discussion with~$D$--$\bar D$~mixing, which receives tree-level corrections of the form 
\beq \label{eq:HeffDeltaC2}
{\cal H}_{\rm eff}^{\Delta C = 2} = -\frac{1}{2 \Lambda^2} \left ( c_{QQ}^1 + \frac{1}{3} \hspace{0.25mm} c_{QQ}^ 8 \right ) \left ( V_{cb}^\ast V_{ub} \right )^2 \hspace{0.25mm} \left ( \bar u \gamma_\mu P_L c \right ) \left ( \bar u \gamma^\mu P_L c \right ) + {\rm h.c.} \,, 
\eeq
due to the CKM rotations acting on the left-handed quark fields present in the operators ${\cal O}_{QQ}^1$ and ${\cal O}_{QQ}^8$. Consequently, $D$--$\bar D$~mixing can be used to constrain the specific combination of Wilson coefficients appearing in~(\ref{eq:HeffDeltaC2}). At the 95\% confidence level~(CL), we find the following limit
\beq \label{eq:Dmixbound}
\frac{\left | c_{QQ}^1 + \frac{1}{3} \hspace{0.25mm} c_{QQ}^ 8 \right |}{\Lambda^2} < \frac{1}{\left ( 200 \, {\rm GeV} \right )^2} \,,
\eeq
using the results on the model-independent constraints on $\Delta F = 2$ operators given in~\cite{LS}.  Comparable bounds have been obtained in~\cite{Aguilar-Saavedra:2018ksv,Allwicher:2023shc}. Note that, although the $D$--$\bar D$~mixing constraint arises at tree level, it is weak due to the strong CKM suppression evident in the LEFT Hamiltonian~(\ref{eq:HeffDeltaC2}). Because $D$--$\bar D$ mixing is the only $D$-meson process occurring at tree level, it offers the strongest constraint from that sector on $c_{QQ}^1$ and $c_{QQ}^8$. 

In the case of down-alignment, top-quark FCNCs are generated first at the one-loop level. To estimate the bounds on the Wilson coefficients of the operators in~(\ref{eq:operators}) arising from such processes, we use the decay $t \to cZ$ as an example. The leading term in the HTE expansion of the branching ratio for the $t \to cZ$ decay takes the following form
\beq \label{eq:BRtcZ}
{\rm BR} \left (t \to cZ \right ) \simeq \frac{25}{128 \pi^4 c_w^2} \frac{m_t^4}{\Lambda^4} \left | V_{tb}^\ast V_{ts} \right |^2 \left | c_{QQ}^1 - \frac{4}{15} \hspace{0.25mm} c_{QQ}^8 - \frac{6}{5} \hspace{0.25mm} c_{Qt}^1 \right |^2 \,,
\eeq
if the renormalisation scale $\mu$ is fixed to $m_t$. We add that~(\ref{eq:BRtcZ}) only involves the left-handed $tcZ$ coupling. In this case, the ATLAS measurement~\cite{ATLAS:2023qzr} sets a 95\%~CL limit of ${\rm BR} \left (t \to cZ \right ) < 1.3 \cdot 10^{-4}$. It follows that
\beq \label{eq;boundtcZ}
\frac{\left | c_{QQ}^1 - \frac{4}{15} \hspace{0.25mm} c_{QQ}^8 - \frac{6}{5} \hspace{0.25mm} c_{Qt}^1 \right |}{\Lambda^2} < \frac{1}{\left ( 70 \, {\rm GeV} \right )^2} \,,
\eeq
which, with regard to $c_{QQ}^1$ and $c_{QQ}^8$, is nominally even weaker than~(\ref{eq:Dmixbound}). Note that an estimate of the branching ratio for $t \to uZ$ can be obtained from~(\ref{eq:BRtcZ}) by replacing~$V_{ts}$~with~$V_{td}$. Since experimentally the bounds on ${\rm BR} \left (t \to uZ \right )$ are essentially the same as those on ${\rm BR} \left (t \to cZ \right )$ (see for instance \cite{ATLAS:2023qzr}), the sensitivity of $t \to uZ$ to the Wilson coefficients of the operators in~(\ref{eq:operators}) is less stringent than that shown in~(\ref{eq;boundtcZ}). The above discussion makes it clear that the processes $t \to cZ$ or $t \to uZ$ do not provide relevant constraints due to the combined effects of loop and CKM suppression. Similar arguments apply to all other top-quark FCNCs in the case of down-alignment.


\begin{thebibliography}{106}%
\makeatletter
\providecommand \@ifxundefined [1]{%
 \@ifx{#1\undefined}
}%
\providecommand \@ifnum [1]{%
 \ifnum #1\expandafter \@firstoftwo
 \else \expandafter \@secondoftwo
 \fi
}%
\providecommand \@ifx [1]{%
 \ifx #1\expandafter \@firstoftwo
 \else \expandafter \@secondoftwo
 \fi
}%
\providecommand \natexlab [1]{#1}%
\providecommand \enquote [1]{``#1''}%
\providecommand \bibnamefont [1]{#1}%
\providecommand \bibfnamefont [1]{#1}%
\providecommand \citenamefont [1]{#1}%
\providecommand \href@noop [0]{\@secondoftwo}%
\providecommand \href [0]{\begingroup \@sanitize@url \@href}%
\providecommand \@href[1]{\@@startlink{#1}\@@href}%
\providecommand \@@href[1]{\endgroup#1\@@endlink}%
\providecommand \@sanitize@url [0]{\catcode `\\12\catcode `\$12\catcode
 `\&12\catcode `\#12\catcode `\^12\catcode `\_12\catcode `\%12\relax}%
\providecommand \@@startlink[1]{}%
\providecommand \@@endlink[0]{}%
\providecommand \url [0]{\begingroup\@sanitize@url \@url }%
\providecommand \@url [1]{\endgroup\@href {#1}{\urlprefix }}%
\providecommand \urlprefix [0]{URL }%
\providecommand \Eprint [0]{\href }%
\providecommand \doibase [0]{http://dx.doi.org/}%
\providecommand \selectlanguage [0]{\@gobble}%
\providecommand \bibinfo [0]{\@secondoftwo}%
\providecommand \bibfield [0]{\@secondoftwo}%
\providecommand \translation [1]{[#1]}%
\providecommand \BibitemOpen [0]{}%
\providecommand \bibitemStop [0]{}%
\providecommand \bibitemNoStop [0]{.\EOS\space}%
\providecommand \EOS [0]{\spacefactor3000\relax}%
\providecommand\BibitemShut [1]{\csname bibitem#1\endcsname}%
\let\auto@bib@innerbib\@empty
\bibitem [{\citenamefont {Buchm{\"u}ller}\ and\ \citenamefont
 {Wyler}(1986)}]{Buchmuller:1985jz}%
 \BibitemOpen
 \bibfield {author} {\bibinfo {author} {\bibfnamefont {W.}~\bibnamefont
 {Buchm{\"u}ller}}\ and\ \bibinfo {author} {\bibfnamefont {D.}~\bibnamefont
 {Wyler}},\ }\href {\doibase 10.1016/0550-3213(86)90262-2} {\bibfield
 {journal} {\bibinfo {journal} {Nucl. Phys. B}\ }\textbf {\bibinfo {volume}
 {268}},\ \bibinfo {pages} {621} (\bibinfo {year} {1986})}\BibitemShut
 {NoStop}%
\bibitem [{\citenamefont {Grzadkowski}\ \emph {et~al.}(2010)\citenamefont
 {Grzadkowski}, \citenamefont {Iskrzynski}, \citenamefont {Misiak},\ and\
 \citenamefont {Rosiek}}]{Grzadkowski:2010es}%
 \BibitemOpen
 \bibfield {author} {\bibinfo {author} {\bibfnamefont {B.}~\bibnamefont
 {Grzadkowski}}, \bibinfo {author} {\bibfnamefont {M.}~\bibnamefont
 {Iskrzynski}}, \bibinfo {author} {\bibfnamefont {M.}~\bibnamefont {Misiak}},
 \ and\ \bibinfo {author} {\bibfnamefont {J.}~\bibnamefont {Rosiek}},\ }\href
 {\doibase 10.1007/JHEP10(2010)085} {\bibfield {journal} {\bibinfo {journal}
 {JHEP}\ }\textbf {\bibinfo {volume} {10}},\ \bibinfo {pages} {085} (\bibinfo
 {year} {2010})},\ \Eprint {http://arxiv.org/abs/1008.4884} {arXiv:1008.4884
 [hep-ph]}\BibitemShut {NoStop}%
\bibitem [{\citenamefont {Brivio}\ and\ \citenamefont
 {Trott}(2019)}]{Brivio:2017vri}%
 \BibitemOpen
 \bibfield {author} {\bibinfo {author} {\bibfnamefont {I.}~\bibnamefont
 {Brivio}}\ and\ \bibinfo {author} {\bibfnamefont {M.}~\bibnamefont {Trott}},\
 }\href {\doibase 10.1016/j.physrep.2018.11.002} {\bibfield {journal}
 {\bibinfo {journal} {Phys. Rept.}\ }\textbf {\bibinfo {volume} {793}},\
 \bibinfo {pages} {1} (\bibinfo {year} {2019})},\ \Eprint
 {http://arxiv.org/abs/1706.08945} {arXiv:1706.08945 [hep-ph]}\BibitemShut
 {NoStop}%
\bibitem [{\citenamefont {Isidori}\ \emph {et~al.}(2024)\citenamefont
 {Isidori}, \citenamefont {Wilsch},\ and\ \citenamefont
 {Wyler}}]{Isidori:2023pyp}%
 \BibitemOpen
 \bibfield {author} {\bibinfo {author} {\bibfnamefont {G.}~\bibnamefont
 {Isidori}}, \bibinfo {author} {\bibfnamefont {F.}~\bibnamefont {Wilsch}}, \
 and\ \bibinfo {author} {\bibfnamefont {D.}~\bibnamefont {Wyler}},\ }\href
 {\doibase 10.1103/RevModPhys.96.015006} {\bibfield {journal} {\bibinfo
 {journal} {Rev. Mod. Phys.}\ }\textbf {\bibinfo {volume} {96}},\ \bibinfo
 {pages} {015006} (\bibinfo {year} {2024})},\ \Eprint
 {http://arxiv.org/abs/2303.16922} {arXiv:2303.16922 [hep-ph]}\BibitemShut
 {NoStop}%
\bibitem [{\citenamefont {Brivio}\ \emph {et~al.}(2020)\citenamefont {Brivio},
 \citenamefont {Bruggisser}, \citenamefont {Maltoni}, \citenamefont
 {Moutafis}, \citenamefont {Plehn}, \citenamefont {Vryonidou}, \citenamefont
 {Westhoff},\ and\ \citenamefont {Zhang}}]{Brivio:2019ius}%
 \BibitemOpen
 \bibfield {author} {\bibinfo {author} {\bibfnamefont {I.}~\bibnamefont
 {Brivio}}, \bibinfo {author} {\bibfnamefont {S.}~\bibnamefont {Bruggisser}},
 \bibinfo {author} {\bibfnamefont {F.}~\bibnamefont {Maltoni}}, \bibinfo
 {author} {\bibfnamefont {R.}~\bibnamefont {Moutafis}}, \bibinfo {author}
 {\bibfnamefont {T.}~\bibnamefont {Plehn}}, \bibinfo {author} {\bibfnamefont
 {E.}~\bibnamefont {Vryonidou}}, \bibinfo {author} {\bibfnamefont
 {S.}~\bibnamefont {Westhoff}}, \ and\ \bibinfo {author} {\bibfnamefont
 {C.}~\bibnamefont {Zhang}},\ }\href {\doibase 10.1007/JHEP02(2020)131}
 {\bibfield {journal} {\bibinfo {journal} {JHEP}\ }\textbf {\bibinfo
 {volume} {02}},\ \bibinfo {pages} {131} (\bibinfo {year} {2020})},\ \Eprint
 {http://arxiv.org/abs/1910.03606} {arXiv:1910.03606 [hep-ph]}\BibitemShut
 {NoStop}%
\bibitem [{\citenamefont {Ellis}\ \emph {et~al.}(2021)\citenamefont {Ellis},
 \citenamefont {Madigan}, \citenamefont {Mimasu}, \citenamefont {Sanz},\ and\
 \citenamefont {You}}]{Ellis:2020unq}%
 \BibitemOpen
 \bibfield {author} {\bibinfo {author} {\bibfnamefont {J.}~\bibnamefont
 {Ellis}}, \bibinfo {author} {\bibfnamefont {M.}~\bibnamefont {Madigan}},
 \bibinfo {author} {\bibfnamefont {K.}~\bibnamefont {Mimasu}}, \bibinfo
 {author} {\bibfnamefont {V.}~\bibnamefont {Sanz}}, \ and\ \bibinfo {author}
 {\bibfnamefont {T.}~\bibnamefont {You}},\ }\href {\doibase
 10.1007/JHEP04(2021)279} {\bibfield {journal} {\bibinfo {journal} {JHEP}\
 }\textbf {\bibinfo {volume} {04}},\ \bibinfo {pages} {279} (\bibinfo {year}
 {2021})},\ \Eprint {http://arxiv.org/abs/2012.02779} {arXiv:2012.02779
 [hep-ph]}\BibitemShut {NoStop}%
\bibitem [{\citenamefont {Ethier}\ \emph {et~al.}(2021)\citenamefont {Ethier},
 \citenamefont {Magni}, \citenamefont {Maltoni}, \citenamefont {Mantani},
 \citenamefont {Nocera}, \citenamefont {Rojo}, \citenamefont {Slade},
 \citenamefont {Vryonidou},\ and\ \citenamefont {Zhang}}]{Ethier:2021bye}%
 \BibitemOpen
 \bibfield {author} {\bibinfo {author} {\bibfnamefont {J.~J.}\ \bibnamefont
 {Ethier}}, \bibinfo {author} {\bibfnamefont {G.}~\bibnamefont {Magni}},
 \bibinfo {author} {\bibfnamefont {F.}~\bibnamefont {Maltoni}}, \bibinfo
 {author} {\bibfnamefont {L.}~\bibnamefont {Mantani}}, \bibinfo {author}
 {\bibfnamefont {E.~R.}\ \bibnamefont {Nocera}}, \bibinfo {author}
 {\bibfnamefont {J.}~\bibnamefont {Rojo}}, \bibinfo {author} {\bibfnamefont
 {E.}~\bibnamefont {Slade}}, \bibinfo {author} {\bibfnamefont
 {E.}~\bibnamefont {Vryonidou}}, \ and\ \bibinfo {author} {\bibfnamefont
 {C.}~\bibnamefont {Zhang}} (\bibinfo {collaboration} {SMEFiT}),\ }\href
 {\doibase 10.1007/JHEP11(2021)089} {\bibfield {journal} {\bibinfo {journal}
 {JHEP}\ }\textbf {\bibinfo {volume} {11}},\ \bibinfo {pages} {089} (\bibinfo
 {year} {2021})},\ \Eprint {http://arxiv.org/abs/2105.00006} {arXiv:2105.00006
 [hep-ph]}\BibitemShut {NoStop}%
\bibitem [{ATL(2022)}]{ATL-PHYS-PUB-2022-037}%
 \BibitemOpen
 \href {https://cds.cern.ch/record/2816369} {\emph {\bibinfo {title}
 {{Combined effective field theory interpretation of Higgs boson and weak
 boson production and decay with ATLAS data and electroweak precision
 observables}}}},\ \bibinfo {type} {Tech. Rep.}\ (\bibinfo {institution}
 {CERN},\ \bibinfo {address} {Geneva},\ \bibinfo {year} {2022})\BibitemShut
 {NoStop}%
\bibitem [{\citenamefont {Celada}\ \emph {et~al.}(2024)\citenamefont {Celada},
 \citenamefont {Giani}, \citenamefont {ter Hoeve}, \citenamefont {Mantani},
 \citenamefont {Rojo}, \citenamefont {Rossia}, \citenamefont {Thomas},\ and\
 \citenamefont {Vryonidou}}]{Celada:2024mcf}%
 \BibitemOpen
 \bibfield {author} {\bibinfo {author} {\bibfnamefont {E.}~\bibnamefont
 {Celada}}, \bibinfo {author} {\bibfnamefont {T.}~\bibnamefont {Giani}},
 \bibinfo {author} {\bibfnamefont {J.}~\bibnamefont {ter Hoeve}}, \bibinfo
 {author} {\bibfnamefont {L.}~\bibnamefont {Mantani}}, \bibinfo {author}
 {\bibfnamefont {J.}~\bibnamefont {Rojo}}, \bibinfo {author} {\bibfnamefont
 {A.~N.}\ \bibnamefont {Rossia}}, \bibinfo {author} {\bibfnamefont {M.~O.~A.}\
 \bibnamefont {Thomas}}, \ and\ \bibinfo {author} {\bibfnamefont
 {E.}~\bibnamefont {Vryonidou}},\ }\href@noop {} {\ (\bibinfo {year}
 {2024})},\ \Eprint {http://arxiv.org/abs/2404.12809} {arXiv:2404.12809
 [hep-ph]}\BibitemShut {NoStop}%
\bibitem [{\citenamefont {Sirunyan}\ \emph
 {et~al.}(2020{\natexlab{a}})\citenamefont {Sirunyan} \emph
 {et~al.}}]{CMS:2019rvj}%
 \BibitemOpen
 \bibfield {author} {\bibinfo {author} {\bibfnamefont {A.~M.}\ \bibnamefont
 {Sirunyan}} \emph {et~al.} (\bibinfo {collaboration} {CMS}),\ }\href
 {\doibase 10.1140/epjc/s10052-019-7593-7} {\bibfield {journal} {\bibinfo
 {journal} {Eur. Phys. J. C}\ }\textbf {\bibinfo {volume} {80}},\ \bibinfo
 {pages} {75} (\bibinfo {year} {2020}{\natexlab{a}})},\ \Eprint
 {http://arxiv.org/abs/1908.06463} {arXiv:1908.06463 [hep-ex]}\BibitemShut
 {NoStop}%
\bibitem [{\citenamefont {Aad}\ \emph {et~al.}(2020)\citenamefont {Aad} \emph
 {et~al.}}]{ATLAS:2020hpj}%
 \BibitemOpen
 \bibfield {author} {\bibinfo {author} {\bibfnamefont {G.}~\bibnamefont
 {Aad}} \emph {et~al.} (\bibinfo {collaboration} {ATLAS}),\ }\href {\doibase
 10.1140/epjc/s10052-020-08509-3} {\bibfield {journal} {\bibinfo {journal}
 {Eur. Phys. J. C}\ }\textbf {\bibinfo {volume} {80}},\ \bibinfo {pages}
 {1085} (\bibinfo {year} {2020})},\ \Eprint {http://arxiv.org/abs/2007.14858}
 {arXiv:2007.14858 [hep-ex]}\BibitemShut {NoStop}%
\bibitem [{\citenamefont {Aad}\ \emph {et~al.}(2021)\citenamefont {Aad} \emph
 {et~al.}}]{ATLAS:2021kqb}%
 \BibitemOpen
 \bibfield {author} {\bibinfo {author} {\bibfnamefont {G.}~\bibnamefont
 {Aad}} \emph {et~al.} (\bibinfo {collaboration} {ATLAS}),\ }\href {\doibase
 10.1007/JHEP11(2021)118} {\bibfield {journal} {\bibinfo {journal} {JHEP}\
 }\textbf {\bibinfo {volume} {11}},\ \bibinfo {pages} {118} (\bibinfo {year}
 {2021})},\ \Eprint {http://arxiv.org/abs/2106.11683} {arXiv:2106.11683
 [hep-ex]}\BibitemShut {NoStop}%
\bibitem [{\citenamefont {Tumasyan}\ \emph {et~al.}(2023)\citenamefont
 {Tumasyan} \emph {et~al.}}]{CMS:2023zdh}%
 \BibitemOpen
 \bibfield {author} {\bibinfo {author} {\bibfnamefont {A.}~\bibnamefont
 {Tumasyan}} \emph {et~al.} (\bibinfo {collaboration} {CMS}),\ }\href
 {\doibase 10.1016/j.physletb.2023.138076} {\bibfield {journal} {\bibinfo
 {journal} {Phys. Lett. B}\ }\textbf {\bibinfo {volume} {844}},\ \bibinfo
 {pages} {138076} (\bibinfo {year} {2023})},\ \Eprint
 {http://arxiv.org/abs/2303.03864} {arXiv:2303.03864 [hep-ex]}\BibitemShut
 {NoStop}%
\bibitem [{\citenamefont {Aad}\ \emph {et~al.}(2023)\citenamefont {Aad} \emph
 {et~al.}}]{ATLAS:2023ajo}%
 \BibitemOpen
 \bibfield {author} {\bibinfo {author} {\bibfnamefont {G.}~\bibnamefont
 {Aad}} \emph {et~al.} (\bibinfo {collaboration} {ATLAS}),\ }\href {\doibase
 10.1140/epjc/s10052-023-11573-0} {\bibfield {journal} {\bibinfo {journal}
 {Eur. Phys. J. C}\ }\textbf {\bibinfo {volume} {83}},\ \bibinfo {pages} {496}
 (\bibinfo {year} {2023})},\ \Eprint {http://arxiv.org/abs/2303.15061}
 {arXiv:2303.15061 [hep-ex]}\BibitemShut {NoStop}%
\bibitem [{\citenamefont {Aaboud}\ \emph {et~al.}(2019)\citenamefont {Aaboud}
 \emph {et~al.}}]{ATLAS:2018fwl}%
 \BibitemOpen
 \bibfield {author} {\bibinfo {author} {\bibfnamefont {M.}~\bibnamefont
 {Aaboud}} \emph {et~al.} (\bibinfo {collaboration} {ATLAS}),\ }\href
 {\doibase 10.1007/JHEP04(2019)046} {\bibfield {journal} {\bibinfo {journal}
 {JHEP}\ }\textbf {\bibinfo {volume} {04}},\ \bibinfo {pages} {046} (\bibinfo
 {year} {2019})},\ \Eprint {http://arxiv.org/abs/1811.12113} {arXiv:1811.12113
 [hep-ex]}\BibitemShut {NoStop}%
\bibitem [{\citenamefont {Sirunyan}\ \emph
 {et~al.}(2020{\natexlab{b}})\citenamefont {Sirunyan} \emph
 {et~al.}}]{CMS:2019eih}%
 \BibitemOpen
 \bibfield {author} {\bibinfo {author} {\bibfnamefont {A.~M.}\ \bibnamefont
 {Sirunyan}} \emph {et~al.} (\bibinfo {collaboration} {CMS}),\ }\href
 {\doibase 10.1016/j.physletb.2020.135285} {\bibfield {journal} {\bibinfo
 {journal} {Phys. Lett. B}\ }\textbf {\bibinfo {volume} {803}},\ \bibinfo
 {pages} {135285} (\bibinfo {year} {2020}{\natexlab{b}})},\ \Eprint
 {http://arxiv.org/abs/1909.05306} {arXiv:1909.05306 [hep-ex]}\BibitemShut
 {NoStop}%
\bibitem [{\citenamefont {Hartland}\ \emph {et~al.}(2019)\citenamefont
 {Hartland}, \citenamefont {Maltoni}, \citenamefont {Nocera}, \citenamefont
 {Rojo}, \citenamefont {Slade}, \citenamefont {Vryonidou},\ and\ \citenamefont
 {Zhang}}]{Hartland:2019bjb}%
 \BibitemOpen
 \bibfield {author} {\bibinfo {author} {\bibfnamefont {N.~P.}\ \bibnamefont
 {Hartland}}, \bibinfo {author} {\bibfnamefont {F.}~\bibnamefont {Maltoni}},
 \bibinfo {author} {\bibfnamefont {E.~R.}\ \bibnamefont {Nocera}}, \bibinfo
 {author} {\bibfnamefont {J.}~\bibnamefont {Rojo}}, \bibinfo {author}
 {\bibfnamefont {E.}~\bibnamefont {Slade}}, \bibinfo {author} {\bibfnamefont
 {E.}~\bibnamefont {Vryonidou}}, \ and\ \bibinfo {author} {\bibfnamefont
 {C.}~\bibnamefont {Zhang}},\ }\href {\doibase 10.1007/JHEP04(2019)100}
 {\bibfield {journal} {\bibinfo {journal} {JHEP}\ }\textbf {\bibinfo
 {volume} {04}},\ \bibinfo {pages} {100} (\bibinfo {year} {2019})},\ \Eprint
 {http://arxiv.org/abs/1901.05965} {arXiv:1901.05965 [hep-ph]}\BibitemShut
 {NoStop}%
\bibitem [{\citenamefont {Degrande}\ \emph {et~al.}(2024)\citenamefont
 {Degrande}, \citenamefont {Rosenfeld},\ and\ \citenamefont
 {Vasquez}}]{Degrande:2024mbg}%
 \BibitemOpen
 \bibfield {author} {\bibinfo {author} {\bibfnamefont {C.}~\bibnamefont
 {Degrande}}, \bibinfo {author} {\bibfnamefont {R.}~\bibnamefont {Rosenfeld}},
 \ and\ \bibinfo {author} {\bibfnamefont {A.}~\bibnamefont {Vasquez}},\
 }\href@noop {} {\ (\bibinfo {year} {2024})},\ \Eprint
 {http://arxiv.org/abs/2402.06528} {arXiv:2402.06528 [hep-ph]}\BibitemShut
 {NoStop}%
\bibitem [{\citenamefont {de~Blas}\ \emph {et~al.}(2015)\citenamefont
 {de~Blas}, \citenamefont {Chala},\ and\ \citenamefont
 {Santiago}}]{deBlas:2015aea}%
 \BibitemOpen
 \bibfield {author} {\bibinfo {author} {\bibfnamefont {J.}~\bibnamefont
 {de~Blas}}, \bibinfo {author} {\bibfnamefont {M.}~\bibnamefont {Chala}}, \
 and\ \bibinfo {author} {\bibfnamefont {J.}~\bibnamefont {Santiago}},\ }\href
 {\doibase 10.1007/JHEP09(2015)189} {\bibfield {journal} {\bibinfo {journal}
 {JHEP}\ }\textbf {\bibinfo {volume} {09}},\ \bibinfo {pages} {189} (\bibinfo
 {year} {2015})},\ \Eprint {http://arxiv.org/abs/1507.00757} {arXiv:1507.00757
 [hep-ph]}\BibitemShut {NoStop}%
\bibitem [{\citenamefont {Degrande}\ \emph {et~al.}(2021)\citenamefont
 {Degrande}, \citenamefont {Durieux}, \citenamefont {Maltoni}, \citenamefont
 {Mimasu}, \citenamefont {Vryonidou},\ and\ \citenamefont
 {Zhang}}]{Degrande:2020evl}%
 \BibitemOpen
 \bibfield {author} {\bibinfo {author} {\bibfnamefont {C.}~\bibnamefont
 {Degrande}}, \bibinfo {author} {\bibfnamefont {G.}~\bibnamefont {Durieux}},
 \bibinfo {author} {\bibfnamefont {F.}~\bibnamefont {Maltoni}}, \bibinfo
 {author} {\bibfnamefont {K.}~\bibnamefont {Mimasu}}, \bibinfo {author}
 {\bibfnamefont {E.}~\bibnamefont {Vryonidou}}, \ and\ \bibinfo {author}
 {\bibfnamefont {C.}~\bibnamefont {Zhang}},\ }\href {\doibase
 10.1103/PhysRevD.103.096024} {\bibfield {journal} {\bibinfo {journal}
 {Phys. Rev. D}\ }\textbf {\bibinfo {volume} {103}},\ \bibinfo {pages}
 {096024} (\bibinfo {year} {2021})},\ \Eprint
 {http://arxiv.org/abs/2008.11743} {arXiv:2008.11743 [hep-ph]}\BibitemShut
 {NoStop}%
\bibitem [{\citenamefont {Boughezal}\ \emph {et~al.}(2019)\citenamefont
 {Boughezal}, \citenamefont {Chen}, \citenamefont {Petriello},\ and\
 \citenamefont {Wiegand}}]{Boughezal:2019xpp}%
 \BibitemOpen
 \bibfield {author} {\bibinfo {author} {\bibfnamefont {R.}~\bibnamefont
 {Boughezal}}, \bibinfo {author} {\bibfnamefont {C.-Y.}\ \bibnamefont {Chen}},
 \bibinfo {author} {\bibfnamefont {F.}~\bibnamefont {Petriello}}, \ and\
 \bibinfo {author} {\bibfnamefont {D.}~\bibnamefont {Wiegand}},\ }\href
 {\doibase 10.1103/PhysRevD.100.056023} {\bibfield {journal} {\bibinfo
 {journal} {Phys. Rev. D}\ }\textbf {\bibinfo {volume} {100}},\ \bibinfo
 {pages} {056023} (\bibinfo {year} {2019})},\ \Eprint
 {http://arxiv.org/abs/1907.00997} {arXiv:1907.00997 [hep-ph]}\BibitemShut
 {NoStop}%
\bibitem [{\citenamefont {Hartmann}\ \emph {et~al.}(2017)\citenamefont
 {Hartmann}, \citenamefont {Shepherd},\ and\ \citenamefont
 {Trott}}]{Hartmann:2016pil}%
 \BibitemOpen
 \bibfield {author} {\bibinfo {author} {\bibfnamefont {C.}~\bibnamefont
 {Hartmann}}, \bibinfo {author} {\bibfnamefont {W.}~\bibnamefont {Shepherd}},
 \ and\ \bibinfo {author} {\bibfnamefont {M.}~\bibnamefont {Trott}},\ }\href
 {\doibase 10.1007/JHEP03(2017)060} {\bibfield {journal} {\bibinfo {journal}
 {JHEP}\ }\textbf {\bibinfo {volume} {03}},\ \bibinfo {pages} {060} (\bibinfo
 {year} {2017})},\ \Eprint {http://arxiv.org/abs/1611.09879} {arXiv:1611.09879
 [hep-ph]}\BibitemShut {NoStop}%
\bibitem [{\citenamefont {Dawson}\ and\ \citenamefont
 {Giardino}(2022)}]{Dawson:2022bxd}%
 \BibitemOpen
 \bibfield {author} {\bibinfo {author} {\bibfnamefont {S.}~\bibnamefont
 {Dawson}}\ and\ \bibinfo {author} {\bibfnamefont {P.~P.}\ \bibnamefont
 {Giardino}},\ }\href {\doibase 10.1103/PhysRevD.105.073006} {\bibfield
 {journal} {\bibinfo {journal} {Phys. Rev. D}\ }\textbf {\bibinfo {volume}
 {105}},\ \bibinfo {pages} {073006} (\bibinfo {year} {2022})},\ \Eprint
 {http://arxiv.org/abs/2201.09887} {arXiv:2201.09887 [hep-ph]}\BibitemShut
 {NoStop}%
 \bibitem [{\citenamefont {Haisch}\ and\ \citenamefont
 {Schnell}(2025)}]{inprep}%
 \BibitemOpen
 \bibfield {author} {\bibinfo {author} {\bibfnamefont {U.}~\bibnamefont
 {Haisch}}\ and\ \bibinfo {author} {\bibfnamefont {L.}\ \bibnamefont
 {Schnell}},} in preparation\BibitemShut
 {NoStop}%
\bibitem [{\citenamefont {Gauld}\ \emph
 {et~al.}(2016{\natexlab{a}})\citenamefont {Gauld}, \citenamefont {Pecjak},\
 and\ \citenamefont {Scott}}]{Gauld:2015lmb}%
 \BibitemOpen
 \bibfield {author} {\bibinfo {author} {\bibfnamefont {R.}~\bibnamefont
 {Gauld}}, \bibinfo {author} {\bibfnamefont {B.~D.}\ \bibnamefont {Pecjak}}, \
 and\ \bibinfo {author} {\bibfnamefont {D.~J.}\ \bibnamefont {Scott}},\ }\href
 {\doibase 10.1007/JHEP05(2016)080} {\bibfield {journal} {\bibinfo {journal}
 {JHEP}\ }\textbf {\bibinfo {volume} {05}},\ \bibinfo {pages} {080} (\bibinfo
 {year} {2016}{\natexlab{a}})},\ \Eprint {http://arxiv.org/abs/1512.02508}
 {arXiv:1512.02508 [hep-ph]}\BibitemShut {NoStop}%
\bibitem [{\citenamefont {Alasfar}\ \emph {et~al.}(2022)\citenamefont
 {Alasfar}, \citenamefont {de~Blas},\ and\ \citenamefont
 {Gr\"ober}}]{Alasfar:2022zyr}%
 \BibitemOpen
 \bibfield {author} {\bibinfo {author} {\bibfnamefont {L.}~\bibnamefont
 {Alasfar}}, \bibinfo {author} {\bibfnamefont {J.}~\bibnamefont {de~Blas}}, \
 and\ \bibinfo {author} {\bibfnamefont {R.}~\bibnamefont {Gr\"ober}},\ }\href
 {\doibase 10.1007/JHEP05(2022)111} {\bibfield {journal} {\bibinfo {journal}
 {JHEP}\ }\textbf {\bibinfo {volume} {05}},\ \bibinfo {pages} {111} (\bibinfo
 {year} {2022})},\ \Eprint {http://arxiv.org/abs/2202.02333} {arXiv:2202.02333
 [hep-ph]}\BibitemShut {NoStop}%
\bibitem [{\citenamefont {Di~Noi}\ \emph {et~al.}(2024)\citenamefont {Di~Noi},
 \citenamefont {Gr\"ober}, \citenamefont {Heinrich}, \citenamefont {Lang},\
 and\ \citenamefont {Vitti}}]{DiNoi:2023ygk}%
 \BibitemOpen
 \bibfield {author} {\bibinfo {author} {\bibfnamefont {S.}~\bibnamefont
 {Di~Noi}}, \bibinfo {author} {\bibfnamefont {R.}~\bibnamefont {Gr\"ober}},
 \bibinfo {author} {\bibfnamefont {G.}~\bibnamefont {Heinrich}}, \bibinfo
 {author} {\bibfnamefont {J.}~\bibnamefont {Lang}}, \ and\ \bibinfo {author}
 {\bibfnamefont {M.}~\bibnamefont {Vitti}},\ }\href {\doibase
 10.1103/PhysRevD.109.095024} {\bibfield {journal} {\bibinfo {journal}
 {Phys. Rev. D}\ }\textbf {\bibinfo {volume} {109}},\ \bibinfo {pages}
 {095024} (\bibinfo {year} {2024})},\ \Eprint
 {http://arxiv.org/abs/2310.18221} {arXiv:2310.18221 [hep-ph]}\BibitemShut
 {NoStop}%
\bibitem [{\citenamefont {Heinrich}\ and\ \citenamefont
 {Lang}(2024)}]{Heinrich:2023rsd}%
 \BibitemOpen
 \bibfield {author} {\bibinfo {author} {\bibfnamefont {G.}~\bibnamefont
 {Heinrich}}\ and\ \bibinfo {author} {\bibfnamefont {J.}~\bibnamefont
 {Lang}},\ }\href {\doibase 10.1007/JHEP05(2024)121} {\bibfield {journal}
 {\bibinfo {journal} {JHEP}\ }\textbf {\bibinfo {volume} {05}},\ \bibinfo
 {pages} {121} (\bibinfo {year} {2024})},\ \Eprint
 {http://arxiv.org/abs/2311.15004} {arXiv:2311.15004 [hep-ph]}\BibitemShut
 {NoStop}%
\bibitem [{\citenamefont {Peskin}\ and\ \citenamefont
 {Takeuchi}(1992)}]{Peskin:1991sw}%
 \BibitemOpen
 \bibfield {author} {\bibinfo {author} {\bibfnamefont {M.~E.}\ \bibnamefont
 {Peskin}}\ and\ \bibinfo {author} {\bibfnamefont {T.}~\bibnamefont
 {Takeuchi}},\ }\href {\doibase 10.1103/PhysRevD.46.381} {\bibfield {journal}
 {\bibinfo {journal} {Phys. Rev. D}\ }\textbf {\bibinfo {volume} {46}},\
 \bibinfo {pages} {381} (\bibinfo {year} {1992})}\BibitemShut {NoStop}%
\bibitem [{\citenamefont {Barducci}\ \emph {et~al.}(2018)\citenamefont
 {Barducci} \emph {et~al.}}]{Aguilar-Saavedra:2018ksv}%
 \BibitemOpen
 \bibfield {author} {\bibinfo {author} {\bibfnamefont {D.}~\bibnamefont
 {Barducci}} \emph {et~al.},\ }\href@noop {} {\ (\bibinfo {year} {2018})},\
 \Eprint {http://arxiv.org/abs/1802.07237} {arXiv:1802.07237 [hep-ph]}\BibitemShut {NoStop}%
\bibitem [{\citenamefont {Alloul}\ \emph {et~al.}(2014)\citenamefont {Alloul},
 \citenamefont {Christensen}, \citenamefont {Degrande}, \citenamefont {Duhr},\
 and\ \citenamefont {Fuks}}]{Alloul:2013bka}%
 \BibitemOpen
 \bibfield {author} {\bibinfo {author} {\bibfnamefont {A.}~\bibnamefont
 {Alloul}}, \bibinfo {author} {\bibfnamefont {N.~D.}\ \bibnamefont
 {Christensen}}, \bibinfo {author} {\bibfnamefont {C.}~\bibnamefont
 {Degrande}}, \bibinfo {author} {\bibfnamefont {C.}~\bibnamefont {Duhr}}, \
 and\ \bibinfo {author} {\bibfnamefont {B.}~\bibnamefont {Fuks}},\ }\href
 {\doibase 10.1016/j.cpc.2014.04.012} {\bibfield {journal} {\bibinfo
 {journal} {Comput. Phys. Commun.}\ }\textbf {\bibinfo {volume} {185}},\
 \bibinfo {pages} {2250} (\bibinfo {year} {2014})},\ \Eprint
 {http://arxiv.org/abs/1310.1921} {arXiv:1310.1921 [hep-ph]}\BibitemShut
 {NoStop}%
\bibitem [{\citenamefont {Durieux}\ and\ \citenamefont {Zhang}()}]{dim6top}%
 \BibitemOpen
 \bibfield {author} {\bibinfo {author} {\bibfnamefont {G.}~\bibnamefont
 {Durieux}}\ and\ \bibinfo {author} {\bibfnamefont {C.}~\bibnamefont
 {Zhang}},\ } (\bibinfo {year} {2018}), \href {https://feynrules.irmp.ucl.ac.be/wiki/dim6top} {\emph
 {\bibinfo {title} {{\tt dim6top}}}}.\BibitemShut {Stop}%
 \bibitem [{\citenamefont {Haisch}\ and\ \citenamefont {Schnell}()}]{analyticresults}%
 \BibitemOpen
 \bibfield {author} {\bibinfo {author} {\bibfnamefont {U.}~\bibnamefont
 {Haisch} \ and \bibfnamefont {L.}~\bibnamefont {Schnell}, } (\bibinfo {year} {2024}), \href
 {https://gitlab.com/lucschnell/4-quark-top}
 {\emph {\bibinfo {title} {{Precision tests of third-generation four-quark operators (GitLab)}}}}\BibitemShut {NoStop}%
\bibitem [{\citenamefont {Hahn}(2001)}]{Hahn:2000kx}%
 \BibitemOpen
 \bibfield {author} {\bibinfo {author} {\bibfnamefont {T.}~\bibnamefont
 {Hahn}},\ }\href {\doibase 10.1016/S0010-4655(01)00290-9} {\bibfield
 {journal} {\bibinfo {journal} {Comput. Phys. Commun.}\ }\textbf {\bibinfo
 {volume} {140}},\ \bibinfo {pages} {418} (\bibinfo {year} {2001})},\ \Eprint
 {http://arxiv.org/abs/hep-ph/0012260} {arXiv:hep-ph/0012260}\BibitemShut
 {NoStop}%
\bibitem [{\citenamefont {Shtabovenko}\ \emph {et~al.}(2020)\citenamefont
 {Shtabovenko}, \citenamefont {Mertig},\ and\ \citenamefont
 {Orellana}}]{Shtabovenko:2020gxv}%
 \BibitemOpen
 \bibfield {author} {\bibinfo {author} {\bibfnamefont {V.}~\bibnamefont
 {Shtabovenko}}, \bibinfo {author} {\bibfnamefont {R.}~\bibnamefont {Mertig}},
 \ and\ \bibinfo {author} {\bibfnamefont {F.}~\bibnamefont {Orellana}},\
 }\href {\doibase 10.1016/j.cpc.2020.107478} {\bibfield {journal} {\bibinfo
 {journal} {Comput. Phys. Commun.}\ }\textbf {\bibinfo {volume} {256}},\
 \bibinfo {pages} {107478} (\bibinfo {year} {2020})},\ \Eprint
 {http://arxiv.org/abs/2001.04407} {arXiv:2001.04407 [hep-ph]}\BibitemShut
 {NoStop}%
\bibitem [{\citenamefont {Hahn}\ \emph {et~al.}(2016)\citenamefont {Hahn},
 \citenamefont {Pa\ss{}ehr},\ and\ \citenamefont
 {Schappacher}}]{Hahn:2016ebn}%
 \BibitemOpen
 \bibfield {author} {\bibinfo {author} {\bibfnamefont {T.}~\bibnamefont
 {Hahn}}, \bibinfo {author} {\bibfnamefont {S.}~\bibnamefont {Pa\ss{}ehr}}, \
 and\ \bibinfo {author} {\bibfnamefont {C.}~\bibnamefont {Schappacher}},\
 }\href {\doibase 10.1088/1742-6596/762/1/012065} {\bibfield {journal}
 {\bibinfo {journal} {PoS}\ }\textbf {\bibinfo {volume} {LL2016}},\ \bibinfo
 {pages} {068} (\bibinfo {year} {2016})},\ \Eprint
 {http://arxiv.org/abs/1604.04611} {arXiv:1604.04611 [hep-ph]}\BibitemShut
 {NoStop}%
\bibitem [{\citenamefont {Lee}(2014)}]{Lee:2013mka}%
 \BibitemOpen
 \bibfield {author} {\bibinfo {author} {\bibfnamefont {R.~N.}\ \bibnamefont
 {Lee}},\ }\href {\doibase 10.1088/1742-6596/523/1/012059} {\bibfield
 {journal} {\bibinfo {journal} {J. Phys. Conf. Ser.}\ }\textbf {\bibinfo
 {volume} {523}},\ \bibinfo {pages} {012059} (\bibinfo {year} {2014})},\
 \Eprint {http://arxiv.org/abs/1310.1145} {arXiv:1310.1145 [hep-ph]}
\BibitemShut {NoStop}%
\bibitem [{\citenamefont {Patel}(2015)}]{Patel:2015tea}%
 \BibitemOpen
 \bibfield {author} {\bibinfo {author} {\bibfnamefont {H.~H.}\ \bibnamefont
 {Patel}},\ }\href {\doibase 10.1016/j.cpc.2015.08.017} {\bibfield {journal}
 {\bibinfo {journal} {Comput. Phys. Commun.}\ }\textbf {\bibinfo {volume}
 {197}},\ \bibinfo {pages} {276} (\bibinfo {year} {2015})},\ \Eprint
 {http://arxiv.org/abs/1503.01469} {arXiv:1503.01469 [hep-ph]}\BibitemShut
 {NoStop}%
 \bibitem [{\citenamefont {Beneke}\ and\ \citenamefont
 {Smirnov}(1998)}]{Beneke:1997zp}%
 \BibitemOpen
 \bibfield {author} {\bibinfo {author} {\bibfnamefont {M.}~\bibnamefont
 {Beneke}}\ and\ \bibinfo {author} {\bibfnamefont {V.~A.}\ \bibnamefont
 {Smirnov}},\ }\href {\doibase 10.1016/S0550-3213(98)00138-2} {\bibfield
 {journal} {\bibinfo {journal} {Nucl. Phys. B}\ }\textbf {\bibinfo {volume}
 {522}},\ \bibinfo {pages} {321} (\bibinfo {year} {1998})},\ \Eprint
 {http://arxiv.org/abs/hep-ph/9711391} {arXiv:hep-ph/9711391}\BibitemShut
 {NoStop}%
\bibitem [{\citenamefont {Smirnov}(2002)}]{Smirnov:2002pj}%
 \BibitemOpen
 \bibfield {author} {\bibinfo {author} {\bibfnamefont {V.~A.}\ \bibnamefont
 {Smirnov}},\ }\href@noop {} {\bibfield {journal} {\bibinfo {journal}
 {Springer Tracts Mod. Phys.}\ }\textbf {\bibinfo {volume} {177}},\ \bibinfo
 {pages} {1} (\bibinfo {year} {2002})}\BibitemShut {NoStop}%
\bibitem [{\citenamefont {Jantzen}(2011)}]{Jantzen:2011nz}%
 \BibitemOpen
 \bibfield {author} {\bibinfo {author} {\bibfnamefont {B.}~\bibnamefont
 {Jantzen}},\ }\href {\doibase 10.1007/JHEP12(2011)076} {\bibfield {journal}
 {\bibinfo {journal} {JHEP}\ }\textbf {\bibinfo {volume} {12}},\ \bibinfo
 {pages} {076} (\bibinfo {year} {2011})},\ \Eprint
 {http://arxiv.org/abs/1111.2589} {arXiv:1111.2589 [hep-ph]}\BibitemShut
 {NoStop}%
\bibitem [{\citenamefont {Fleischer}\ and\ \citenamefont
 {Tarasov}(1994)}]{Fleischer:1994ef}%
 \BibitemOpen
 \bibfield {author} {\bibinfo {author} {\bibfnamefont {J.}~\bibnamefont
 {Fleischer}}\ and\ \bibinfo {author} {\bibfnamefont {O.~V.}\ \bibnamefont
 {Tarasov}},\ }\href {\doibase 10.1007/BF01560102} {\bibfield {journal}
 {\bibinfo {journal} {Z. Phys. C}\ }\textbf {\bibinfo {volume} {64}},\
 \bibinfo {pages} {413} (\bibinfo {year} {1994})},\ \Eprint
 {http://arxiv.org/abs/hep-ph/9403230} {arXiv:hep-ph/9403230}\BibitemShut
 {NoStop}%
\bibitem [{\citenamefont {Tarasov}(1995)}]{Tarasov:1995jf}%
 \BibitemOpen
 \bibfield {author} {\bibinfo {author} {\bibfnamefont {O.~V.}\ \bibnamefont
 {Tarasov}},\ }in\ \href@noop {} {\emph {\bibinfo {booktitle} {{4th
 International Workshop on Software Engineering and Artificial Intelligence
 for High-energy and Nuclear Physics}}}}\ (\bibinfo {year} {1995})\ \Eprint
 {http://arxiv.org/abs/hep-ph/9505277} {arXiv:hep-ph/9505277}\BibitemShut
 {NoStop}%
\bibitem [{\citenamefont {Davydychev}\ and\ \citenamefont
 {Tausk}(1993)}]{Davydychev:1992mt}%
 \BibitemOpen
 \bibfield {author} {\bibinfo {author} {\bibfnamefont {A.~I.}\ \bibnamefont
 {Davydychev}}\ and\ \bibinfo {author} {\bibfnamefont {J.~B.}\ \bibnamefont
 {Tausk}},\ }\href {\doibase 10.1016/0550-3213(93)90338-P} {\bibfield
 {journal} {\bibinfo {journal} {Nucl. Phys. B}\ }\textbf {\bibinfo {volume}
 {397}},\ \bibinfo {pages} {123} (\bibinfo {year} {1993})}\BibitemShut
 {NoStop}%
\bibitem [{\citenamefont {Korner}\ \emph {et~al.}(2005)\citenamefont {Korner},
 \citenamefont {Merebashvili},\ and\ \citenamefont {Rogal}}]{Korner:2004rr}%
 \BibitemOpen
 \bibfield {author} {\bibinfo {author} {\bibfnamefont {J.~G.}\ \bibnamefont
 {K{\"o}rner}}, \bibinfo {author} {\bibfnamefont {Z.}~\bibnamefont
 {Merebashvili}}, \ and\ \bibinfo {author} {\bibfnamefont {M.}~\bibnamefont
 {Rogal}},\ }\href {\doibase 10.1103/PhysRevD.71.054028} {\bibfield {journal}
 {\bibinfo {journal} {Phys. Rev. D}\ }\textbf {\bibinfo {volume} {71}},\
 \bibinfo {pages} {054028} (\bibinfo {year} {2005})},\ \Eprint
 {http://arxiv.org/abs/hep-ph/0412088} {arXiv:hep-ph/0412088}\BibitemShut
 {NoStop}%
 \bibitem [{\citenamefont {Jenkins}\ \emph {et~al.}(2014)\citenamefont
 {Jenkins}, \citenamefont {Manohar},\ and\ \citenamefont
 {Trott}}]{Jenkins:2013wua}%
 \BibitemOpen
 \bibfield {author} {\bibinfo {author} {\bibfnamefont {E.~E.}\ \bibnamefont
 {Jenkins}}, \bibinfo {author} {\bibfnamefont {A.~V.}\ \bibnamefont
 {Manohar}}, \ and\ \bibinfo {author} {\bibfnamefont {M.}~\bibnamefont
 {Trott}},\ }\href {\doibase 10.1007/JHEP01(2014)035} {\bibfield {journal}
 {\bibinfo {journal} {JHEP}\ }\textbf {\bibinfo {volume} {01}},\ \bibinfo
 {pages} {035} (\bibinfo {year} {2014})},\ \Eprint
 {http://arxiv.org/abs/1310.4838} {arXiv:1310.4838 [hep-ph]}\BibitemShut
 {NoStop}%
\bibitem [{\citenamefont {Alonso}\ \emph {et~al.}(2014)\citenamefont {Alonso},
 \citenamefont {Jenkins}, \citenamefont {Manohar},\ and\ \citenamefont
 {Trott}}]{Alonso:2013hga}%
 \BibitemOpen
 \bibfield {author} {\bibinfo {author} {\bibfnamefont {R.}~\bibnamefont
 {Alonso}}, \bibinfo {author} {\bibfnamefont {E.~E.}\ \bibnamefont {Jenkins}},
 \bibinfo {author} {\bibfnamefont {A.~V.}\ \bibnamefont {Manohar}}, \ and\
 \bibinfo {author} {\bibfnamefont {M.}~\bibnamefont {Trott}},\ }\href
 {\doibase 10.1007/JHEP04(2014)159} {\bibfield {journal} {\bibinfo {journal}
 {JHEP}\ }\textbf {\bibinfo {volume} {04}},\ \bibinfo {pages} {159} (\bibinfo
 {year} {2014})},\ \Eprint {http://arxiv.org/abs/1312.2014} {arXiv:1312.2014
 [hep-ph]}\BibitemShut {NoStop}%
\bibitem [{\citenamefont {Gauld}\ \emph
 {et~al.}(2016{\natexlab{b}})\citenamefont {Gauld}, \citenamefont {Pecjak},\
 and\ \citenamefont {Scott}}]{Gauld:2016kuu}%
 \BibitemOpen
 \bibfield {author} {\bibinfo {author} {\bibfnamefont {R.}~\bibnamefont
 {Gauld}}, \bibinfo {author} {\bibfnamefont {B.~D.}\ \bibnamefont {Pecjak}}, \
 and\ \bibinfo {author} {\bibfnamefont {D.~J.}\ \bibnamefont {Scott}},\ }\href
 {\doibase 10.1103/PhysRevD.94.074045} {\bibfield {journal} {\bibinfo
 {journal} {Phys. Rev. D}\ }\textbf {\bibinfo {volume} {94}},\ \bibinfo
 {pages} {074045} (\bibinfo {year} {2016}{\natexlab{b}})},\ \Eprint
 {http://arxiv.org/abs/1607.06354} {arXiv:1607.06354 [hep-ph]}\BibitemShut
 {NoStop}%
\bibitem [{\citenamefont {Dawson}\ and\ \citenamefont
 {Giardino}(2018)}]{Dawson:2018pyl}%
 \BibitemOpen
 \bibfield {author} {\bibinfo {author} {\bibfnamefont {S.}~\bibnamefont
 {Dawson}}\ and\ \bibinfo {author} {\bibfnamefont {P.~P.}\ \bibnamefont
 {Giardino}},\ }\href {\doibase 10.1103/PhysRevD.97.093003} {\bibfield
 {journal} {\bibinfo {journal} {Phys. Rev. D}\ }\textbf {\bibinfo {volume}
 {97}},\ \bibinfo {pages} {093003} (\bibinfo {year} {2018})},\ \Eprint
 {http://arxiv.org/abs/1801.01136} {arXiv:1801.01136 [hep-ph]}\BibitemShut
 {NoStop}%
\bibitem [{\citenamefont {Cullen}\ \emph {et~al.}(2019)\citenamefont {Cullen},
 \citenamefont {Pecjak},\ and\ \citenamefont {Scott}}]{Cullen:2019nnr}%
 \BibitemOpen
 \bibfield {author} {\bibinfo {author} {\bibfnamefont {J.~M.}\ \bibnamefont
 {Cullen}}, \bibinfo {author} {\bibfnamefont {B.~D.}\ \bibnamefont {Pecjak}},
 \ and\ \bibinfo {author} {\bibfnamefont {D.~J.}\ \bibnamefont {Scott}},\
 }\href {\doibase 10.1007/JHEP08(2019)173} {\bibfield {journal} {\bibinfo
 {journal} {JHEP}\ }\textbf {\bibinfo {volume} {08}},\ \bibinfo {pages} {173}
 (\bibinfo {year} {2019})},\ \Eprint {http://arxiv.org/abs/1904.06358}
 {arXiv:1904.06358 [hep-ph]}\BibitemShut {NoStop}%
\bibitem [{\citenamefont {Cullen}\ and\ \citenamefont
 {Pecjak}(2020)}]{Cullen:2020zof}%
 \BibitemOpen
 \bibfield {author} {\bibinfo {author} {\bibfnamefont {J.~M.}\ \bibnamefont
 {Cullen}}\ and\ \bibinfo {author} {\bibfnamefont {B.~D.}\ \bibnamefont
 {Pecjak}},\ }\href {\doibase 10.1007/JHEP11(2020)079} {\bibfield {journal}
 {\bibinfo {journal} {JHEP}\ }\textbf {\bibinfo {volume} {11}},\ \bibinfo
 {pages} {079} (\bibinfo {year} {2020})},\ \Eprint
 {http://arxiv.org/abs/2007.15238} {arXiv:2007.15238 [hep-ph]}\BibitemShut
 {NoStop}%
\bibitem [{\citenamefont {Chetyrkin}\ \emph
 {et~al.}(1998{\natexlab{a}})\citenamefont {Chetyrkin}, \citenamefont
 {Misiak},\ and\ \citenamefont {M{\"u}nz}}]{Chetyrkin:1997fm}%
 \BibitemOpen
 \bibfield {author} {\bibinfo {author} {\bibfnamefont {K.~G.}\ \bibnamefont
 {Chetyrkin}}, \bibinfo {author} {\bibfnamefont {M.}~\bibnamefont {Misiak}}, \
 and\ \bibinfo {author} {\bibfnamefont {M.}~\bibnamefont {M{\"u}nz}},\ }\href
 {\doibase 10.1016/S0550-3213(98)00122-9} {\bibfield {journal} {\bibinfo
 {journal} {Nucl. Phys. B}\ }\textbf {\bibinfo {volume} {518}},\ \bibinfo
 {pages} {473} (\bibinfo {year} {1998}{\natexlab{a}})},\ \Eprint
 {http://arxiv.org/abs/hep-ph/9711266} {arXiv:hep-ph/9711266}\BibitemShut
 {NoStop}%
\bibitem [{\citenamefont {Gambino}\ \emph {et~al.}(2003)\citenamefont
 {Gambino}, \citenamefont {Gorbahn},\ and\ \citenamefont
 {Haisch}}]{Gambino:2003zm}%
 \BibitemOpen
 \bibfield {author} {\bibinfo {author} {\bibfnamefont {P.}~\bibnamefont
 {Gambino}}, \bibinfo {author} {\bibfnamefont {M.}~\bibnamefont {Gorbahn}}, \
 and\ \bibinfo {author} {\bibfnamefont {U.}~\bibnamefont {Haisch}},\ }\href
 {\doibase 10.1016/j.nuclphysb.2003.09.024} {\bibfield {journal} {\bibinfo
 {journal} {Nucl. Phys. B}\ }\textbf {\bibinfo {volume} {673}},\ \bibinfo
 {pages} {238} (\bibinfo {year} {2003})},\ \Eprint
 {http://arxiv.org/abs/hep-ph/0306079} {arXiv:hep-ph/0306079}\BibitemShut
 {NoStop}%
\bibitem [{\citenamefont {Allwicher}\ \emph {et~al.}(2024)\citenamefont
 {Allwicher}, \citenamefont {Cornella}, \citenamefont {Isidori},\ and\
 \citenamefont {Stefanek}}]{Allwicher:2023shc}%
 \BibitemOpen
 \bibfield {author} {\bibinfo {author} {\bibfnamefont {L.}~\bibnamefont
 {Allwicher}}, \bibinfo {author} {\bibfnamefont {C.}~\bibnamefont {Cornella}},
 \bibinfo {author} {\bibfnamefont {G.}~\bibnamefont {Isidori}}, \ and\
 \bibinfo {author} {\bibfnamefont {B.~A.}\ \bibnamefont {Stefanek}},\ }\href
 {\doibase 10.1007/JHEP03(2024)049} {\bibfield {journal} {\bibinfo {journal}
 {JHEP}\ }\textbf {\bibinfo {volume} {03}},\ \bibinfo {pages} {049} (\bibinfo
 {year} {2024})},\ \Eprint {http://arxiv.org/abs/2311.00020} {arXiv:2311.00020
 [hep-ph]}\BibitemShut {NoStop}%
\bibitem [{\citenamefont {Bobeth}\ \emph {et~al.}(2000)\citenamefont {Bobeth},
 \citenamefont {Misiak},\ and\ \citenamefont {Urban}}]{Bobeth:1999mk}%
 \BibitemOpen
 \bibfield {author} {\bibinfo {author} {\bibfnamefont {C.}~\bibnamefont
 {Bobeth}}, \bibinfo {author} {\bibfnamefont {M.}~\bibnamefont {Misiak}}, \
 and\ \bibinfo {author} {\bibfnamefont {J.}~\bibnamefont {Urban}},\ }\href
 {\doibase 10.1016/S0550-3213(00)00007-9} {\bibfield {journal} {\bibinfo
 {journal} {Nucl. Phys. B}\ }\textbf {\bibinfo {volume} {574}},\ \bibinfo
 {pages} {291} (\bibinfo {year} {2000})},\ \Eprint
 {http://arxiv.org/abs/hep-ph/9910220} {arXiv:hep-ph/9910220}\BibitemShut
 {NoStop}%
\bibitem [{\citenamefont {Liu}\ and\ \citenamefont {Ma}(2023)}]{Liu:2022chg}%
 \BibitemOpen
 \bibfield {author} {\bibinfo {author} {\bibfnamefont {X.}~\bibnamefont
 {Liu}}\ and\ \bibinfo {author} {\bibfnamefont {Y.-Q.}\ \bibnamefont {Ma}},\
 }\href {\doibase 10.1016/j.cpc.2022.108565} {\bibfield {journal} {\bibinfo
 {journal} {Comput. Phys. Commun.}\ }\textbf {\bibinfo {volume} {283}},\
 \bibinfo {pages} {108565} (\bibinfo {year} {2023})},\ \Eprint
 {http://arxiv.org/abs/2201.11669} {arXiv:2201.11669 [hep-ph]}\BibitemShut
 {NoStop}%
\bibitem [{\citenamefont {Chetyrkin}\ \emph {et~al.}(1997)\citenamefont
 {Chetyrkin}, \citenamefont {Misiak},\ and\ \citenamefont
 {M{\"u}nz}}]{Chetyrkin:1996vx}%
 \BibitemOpen
 \bibfield {author} {\bibinfo {author} {\bibfnamefont {K.~G.}\ \bibnamefont
 {Chetyrkin}}, \bibinfo {author} {\bibfnamefont {M.}~\bibnamefont {Misiak}}, \
 and\ \bibinfo {author} {\bibfnamefont {M.}~\bibnamefont {M{\"u}nz}},\ }\href
 {\doibase 10.1016/S0370-2693(97)00324-9} {\bibfield {journal} {\bibinfo
 {journal} {Phys. Lett. B}\ }\textbf {\bibinfo {volume} {400}},\ \bibinfo
 {pages} {206} (\bibinfo {year} {1997})},\ \bibinfo {note} {[Erratum:
 Phys. Lett. B {\bf 425}, 414 (1998)]},\ \Eprint {http://arxiv.org/abs/hep-ph/9612313}
 {arXiv:hep-ph/9612313}\BibitemShut {NoStop}%
\bibitem [{\citenamefont {Chetyrkin}\ \emph
 {et~al.}(1998{\natexlab{b}})\citenamefont {Chetyrkin}, \citenamefont
 {Misiak},\ and\ \citenamefont {M{\"u}nz}}]{Chetyrkin:1997gb}%
 \BibitemOpen
 \bibfield {author} {\bibinfo {author} {\bibfnamefont {K.~G.}\ \bibnamefont
 {Chetyrkin}}, \bibinfo {author} {\bibfnamefont {M.}~\bibnamefont {Misiak}}, \
 and\ \bibinfo {author} {\bibfnamefont {M.}~\bibnamefont {M{\"u}nz}},\ }\href
 {\doibase 10.1016/S0550-3213(98)00131-X} {\bibfield {journal} {\bibinfo
 {journal} {Nucl. Phys. B}\ }\textbf {\bibinfo {volume} {520}},\ \bibinfo
 {pages} {279} (\bibinfo {year} {1998}{\natexlab{b}})},\ \Eprint
 {http://arxiv.org/abs/hep-ph/9711280} {arXiv:hep-ph/9711280}\BibitemShut
 {NoStop}%
\bibitem [{\citenamefont {Misiak}\ and\ \citenamefont
 {Urban}(1999)}]{Misiak:1999yg}%
 \BibitemOpen
 \bibfield {author} {\bibinfo {author} {\bibfnamefont {M.}~\bibnamefont
 {Misiak}}\ and\ \bibinfo {author} {\bibfnamefont {J.}~\bibnamefont {Urban}},\
 }\href {\doibase 10.1016/S0370-2693(99)00150-1} {\bibfield {journal}
 {\bibinfo {journal} {Phys. Lett. B}\ }\textbf {\bibinfo {volume} {451}},\
 \bibinfo {pages} {161} (\bibinfo {year} {1999})},\ \Eprint
 {http://arxiv.org/abs/hep-ph/9901278} {arXiv:hep-ph/9901278}\BibitemShut
 {NoStop}%
\bibitem [{\citenamefont {Gorbahn}\ and\ \citenamefont
 {Haisch}(2005)}]{Gorbahn:2004my}%
 \BibitemOpen
 \bibfield {author} {\bibinfo {author} {\bibfnamefont {M.}~\bibnamefont
 {Gorbahn}}\ and\ \bibinfo {author} {\bibfnamefont {U.}~\bibnamefont
 {Haisch}},\ }\href {\doibase 10.1016/j.nuclphysb.2005.01.047} {\bibfield
 {journal} {\bibinfo {journal} {Nucl. Phys. B}\ }\textbf {\bibinfo {volume}
 {713}},\ \bibinfo {pages} {291} (\bibinfo {year} {2005})},\ \Eprint
 {http://arxiv.org/abs/hep-ph/0411071} {arXiv:hep-ph/0411071}\BibitemShut
 {NoStop}%
\bibitem [{\citenamefont {Buras}\ \emph {et~al.}(1990)\citenamefont {Buras},
 \citenamefont {Jamin},\ and\ \citenamefont {Weisz}}]{Buras:1990fn}%
 \BibitemOpen
 \bibfield {author} {\bibinfo {author} {\bibfnamefont {A.~J.}\ \bibnamefont
 {Buras}}, \bibinfo {author} {\bibfnamefont {M.}~\bibnamefont {Jamin}}, \ and\
 \bibinfo {author} {\bibfnamefont {P.~H.}\ \bibnamefont {Weisz}},\ }\href
 {\doibase 10.1016/0550-3213(90)90373-L} {\bibfield {journal} {\bibinfo
 {journal} {Nucl. Phys. B}\ }\textbf {\bibinfo {volume} {347}},\ \bibinfo
 {pages} {491} (\bibinfo {year} {1990})}\BibitemShut {NoStop}%
\bibitem [{\citenamefont {Herrlich}\ and\ \citenamefont
 {Nierste}(1996)}]{Herrlich:1996vf}%
 \BibitemOpen
 \bibfield {author} {\bibinfo {author} {\bibfnamefont {S.}~\bibnamefont
 {Herrlich}}\ and\ \bibinfo {author} {\bibfnamefont {U.}~\bibnamefont
 {Nierste}},\ }\href {\doibase 10.1016/0550-3213(96)00324-0} {\bibfield
 {journal} {\bibinfo {journal} {Nucl. Phys. B}\ }\textbf {\bibinfo {volume}
 {476}},\ \bibinfo {pages} {27} (\bibinfo {year} {1996})},\ \Eprint
 {http://arxiv.org/abs/hep-ph/9604330} {arXiv:hep-ph/9604330}\BibitemShut
 {NoStop}%
\bibitem [{\citenamefont {Cella}\ \emph {et~al.}(1994)\citenamefont {Cella},
 \citenamefont {Curci}, \citenamefont {Ricciardi},\ and\ \citenamefont
 {Vicere}}]{Cella:1994np}%
 \BibitemOpen
 \bibfield {author} {\bibinfo {author} {\bibfnamefont {G.}~\bibnamefont
 {Cella}}, \bibinfo {author} {\bibfnamefont {G.}~\bibnamefont {Curci}},
 \bibinfo {author} {\bibfnamefont {G.}~\bibnamefont {Ricciardi}}, \ and\
 \bibinfo {author} {\bibfnamefont {A.}~\bibnamefont {Vicere}},\ }\href
 {\doibase 10.1016/0550-3213(94)90212-7} {\bibfield {journal} {\bibinfo
 {journal} {Nucl. Phys. B}\ }\textbf {\bibinfo {volume} {431}},\ \bibinfo
 {pages} {417} (\bibinfo {year} {1994})},\ \Eprint
 {http://arxiv.org/abs/hep-ph/9406203} {arXiv:hep-ph/9406203}\BibitemShut
 {NoStop}%
\bibitem [{\citenamefont {Haisch}(2002)}]{Haisch:2002zz}%
 \BibitemOpen
 \bibfield {author} {\bibinfo {author} {\bibfnamefont {U.}~\bibnamefont
 {Haisch}},\ } (\bibinfo {year} {2002}), \href {https://mediatum.ub.tum.de/doc/602925/602925.pdf} {\emph
 {\bibinfo {title} {{The Inclusive Radiative $B \to X_s \gamma$ Decay in the Standard Model}}}}\BibitemShut {NoStop}%
\bibitem [{\citenamefont {Burdman}\ \emph {et~al.}(2000)\citenamefont
 {Burdman}, \citenamefont {Chivukula},\ and\ \citenamefont
 {Evans}}]{Burdman:1999us}%
 \BibitemOpen
 \bibfield {author} {\bibinfo {author} {\bibfnamefont {G.}~\bibnamefont
 {Burdman}}, \bibinfo {author} {\bibfnamefont {R.~S.}\ \bibnamefont
 {Chivukula}}, \ and\ \bibinfo {author} {\bibfnamefont {N.~J.}\ \bibnamefont
 {Evans}},\ }\href {\doibase 10.1103/PhysRevD.61.035009} {\bibfield {journal}
 {\bibinfo {journal} {Phys. Rev. D}\ }\textbf {\bibinfo {volume} {61}},\
 \bibinfo {pages} {035009} (\bibinfo {year} {2000})},\ \Eprint
 {http://arxiv.org/abs/hep-ph/9906292} {arXiv:hep-ph/9906292}\BibitemShut
 {NoStop}%
\bibitem [{\citenamefont {Haisch}\ and\ \citenamefont
 {Westhoff}(2011)}]{Haisch:2011up}%
 \BibitemOpen
 \bibfield {author} {\bibinfo {author} {\bibfnamefont {U.}~\bibnamefont
 {Haisch}}\ and\ \bibinfo {author} {\bibfnamefont {S.}~\bibnamefont
 {Westhoff}},\ }\href {\doibase 10.1007/JHEP08(2011)088} {\bibfield {journal}
 {\bibinfo {journal} {JHEP}\ }\textbf {\bibinfo {volume} {08}},\ \bibinfo
 {pages} {088} (\bibinfo {year} {2011})},\ \Eprint
 {http://arxiv.org/abs/1106.0529} {arXiv:1106.0529 [hep-ph]}\BibitemShut
 {NoStop}%
\bibitem [{\citenamefont {Allwicher}\ \emph {et~al.}(2023)\citenamefont
 {Allwicher}, \citenamefont {Isidori}, \citenamefont {Lizana}, \citenamefont
 {Selimovic},\ and\ \citenamefont {Stefanek}}]{Allwicher:2023aql}%
 \BibitemOpen
 \bibfield {author} {\bibinfo {author} {\bibfnamefont {L.}~\bibnamefont
 {Allwicher}}, \bibinfo {author} {\bibfnamefont {G.}~\bibnamefont {Isidori}},
 \bibinfo {author} {\bibfnamefont {J.~M.}\ \bibnamefont {Lizana}}, \bibinfo
 {author} {\bibfnamefont {N.}~\bibnamefont {Selimovic}}, \ and\ \bibinfo
 {author} {\bibfnamefont {B.~A.}\ \bibnamefont {Stefanek}},\ }\href {\doibase
 10.1007/JHEP05(2023)179} {\bibfield {journal} {\bibinfo {journal} {JHEP}\
 }\textbf {\bibinfo {volume} {05}},\ \bibinfo {pages} {179} (\bibinfo {year}
 {2023})},\ \Eprint {http://arxiv.org/abs/2302.11584} {arXiv:2302.11584
 [hep-ph]}\BibitemShut {NoStop}%
\bibitem [{\citenamefont {Garosi}\ \emph {et~al.}(2023)\citenamefont {Garosi},
 \citenamefont {Marzocca}, \citenamefont {Rodriguez-Sanchez},\ and\
 \citenamefont {Stanzione}}]{Garosi:2023yxg}%
 \BibitemOpen
 \bibfield {author} {\bibinfo {author} {\bibfnamefont {F.}~\bibnamefont
 {Garosi}}, \bibinfo {author} {\bibfnamefont {D.}~\bibnamefont {Marzocca}},
 \bibinfo {author} {\bibfnamefont {A.}~\bibnamefont {Rodriguez-Sanchez}}, \
 and\ \bibinfo {author} {\bibfnamefont {A.}~\bibnamefont {Stanzione}},\ }\href
 {\doibase 10.1007/JHEP12(2023)129} {\bibfield {journal} {\bibinfo {journal}
 {JHEP}\ }\textbf {\bibinfo {volume} {12}},\ \bibinfo {pages} {129} (\bibinfo
 {year} {2023})},\ \Eprint {http://arxiv.org/abs/2310.00047} {arXiv:2310.00047
 [hep-ph]}\BibitemShut {NoStop}%
 \bibitem [{\citenamefont {Stefanek}(2024)}]{Stefanek:2024kds}%
 \BibitemOpen
 \bibfield {author} {\bibinfo {author} {\bibfnamefont {B.~A.}\ \bibnamefont
 {Stefanek}},\ }\href {\doibase 10.1007/JHEP09(2024)103} {\bibfield {journal}
 {\bibinfo {journal} {JHEP}\ }\textbf {\bibinfo {volume} {09}},\ \bibinfo
 {pages} {103} (\bibinfo {year} {2024})},\ \Eprint
 {http://arxiv.org/abs/2407.09593} {arXiv:2407.09593 [hep-ph]}\BibitemShut
 {NoStop}%
\bibitem [{\citenamefont {Manohar}\ and\ \citenamefont
 {Wise}(2000)}]{Manohar:2000dt}%
 \BibitemOpen
 \bibfield {author} {\bibinfo {author} {\bibfnamefont {A.~V.}\ \bibnamefont
 {Manohar}}\ and\ \bibinfo {author} {\bibfnamefont {M.~B.}\ \bibnamefont
 {Wise}},\ }\href@noop {} {\emph {\bibinfo {title} {{Heavy quark physics}}}},\
 Vol.~\bibinfo {volume} {10}\ (\bibinfo {year} {2000})\BibitemShut {NoStop}%
\bibitem [{\citenamefont {Sirlin}(1982)}]{Sirlin:1981ie}%
 \BibitemOpen
 \bibfield {author} {\bibinfo {author} {\bibfnamefont {A.}~\bibnamefont
 {Sirlin}},\ }\href {\doibase 10.1016/0550-3213(82)90303-0} {\bibfield
 {journal} {\bibinfo {journal} {Nucl. Phys. B}\ }\textbf {\bibinfo {volume}
 {196}},\ \bibinfo {pages} {83} (\bibinfo {year} {1982})}\BibitemShut
 {NoStop}%
\bibitem [{\citenamefont {Brod}\ and\ \citenamefont
 {Gorbahn}(2008)}]{Brod:2008ss}%
 \BibitemOpen
 \bibfield {author} {\bibinfo {author} {\bibfnamefont {J.}~\bibnamefont
 {Brod}}\ and\ \bibinfo {author} {\bibfnamefont {M.}~\bibnamefont {Gorbahn}},\
 }\href {\doibase 10.1103/PhysRevD.78.034006} {\bibfield {journal} {\bibinfo
 {journal} {Phys. Rev. D}\ }\textbf {\bibinfo {volume} {78}},\ \bibinfo
 {pages} {034006} (\bibinfo {year} {2008})},\ \Eprint
 {http://arxiv.org/abs/0805.4119} {arXiv:0805.4119 [hep-ph]}\BibitemShut
 {NoStop}%
\bibitem [{\citenamefont {Gorbahn}\ \emph {et~al.}(2023)\citenamefont
 {Gorbahn}, \citenamefont {J\"ager}, \citenamefont {Moretti},\ and\
 \citenamefont {van~der Merwe}}]{Gorbahn:2022rgl}%
 \BibitemOpen
 \bibfield {author} {\bibinfo {author} {\bibfnamefont {M.}~\bibnamefont
 {Gorbahn}}, \bibinfo {author} {\bibfnamefont {S.}~\bibnamefont {J\"ager}},
 \bibinfo {author} {\bibfnamefont {F.}~\bibnamefont {Moretti}}, \ and\
 \bibinfo {author} {\bibfnamefont {E.}~\bibnamefont {van~der Merwe}},\ }\href
 {\doibase 10.1007/JHEP01(2023)159} {\bibfield {journal} {\bibinfo {journal}
 {JHEP}\ }\textbf {\bibinfo {volume} {01}},\ \bibinfo {pages} {159} (\bibinfo
 {year} {2023})},\ \Eprint {http://arxiv.org/abs/2209.05289} {arXiv:2209.05289
 [hep-ph]}\BibitemShut {NoStop}%
\bibitem [{\citenamefont {Bigi}\ \emph {et~al.}(2023)\citenamefont {Bigi},
 \citenamefont {Bordone}, \citenamefont {Gambino}, \citenamefont {Haisch},\
 and\ \citenamefont {Piccione}}]{Bigi:2023cbv}%
 \BibitemOpen
 \bibfield {author} {\bibinfo {author} {\bibfnamefont {D.}~\bibnamefont
 {Bigi}}, \bibinfo {author} {\bibfnamefont {M.}~\bibnamefont {Bordone}},
 \bibinfo {author} {\bibfnamefont {P.}~\bibnamefont {Gambino}}, \bibinfo
 {author} {\bibfnamefont {U.}~\bibnamefont {Haisch}}, \ and\ \bibinfo {author}
 {\bibfnamefont {A.}~\bibnamefont {Piccione}},\ }\href {\doibase
 10.1007/JHEP11(2023)163} {\bibfield {journal} {\bibinfo {journal} {JHEP}\
 }\textbf {\bibinfo {volume} {11}},\ \bibinfo {pages} {163} (\bibinfo {year}
 {2023})},\ \Eprint {http://arxiv.org/abs/2309.02849} {arXiv:2309.02849
 [hep-ph]}\BibitemShut {NoStop}%
\bibitem [{\citenamefont {Workman}\ \emph {et~al.}(2022)\citenamefont {Workman}
 \emph {et~al.}}]{ParticleDataGroup:2022pth}%
 \BibitemOpen
 \bibfield {author} {\bibinfo {author} {\bibfnamefont {R.~L.}\ \bibnamefont
 {Workman}} \emph {et~al.} (\bibinfo {collaboration} {Particle Data Group}),\
 }\href {\doibase 10.1093/ptep/ptac097} {\bibfield {journal} {\bibinfo
 {journal} {PTEP}\ }\textbf {\bibinfo {volume} {2022}},\ \bibinfo {pages}
 {083C01} (\bibinfo {year} {2022})}\BibitemShut {NoStop}%
\bibitem [{\citenamefont {Schael}\ \emph {et~al.}(2006)\citenamefont {Schael}
 \emph {et~al.}}]{ALEPH:2005ab}%
 \BibitemOpen
 \bibfield {author} {\bibinfo {author} {\bibfnamefont {S.}~\bibnamefont
 {Schael}} \emph {et~al.} (\bibinfo {collaboration} {ALEPH, DELPHI, L3, OPAL,
 SLD, LEP Electroweak Working Group, SLD Electroweak Group, SLD Heavy Flavour
 Group}),\ }\href {\doibase 10.1016/j.physrep.2005.12.006} {\bibfield
 {journal} {\bibinfo {journal} {Phys. Rept.}\ }\textbf {\bibinfo {volume}
 {427}},\ \bibinfo {pages} {257} (\bibinfo {year} {2006})},\ \Eprint
 {http://arxiv.org/abs/hep-ex/0509008} {arXiv:hep-ex/0509008}\BibitemShut
 {NoStop}%
\bibitem [{\citenamefont {Workman}\ and\ \citenamefont
 {Others}(2022)}]{Workman:2022ynf}%
 \BibitemOpen
 \bibfield {author} {\bibinfo {author} {\bibfnamefont {R.~L.}\ \bibnamefont
 {Workman}}\ and\ \bibinfo {author} {\bibnamefont {Others}} (\bibinfo
 {collaboration} {Particle Data Group}),\ }\href {\doibase
 10.1093/ptep/ptac097} {\bibfield {journal} {\bibinfo {journal} {PTEP}\
 }\textbf {\bibinfo {volume} {2022}},\ \bibinfo {pages} {083C01} (\bibinfo
 {year} {2022})}\BibitemShut {NoStop}%
\bibitem [{\citenamefont {Field}(1998)}]{Field:1997gz}%
 \BibitemOpen
 \bibfield {author} {\bibinfo {author} {\bibfnamefont {J.~H.}\ \bibnamefont
 {Field}},\ }\href {\doibase 10.1142/S0217732398002059} {\bibfield {journal}
 {\bibinfo {journal} {Mod. Phys. Lett. A}\ }\textbf {\bibinfo {volume}
 {13}},\ \bibinfo {pages} {1937} (\bibinfo {year} {1998})},\ \Eprint
 {http://arxiv.org/abs/hep-ph/9801355} {arXiv:hep-ph/9801355}\BibitemShut
 {NoStop}%
\bibitem [{\citenamefont {Chanowitz}(2001)}]{Chanowitz:2001bv}%
 \BibitemOpen
 \bibfield {author} {\bibinfo {author} {\bibfnamefont {M.~S.}\ \bibnamefont
 {Chanowitz}},\ }\href {\doibase 10.1103/PhysRevLett.87.231802} {\bibfield
 {journal} {\bibinfo {journal} {Phys. Rev. Lett.}\ }\textbf {\bibinfo
 {volume} {87}},\ \bibinfo {pages} {231802} (\bibinfo {year} {2001})},\
 \Eprint {http://arxiv.org/abs/hep-ph/0104024} {arXiv:hep-ph/0104024}
\BibitemShut {NoStop}%
\bibitem [{\citenamefont {Dawson}\ and\ \citenamefont {Giardino}()}]{SDPPG}%
 \BibitemOpen
 \bibfield {author} {\bibinfo {author} {\bibfnamefont {S.}~\bibnamefont
 {Dawson}}\ and\ \bibinfo {author} {\bibfnamefont {P.~P.}\ \bibnamefont
 {Giardino}},\ } (\bibinfo {year} {2022}), \href
 {https://journals.aps.org/prd/supplemental/10.1103/PhysRevD.105.073006}
 {Numerical\_Results\_NLO4f.txt}\BibitemShut{NoStop}%
\bibitem [{\citenamefont {Czarnecki}\ \emph {et~al.}(2010)\citenamefont
 {Czarnecki}, \citenamefont {K{\"o}rner},\ and\ \citenamefont
 {Piclum}}]{Czarnecki:2010gb}%
 \BibitemOpen
 \bibfield {author} {\bibinfo {author} {\bibfnamefont {A.}~\bibnamefont
 {Czarnecki}}, \bibinfo {author} {\bibfnamefont {J.~G.}\ \bibnamefont
 {K{\"o}rner}}, \ and\ \bibinfo {author} {\bibfnamefont {J.~H.}\ \bibnamefont
 {Piclum}},\ }\href {\doibase 10.1103/PhysRevD.81.111503} {\bibfield
 {journal} {\bibinfo {journal} {Phys. Rev. D}\ }\textbf {\bibinfo {volume}
 {81}},\ \bibinfo {pages} {111503} (\bibinfo {year} {2010})},\ \Eprint
 {http://arxiv.org/abs/1005.2625} {arXiv:1005.2625 [hep-ph]}\BibitemShut
 {NoStop}%
\bibitem [{\citenamefont {Gao}\ \emph {et~al.}(2013)\citenamefont {Gao},
 \citenamefont {Li},\ and\ \citenamefont {Zhu}}]{Gao:2012ja}%
 \BibitemOpen
 \bibfield {author} {\bibinfo {author} {\bibfnamefont {J.}~\bibnamefont
 {Gao}}, \bibinfo {author} {\bibfnamefont {C.~S.}\ \bibnamefont {Li}}, \ and\
 \bibinfo {author} {\bibfnamefont {H.~X.}\ \bibnamefont {Zhu}},\ }\href
 {\doibase 10.1103/PhysRevLett.110.042001} {\bibfield {journal} {\bibinfo
 {journal} {Phys. Rev. Lett.}\ }\textbf {\bibinfo {volume} {110}},\ \bibinfo
 {pages} {042001} (\bibinfo {year} {2013})},\ \Eprint
 {http://arxiv.org/abs/1210.2808} {arXiv:1210.2808 [hep-ph]}\BibitemShut
 {NoStop}%
\bibitem [{\citenamefont {Fael}\ \emph {et~al.}(2021)\citenamefont {Fael},
 \citenamefont {Sch\"onwald},\ and\ \citenamefont
 {Steinhauser}}]{Fael:2020tow}%
 \BibitemOpen
 \bibfield {author} {\bibinfo {author} {\bibfnamefont {M.}~\bibnamefont
 {Fael}}, \bibinfo {author} {\bibfnamefont {K.}~\bibnamefont {Sch\"onwald}}, \
 and\ \bibinfo {author} {\bibfnamefont {M.}~\bibnamefont {Steinhauser}},\
 }\href {\doibase 10.1103/PhysRevD.104.016003} {\bibfield {journal} {\bibinfo
 {journal} {Phys. Rev. D}\ }\textbf {\bibinfo {volume} {104}},\ \bibinfo
 {pages} {016003} (\bibinfo {year} {2021})},\ \Eprint
 {http://arxiv.org/abs/2011.13654} {arXiv:2011.13654 [hep-ph]}\BibitemShut
 {NoStop}%
\bibitem [{\citenamefont {Bordone}\ \emph {et~al.}(2021)\citenamefont
 {Bordone}, \citenamefont {Capdevila},\ and\ \citenamefont
 {Gambino}}]{Bordone:2021oof}%
 \BibitemOpen
 \bibfield {author} {\bibinfo {author} {\bibfnamefont {M.}~\bibnamefont
 {Bordone}}, \bibinfo {author} {\bibfnamefont {B.}~\bibnamefont {Capdevila}},
 \ and\ \bibinfo {author} {\bibfnamefont {P.}~\bibnamefont {Gambino}},\ }\href
 {\doibase 10.1016/j.physletb.2021.136679} {\bibfield {journal} {\bibinfo
 {journal} {Phys. Lett. B}\ }\textbf {\bibinfo {volume} {822}},\ \bibinfo
 {pages} {136679} (\bibinfo {year} {2021})},\ \Eprint
 {http://arxiv.org/abs/2107.00604} {arXiv:2107.00604 [hep-ph]}\BibitemShut
 {NoStop}%
\bibitem [{\citenamefont {Bobeth}\ \emph {et~al.}(2004)\citenamefont {Bobeth},
 \citenamefont {Gambino}, \citenamefont {Gorbahn},\ and\ \citenamefont
 {Haisch}}]{Bobeth:2003at}%
 \BibitemOpen
 \bibfield {author} {\bibinfo {author} {\bibfnamefont {C.}~\bibnamefont
 {Bobeth}}, \bibinfo {author} {\bibfnamefont {P.}~\bibnamefont {Gambino}},
 \bibinfo {author} {\bibfnamefont {M.}~\bibnamefont {Gorbahn}}, \ and\
 \bibinfo {author} {\bibfnamefont {U.}~\bibnamefont {Haisch}},\ }\href
 {\doibase 10.1088/1126-6708/2004/04/071} {\bibfield {journal} {\bibinfo
 {journal} {JHEP}\ }\textbf {\bibinfo {volume} {04}},\ \bibinfo {pages} {071}
 (\bibinfo {year} {2004})},\ \Eprint {http://arxiv.org/abs/hep-ph/0312090}
 {arXiv:hep-ph/0312090}\BibitemShut {NoStop}%
\bibitem [{\citenamefont {Huber}\ \emph {et~al.}(2006)\citenamefont {Huber},
 \citenamefont {Lunghi}, \citenamefont {Misiak},\ and\ \citenamefont
 {Wyler}}]{Huber:2005ig}%
 \BibitemOpen
 \bibfield {author} {\bibinfo {author} {\bibfnamefont {T.}~\bibnamefont
 {Huber}}, \bibinfo {author} {\bibfnamefont {E.}~\bibnamefont {Lunghi}},
 \bibinfo {author} {\bibfnamefont {M.}~\bibnamefont {Misiak}}, \ and\ \bibinfo
 {author} {\bibfnamefont {D.}~\bibnamefont {Wyler}},\ }\href {\doibase
 10.1016/j.nuclphysb.2006.01.037} {\bibfield {journal} {\bibinfo {journal}
 {Nucl. Phys. B}\ }\textbf {\bibinfo {volume} {740}},\ \bibinfo {pages} {105}
 (\bibinfo {year} {2006})},\ \Eprint {http://arxiv.org/abs/hep-ph/0512066}
 {arXiv:hep-ph/0512066}\BibitemShut {NoStop}%
\bibitem [{\citenamefont {Bobeth}\ \emph
 {et~al.}(2014{\natexlab{a}})\citenamefont {Bobeth}, \citenamefont {Gorbahn},
 \citenamefont {Hermann}, \citenamefont {Misiak}, \citenamefont {Stamou},\
 and\ \citenamefont {Steinhauser}}]{Bobeth:2013uxa}%
 \BibitemOpen
 \bibfield {author} {\bibinfo {author} {\bibfnamefont {C.}~\bibnamefont
 {Bobeth}}, \bibinfo {author} {\bibfnamefont {M.}~\bibnamefont {Gorbahn}},
 \bibinfo {author} {\bibfnamefont {T.}~\bibnamefont {Hermann}}, \bibinfo
 {author} {\bibfnamefont {M.}~\bibnamefont {Misiak}}, \bibinfo {author}
 {\bibfnamefont {E.}~\bibnamefont {Stamou}}, \ and\ \bibinfo {author}
 {\bibfnamefont {M.}~\bibnamefont {Steinhauser}},\ }\href {\doibase
 10.1103/PhysRevLett.112.101801} {\bibfield {journal} {\bibinfo {journal}
 {Phys. Rev. Lett.}\ }\textbf {\bibinfo {volume} {112}},\ \bibinfo {pages}
 {101801} (\bibinfo {year} {2014}{\natexlab{a}})},\ \Eprint
 {http://arxiv.org/abs/1311.0903} {arXiv:1311.0903 [hep-ph]}\BibitemShut
 {NoStop}%
\bibitem [{\citenamefont {Hermann}\ \emph {et~al.}(2013)\citenamefont
 {Hermann}, \citenamefont {Misiak},\ and\ \citenamefont
 {Steinhauser}}]{Hermann:2013kca}%
 \BibitemOpen
 \bibfield {author} {\bibinfo {author} {\bibfnamefont {T.}~\bibnamefont
 {Hermann}}, \bibinfo {author} {\bibfnamefont {M.}~\bibnamefont {Misiak}}, \
 and\ \bibinfo {author} {\bibfnamefont {M.}~\bibnamefont {Steinhauser}},\
 }\href {\doibase 10.1007/JHEP12(2013)097} {\bibfield {journal} {\bibinfo
 {journal} {JHEP}\ }\textbf {\bibinfo {volume} {12}},\ \bibinfo {pages} {097}
 (\bibinfo {year} {2013})},\ \Eprint {http://arxiv.org/abs/1311.1347}
 {arXiv:1311.1347 [hep-ph]}\BibitemShut {NoStop}%
\bibitem [{\citenamefont {Bobeth}\ \emph
 {et~al.}(2014{\natexlab{b}})\citenamefont {Bobeth}, \citenamefont {Gorbahn},\
 and\ \citenamefont {Stamou}}]{Bobeth:2013tba}%
 \BibitemOpen
 \bibfield {author} {\bibinfo {author} {\bibfnamefont {C.}~\bibnamefont
 {Bobeth}}, \bibinfo {author} {\bibfnamefont {M.}~\bibnamefont {Gorbahn}}, \
 and\ \bibinfo {author} {\bibfnamefont {E.}~\bibnamefont {Stamou}},\ }\href
 {\doibase 10.1103/PhysRevD.89.034023} {\bibfield {journal} {\bibinfo
 {journal} {Phys. Rev. D}\ }\textbf {\bibinfo {volume} {89}},\ \bibinfo
 {pages} {034023} (\bibinfo {year} {2014}{\natexlab{b}})},\ \Eprint
 {http://arxiv.org/abs/1311.1348} {arXiv:1311.1348 [hep-ph]}\BibitemShut
 {NoStop}%
\bibitem [{\citenamefont {Beneke}\ \emph {et~al.}(2018)\citenamefont {Beneke},
 \citenamefont {Bobeth},\ and\ \citenamefont {Szafron}}]{Beneke:2017vpq}%
 \BibitemOpen
 \bibfield {author} {\bibinfo {author} {\bibfnamefont {M.}~\bibnamefont
 {Beneke}}, \bibinfo {author} {\bibfnamefont {C.}~\bibnamefont {Bobeth}}, \
 and\ \bibinfo {author} {\bibfnamefont {R.}~\bibnamefont {Szafron}},\ }\href
 {\doibase 10.1103/PhysRevLett.120.011801} {\bibfield {journal} {\bibinfo
 {journal} {Phys. Rev. Lett.}\ }\textbf {\bibinfo {volume} {120}},\ \bibinfo
 {pages} {011801} (\bibinfo {year} {2018})},\ \Eprint
 {http://arxiv.org/abs/1708.09152} {arXiv:1708.09152 [hep-ph]}\BibitemShut
 {NoStop}%
\bibitem [{\citenamefont {Beneke}\ \emph {et~al.}(2019)\citenamefont {Beneke},
 \citenamefont {Bobeth},\ and\ \citenamefont {Szafron}}]{Beneke:2019slt}%
 \BibitemOpen
 \bibfield {author} {\bibinfo {author} {\bibfnamefont {M.}~\bibnamefont
 {Beneke}}, \bibinfo {author} {\bibfnamefont {C.}~\bibnamefont {Bobeth}}, \
 and\ \bibinfo {author} {\bibfnamefont {R.}~\bibnamefont {Szafron}},\ }\href
 {\doibase 10.1007/JHEP10(2019)232} {\bibfield {journal} {\bibinfo {journal}
 {JHEP}\ }\textbf {\bibinfo {volume} {10}},\ \bibinfo {pages} {232} (\bibinfo
 {year} {2019})},\ \bibinfo {note} {[Erratum: JHEP {\bf 11}, 099 (2022)]},\ \Eprint
 {http://arxiv.org/abs/1908.07011} {arXiv:1908.07011 [hep-ph]}\BibitemShut
 {NoStop}%
\bibitem [{\citenamefont {Geng}\ \emph {et~al.}(2021)\citenamefont {Geng},
 \citenamefont {Grinstein}, \citenamefont {J\"ager}, \citenamefont {Li},
 \citenamefont {Martin~Camalich},\ and\ \citenamefont {Shi}}]{Geng:2021nhg}%
 \BibitemOpen
 \bibfield {author} {\bibinfo {author} {\bibfnamefont {L.-S.}\ \bibnamefont
 {Geng}}, \bibinfo {author} {\bibfnamefont {B.}~\bibnamefont {Grinstein}},
 \bibinfo {author} {\bibfnamefont {S.}~\bibnamefont {J\"ager}}, \bibinfo
 {author} {\bibfnamefont {S.-Y.}\ \bibnamefont {Li}}, \bibinfo {author}
 {\bibfnamefont {J.}~\bibnamefont {Martin~Camalich}}, \ and\ \bibinfo {author}
 {\bibfnamefont {R.-X.}\ \bibnamefont {Shi}},\ }\href {\doibase
 10.1103/PhysRevD.104.035029} {\bibfield {journal} {\bibinfo {journal}
 {Phys. Rev. D}\ }\textbf {\bibinfo {volume} {104}},\ \bibinfo {pages}
 {035029} (\bibinfo {year} {2021})},\ \Eprint
 {http://arxiv.org/abs/2103.12738} {arXiv:2103.12738 [hep-ph]}\BibitemShut
 {NoStop}%
\bibitem [{\citenamefont {Hurth}\ \emph {et~al.}(2022)\citenamefont {Hurth},
 \citenamefont {Mahmoudi}, \citenamefont {Santos},\ and\ \citenamefont
 {Neshatpour}}]{Hurth:2021nsi}%
 \BibitemOpen
 \bibfield {author} {\bibinfo {author} {\bibfnamefont {T.}~\bibnamefont
 {Hurth}}, \bibinfo {author} {\bibfnamefont {F.}~\bibnamefont {Mahmoudi}},
 \bibinfo {author} {\bibfnamefont {D.~M.}\ \bibnamefont {Santos}}, \ and\
 \bibinfo {author} {\bibfnamefont {S.}~\bibnamefont {Neshatpour}},\ }\href
 {\doibase 10.1016/j.physletb.2021.136838} {\bibfield {journal} {\bibinfo
 {journal} {Phys. Lett. B}\ }\textbf {\bibinfo {volume} {824}},\ \bibinfo
 {pages} {136838} (\bibinfo {year} {2022})},\ \Eprint
 {http://arxiv.org/abs/2104.10058} {arXiv:2104.10058 [hep-ph]}\BibitemShut
 {NoStop}%
\bibitem [{\citenamefont {Alguer\'o}\ \emph {et~al.}(2022)\citenamefont
 {Alguer\'o}, \citenamefont {Capdevila}, \citenamefont {Descotes-Genon},
 \citenamefont {Matias},\ and\ \citenamefont
 {Novoa-Brunet}}]{Alguero:2021anc}%
 \BibitemOpen
 \bibfield {author} {\bibinfo {author} {\bibfnamefont {M.}~\bibnamefont
 {Alguer\'o}}, \bibinfo {author} {\bibfnamefont {B.}~\bibnamefont
 {Capdevila}}, \bibinfo {author} {\bibfnamefont {S.}~\bibnamefont
 {Descotes-Genon}}, \bibinfo {author} {\bibfnamefont {J.}~\bibnamefont
 {Matias}}, \ and\ \bibinfo {author} {\bibfnamefont {M.}~\bibnamefont
 {Novoa-Brunet}},\ }\href {\doibase 10.1140/epjc/s10052-022-10231-1}
 {\bibfield {journal} {\bibinfo {journal} {Eur. Phys. J. C}\ }\textbf
 {\bibinfo {volume} {82}},\ \bibinfo {pages} {326} (\bibinfo {year} {2022})},\
 \Eprint {http://arxiv.org/abs/2104.08921} {arXiv:2104.08921 [hep-ph]}
\BibitemShut {NoStop}%
\bibitem [{\citenamefont {Ciuchini}\ \emph {et~al.}(2023)\citenamefont
 {Ciuchini}, \citenamefont {Fedele}, \citenamefont {Franco}, \citenamefont
 {Paul}, \citenamefont {Silvestrini},\ and\ \citenamefont
 {Valli}}]{Ciuchini:2021smi}%
 \BibitemOpen
 \bibfield {author} {\bibinfo {author} {\bibfnamefont {M.}~\bibnamefont
 {Ciuchini}}, \bibinfo {author} {\bibfnamefont {M.}~\bibnamefont {Fedele}},
 \bibinfo {author} {\bibfnamefont {E.}~\bibnamefont {Franco}}, \bibinfo
 {author} {\bibfnamefont {A.}~\bibnamefont {Paul}}, \bibinfo {author}
 {\bibfnamefont {L.}~\bibnamefont {Silvestrini}}, \ and\ \bibinfo {author}
 {\bibfnamefont {M.}~\bibnamefont {Valli}},\ }\href {\doibase
 10.1140/epjc/s10052-023-11191-w} {\bibfield {journal} {\bibinfo {journal}
 {Eur. Phys. J. C}\ }\textbf {\bibinfo {volume} {83}},\ \bibinfo {pages} {64}
 (\bibinfo {year} {2023})},\ \Eprint {http://arxiv.org/abs/2110.10126}
 {arXiv:2110.10126 [hep-ph]}\BibitemShut {NoStop}%
\bibitem [{\citenamefont {Gubernari}\ \emph {et~al.}(2022)\citenamefont
 {Gubernari}, \citenamefont {Reboud}, \citenamefont {van Dyk},\ and\
 \citenamefont {Virto}}]{Gubernari:2022hxn}%
 \BibitemOpen
 \bibfield {author} {\bibinfo {author} {\bibfnamefont {N.}~\bibnamefont
 {Gubernari}}, \bibinfo {author} {\bibfnamefont {M.}~\bibnamefont {Reboud}},
 \bibinfo {author} {\bibfnamefont {D.}~\bibnamefont {van Dyk}}, \ and\
 \bibinfo {author} {\bibfnamefont {J.}~\bibnamefont {Virto}},\ }\href
 {\doibase 10.1007/JHEP09(2022)133} {\bibfield {journal} {\bibinfo {journal}
 {JHEP}\ }\textbf {\bibinfo {volume} {09}},\ \bibinfo {pages} {133} (\bibinfo
 {year} {2022})},\ \Eprint {http://arxiv.org/abs/2206.03797} {arXiv:2206.03797
 [hep-ph]}\BibitemShut {NoStop}%
\bibitem [{\citenamefont {Greljo}\ \emph {et~al.}(2023)\citenamefont {Greljo},
 \citenamefont {Salko}, \citenamefont {Smolkovi\v{c}},\ and\ \citenamefont
 {Stangl}}]{Greljo:2022jac}%
 \BibitemOpen
 \bibfield {author} {\bibinfo {author} {\bibfnamefont {A.}~\bibnamefont
 {Greljo}}, \bibinfo {author} {\bibfnamefont {J.}~\bibnamefont {Salko}},
 \bibinfo {author} {\bibfnamefont {A.}~\bibnamefont {Smolkovi\v{c}}}, \ and\
 \bibinfo {author} {\bibfnamefont {P.}~\bibnamefont {Stangl}},\ }\href
 {\doibase 10.1007/JHEP05(2023)087} {\bibfield {journal} {\bibinfo {journal}
 {JHEP}\ }\textbf {\bibinfo {volume} {05}},\ \bibinfo {pages} {087} (\bibinfo
 {year} {2023})},\ \Eprint {http://arxiv.org/abs/2212.10497} {arXiv:2212.10497
 [hep-ph]}\BibitemShut {NoStop}%
\bibitem [{\citenamefont {Albrecht}\ \emph {et~al.}(2024)\citenamefont
 {Albrecht}, \citenamefont {Bernlochner}, \citenamefont {Lenz},\ and\
 \citenamefont {Rusov}}]{Albrecht:2024oyn}%
 \BibitemOpen
 \bibfield {author} {\bibinfo {author} {\bibfnamefont {J.}~\bibnamefont
 {Albrecht}}, \bibinfo {author} {\bibfnamefont {F.}~\bibnamefont
 {Bernlochner}}, \bibinfo {author} {\bibfnamefont {A.}~\bibnamefont {Lenz}}, \
 and\ \bibinfo {author} {\bibfnamefont {A.}~\bibnamefont {Rusov}},\
 }\href@noop {} {\ (\bibinfo {year} {2024})},\ \Eprint
 {http://arxiv.org/abs/2402.04224} {arXiv:2402.04224 [hep-ph]}\BibitemShut
 {NoStop}%
\bibitem [{\citenamefont {Haisch}()}]{UH}%
 \BibitemOpen
 \bibfield {author} {\bibinfo {author} {\bibfnamefont {U.}~\bibnamefont
 {Haisch}},\ } (\bibinfo {year} {2015}), \href
 {https://indico.fnal.gov/event/8961/contributions/111530/attachments/72703/87323/top20_haisch.pdf}
 {\emph {\bibinfo {title} {The role of top in the Standard Model}}}\BibitemShut {NoStop}%
\bibitem [{\citenamefont {Giudice}\ \emph {et~al.}(2015)\citenamefont
 {Giudice}, \citenamefont {Paradisi},\ and\ \citenamefont
 {Strumia}}]{Giudice:2015toa}%
 \BibitemOpen
 \bibfield {author} {\bibinfo {author} {\bibfnamefont {G.~F.}\ \bibnamefont
 {Giudice}}, \bibinfo {author} {\bibfnamefont {P.}~\bibnamefont {Paradisi}}, \
 and\ \bibinfo {author} {\bibfnamefont {A.}~\bibnamefont {Strumia}},\ }\href
 {\doibase 10.1007/JHEP11(2015)192} {\bibfield {journal} {\bibinfo {journal}
 {JHEP}\ }\textbf {\bibinfo {volume} {11}},\ \bibinfo {pages} {192} (\bibinfo
 {year} {2015})},\ \Eprint {http://arxiv.org/abs/1508.05332} {arXiv:1508.05332
 [hep-ph]}\BibitemShut {NoStop}%
\bibitem [{\citenamefont {Bruggisser}\ \emph {et~al.}(2023)\citenamefont
 {Bruggisser}, \citenamefont {van Dyk},\ and\ \citenamefont
 {Westhoff}}]{Bruggisser:2022rhb}%
 \BibitemOpen
 \bibfield {author} {\bibinfo {author} {\bibfnamefont {S.}~\bibnamefont
 {Bruggisser}}, \bibinfo {author} {\bibfnamefont {D.}~\bibnamefont {van Dyk}},
 \ and\ \bibinfo {author} {\bibfnamefont {S.}~\bibnamefont {Westhoff}},\
 }\href {\doibase 10.1007/JHEP02(2023)225} {\bibfield {journal} {\bibinfo
 {journal} {JHEP}\ }\textbf {\bibinfo {volume} {02}},\ \bibinfo {pages} {225}
 (\bibinfo {year} {2023})},\ \Eprint {http://arxiv.org/abs/2212.02532}
 {arXiv:2212.02532 [hep-ph]}\BibitemShut {NoStop}%
\bibitem [{\citenamefont {Bartocci}\ \emph {et~al.}(2024)\citenamefont
 {Bartocci}, \citenamefont {Biek\"otter},\ and\ \citenamefont
 {Hurth}}]{Bartocci:2023nvp}%
 \BibitemOpen
 \bibfield {author} {\bibinfo {author} {\bibfnamefont {R.}~\bibnamefont
 {Bartocci}}, \bibinfo {author} {\bibfnamefont {A.}~\bibnamefont
 {Biek\"otter}}, \ and\ \bibinfo {author} {\bibfnamefont {T.}~\bibnamefont
 {Hurth}},\ }\href {\doibase 10.1007/JHEP05(2024)074} {\bibfield {journal}
 {\bibinfo {journal} {JHEP}\ }\textbf {\bibinfo {volume} {05}},\ \bibinfo
 {pages} {074} (\bibinfo {year} {2024})},\ \Eprint
 {http://arxiv.org/abs/2311.04963} {arXiv:2311.04963 [hep-ph]}\BibitemShut
 {NoStop}%
\bibitem [{\citenamefont {Vryonidou}()}]{EV}%
 \BibitemOpen
 \bibfield {author} {\bibinfo {author} {\bibfnamefont {E.}~\bibnamefont
 {Vryonidou}},\ } (\bibinfo {year} {2022}), \href
 {https://indico.mpp.mpg.de/event/8599/contributions/31903/attachments/18573/22722/Vryonidou_SMEFT.pdf}
 {\emph {\bibinfo {title} {{SMEFT in Monte Carlo}}}}\BibitemShut {NoStop}%
\bibitem [{\citenamefont {Jenkins}\ \emph {et~al.}(2013)\citenamefont
 {Jenkins}, \citenamefont {Manohar},\ and\ \citenamefont
 {Trott}}]{Jenkins:2013zja}%
 \BibitemOpen
 \bibfield {author} {\bibinfo {author} {\bibfnamefont {E.~E.}\ \bibnamefont
 {Jenkins}}, \bibinfo {author} {\bibfnamefont {A.~V.}\ \bibnamefont
 {Manohar}}, \ and\ \bibinfo {author} {\bibfnamefont {M.}~\bibnamefont
 {Trott}},\ }\href {\doibase 10.1007/JHEP10(2013)087} {\bibfield {journal}
 {\bibinfo {journal} {JHEP}\ }\textbf {\bibinfo {volume} {10}},\ \bibinfo
 {pages} {087} (\bibinfo {year} {2013})},\ \Eprint
 {http://arxiv.org/abs/1308.2627} {arXiv:1308.2627 [hep-ph]}\BibitemShut
 {NoStop}%
\bibitem [{\citenamefont
 {Schnell}(2024)}]{ThesisLuc}%
 \BibitemOpen
 \bibfield {author} {\bibinfo {author} {\bibfnamefont {L.}~\bibnamefont
 {Schnell}},} {(\bibinfo
 {year} {2024})}, dissertation\BibitemShut {NoStop}%
\bibitem [{\citenamefont {Bonnefoy}\ \emph {et~al.}(2021)\citenamefont
 {Bonnefoy}, \citenamefont {Di~Luzio}, \citenamefont {Grojean}, \citenamefont
 {Paul},\ and\ \citenamefont {Rossia}}]{Bonnefoy:2020tyv}%
 \BibitemOpen
 \bibfield {author} {\bibinfo {author} {\bibfnamefont {Q.}~\bibnamefont
 {Bonnefoy}}, \bibinfo {author} {\bibfnamefont {L.}~\bibnamefont {Di~Luzio}},
 \bibinfo {author} {\bibfnamefont {C.}~\bibnamefont {Grojean}}, \bibinfo
 {author} {\bibfnamefont {A.}~\bibnamefont {Paul}}, \ and\ \bibinfo {author}
 {\bibfnamefont {A.~N.}\ \bibnamefont {Rossia}},\ }\href {\doibase
 10.1007/JHEP05(2021)153} {\bibfield {journal} {\bibinfo {journal} {JHEP}\
 }\textbf {\bibinfo {volume} {05}},\ \bibinfo {pages} {153} (\bibinfo {year}
 {2021})},\ \Eprint {http://arxiv.org/abs/2012.07740} {arXiv:2012.07740
 [hep-ph]}\BibitemShut {NoStop}%
\bibitem [{\citenamefont {Feruglio}(2021)}]{Feruglio:2020kfq}%
 \BibitemOpen
 \bibfield {author} {\bibinfo {author} {\bibfnamefont {F.}~\bibnamefont
 {Feruglio}},\ }\href {\doibase 10.1007/JHEP03(2021)128} {\bibfield {journal}
 {\bibinfo {journal} {JHEP}\ }\textbf {\bibinfo {volume} {03}},\ \bibinfo
 {pages} {128} (\bibinfo {year} {2021})},\ \Eprint
 {http://arxiv.org/abs/2012.13989} {arXiv:2012.13989 [hep-ph]}\BibitemShut
 {NoStop}%
\bibitem [{\citenamefont {Cornella}\ \emph {et~al.}(2023)\citenamefont
 {Cornella}, \citenamefont {Feruglio},\ and\ \citenamefont
 {Vecchi}}]{Cornella:2022hkc}%
 \BibitemOpen
 \bibfield {author} {\bibinfo {author} {\bibfnamefont {C.}~\bibnamefont
 {Cornella}}, \bibinfo {author} {\bibfnamefont {F.}~\bibnamefont {Feruglio}},
 \ and\ \bibinfo {author} {\bibfnamefont {L.}~\bibnamefont {Vecchi}},\ }\href
 {\doibase 10.1007/JHEP02(2023)244} {\bibfield {journal} {\bibinfo {journal}
 {JHEP}\ }\textbf {\bibinfo {volume} {02}},\ \bibinfo {pages} {244} (\bibinfo
 {year} {2023})},\ \Eprint {http://arxiv.org/abs/2205.10381} {arXiv:2205.10381
 [hep-ph]}\BibitemShut {NoStop}%
\bibitem [{\citenamefont {Cohen}\ \emph
 {et~al.}(2023{\natexlab{a}})\citenamefont {Cohen}, \citenamefont {Lu},\ and\
 \citenamefont {Zhang}}]{Cohen:2023hmq}%
 \BibitemOpen
 \bibfield {author} {\bibinfo {author} {\bibfnamefont {T.}~\bibnamefont
 {Cohen}}, \bibinfo {author} {\bibfnamefont {X.}~\bibnamefont {Lu}}, \ and\
 \bibinfo {author} {\bibfnamefont {Z.}~\bibnamefont {Zhang}},\ }\href
 {\doibase 10.1103/PhysRevD.107.116015} {\bibfield {journal} {\bibinfo
 {journal} {Phys. Rev. D}\ }\textbf {\bibinfo {volume} {107}},\ \bibinfo
 {pages} {116015} (\bibinfo {year} {2023}{\natexlab{a}})},\ \Eprint
 {http://arxiv.org/abs/2301.00821} {arXiv:2301.00821 [hep-ph]}\BibitemShut
 {NoStop}%
\bibitem [{\citenamefont {Cohen}\ \emph
 {et~al.}(2023{\natexlab{b}})\citenamefont {Cohen}, \citenamefont {Lu},\ and\
 \citenamefont {Zhang}}]{Cohen:2023gap}%
 \BibitemOpen
 \bibfield {author} {\bibinfo {author} {\bibfnamefont {T.}~\bibnamefont
 {Cohen}}, \bibinfo {author} {\bibfnamefont {X.}~\bibnamefont {Lu}}, \ and\
 \bibinfo {author} {\bibfnamefont {Z.}~\bibnamefont {Zhang}},\ }\href
 {\doibase 10.1103/PhysRevD.108.056027} {\bibfield {journal} {\bibinfo
 {journal} {Phys. Rev. D}\ }\textbf {\bibinfo {volume} {108}},\ \bibinfo
 {pages} {056027} (\bibinfo {year} {2023}{\natexlab{b}})},\ \Eprint
 {http://arxiv.org/abs/2301.00827} {arXiv:2301.00827 [hep-ph]}\BibitemShut
 {NoStop}%
\bibitem [{\citenamefont {Wess}\ and\ \citenamefont
 {Zumino}(1971)}]{Wess:1971yu}%
 \BibitemOpen
 \bibfield {author} {\bibinfo {author} {\bibfnamefont {J.}~\bibnamefont
 {Wess}}\ and\ \bibinfo {author} {\bibfnamefont {B.}~\bibnamefont {Zumino}},\
 }\href {\doibase 10.1016/0370-2693(71)90582-X} {\bibfield {journal}
 {\bibinfo {journal} {Phys. Lett. B}\ }\textbf {\bibinfo {volume} {37}},\
 \bibinfo {pages} {95} (\bibinfo {year} {1971})}\BibitemShut {NoStop}%
\bibitem [{\citenamefont {Durieux}\ \emph {et~al.}(2018)\citenamefont
 {Durieux}, \citenamefont {Gu}, \citenamefont {Vryonidou},\ and\ \citenamefont
 {Zhang}}]{Durieux:2018ggn}%
 \BibitemOpen
 \bibfield {author} {\bibinfo {author} {\bibfnamefont {G.}~\bibnamefont
 {Durieux}}, \bibinfo {author} {\bibfnamefont {J.}~\bibnamefont {Gu}},
 \bibinfo {author} {\bibfnamefont {E.}~\bibnamefont {Vryonidou}}, \ and\
 \bibinfo {author} {\bibfnamefont {C.}~\bibnamefont {Zhang}},\ }\href
 {\doibase 10.1088/1674-1137/42/12/123107} {\bibfield {journal} {\bibinfo
 {journal} {Chin. Phys. C}\ }\textbf {\bibinfo {volume} {42}},\ \bibinfo
 {pages} {123107} (\bibinfo {year} {2018})},\ \Eprint
 {http://arxiv.org/abs/1809.03520} {arXiv:1809.03520 [hep-ph]}\BibitemShut
 {NoStop}%
\bibitem [{\citenamefont {'t~Hooft}\ and\ \citenamefont
 {Veltman}(1972)}]{tHooft:1972tcz}%
 \BibitemOpen
 \bibfield {author} {\bibinfo {author} {\bibfnamefont {G.}~\bibnamefont
 {'t~Hooft}}\ and\ \bibinfo {author} {\bibfnamefont {M.~J.~G.}\ \bibnamefont
 {Veltman}},\ }\href {\doibase 10.1016/0550-3213(72)90279-9} {\bibfield
 {journal} {\bibinfo {journal} {Nucl. Phys. B}\ }\textbf {\bibinfo {volume}
 {44}},\ \bibinfo {pages} {189} (\bibinfo {year} {1972})}\BibitemShut
 {NoStop}%
\bibitem [{\citenamefont {Breitenlohner}\ and\ \citenamefont
 {Maison}(1977{\natexlab{a}})}]{Breitenlohner:1977hr}%
 \BibitemOpen
 \bibfield {author} {\bibinfo {author} {\bibfnamefont {P.}~\bibnamefont
 {Breitenlohner}}\ and\ \bibinfo {author} {\bibfnamefont {D.}~\bibnamefont
 {Maison}},\ }\href {\doibase 10.1007/BF01609069} {\bibfield {journal}
 {\bibinfo {journal} {Commun. Math. Phys.}\ }\textbf {\bibinfo {volume}
 {52}},\ \bibinfo {pages} {11} (\bibinfo {year}
 {1977}{\natexlab{a}})}\BibitemShut {NoStop}%
\bibitem [{\citenamefont {Breitenlohner}\ and\ \citenamefont
 {Maison}(1977{\natexlab{b}})}]{Breitenlohner:1975hg}%
 \BibitemOpen
 \bibfield {author} {\bibinfo {author} {\bibfnamefont {P.}~\bibnamefont
 {Breitenlohner}}\ and\ \bibinfo {author} {\bibfnamefont {D.}~\bibnamefont
 {Maison}},\ }\href {\doibase 10.1007/BF01609070} {\bibfield {journal}
 {\bibinfo {journal} {Commun. Math. Phys.}\ }\textbf {\bibinfo {volume}
 {52}},\ \bibinfo {pages} {39} (\bibinfo {year}
 {1977}{\natexlab{b}})}\BibitemShut {NoStop}%
\bibitem [{\citenamefont {Breitenlohner}\ and\ \citenamefont
 {Maison}(1977{\natexlab{c}})}]{Breitenlohner:1976te}%
 \BibitemOpen
 \bibfield {author} {\bibinfo {author} {\bibfnamefont {P.}~\bibnamefont
 {Breitenlohner}}\ and\ \bibinfo {author} {\bibfnamefont {D.}~\bibnamefont
 {Maison}},\ }\href {\doibase 10.1007/BF01609071} {\bibfield {journal}
 {\bibinfo {journal} {Commun. Math. Phys.}\ }\textbf {\bibinfo {volume}
 {52}},\ \bibinfo {pages} {55} (\bibinfo {year}
 {1977}{\natexlab{c}})}\BibitemShut {NoStop}%
 \bibitem [{\citenamefont {Silvestrii}()}]{LS}%
 \BibitemOpen
 \bibfield {author} {\bibinfo {author} {\bibfnamefont {L.}~\bibnamefont
 {Silvestrini}},\ } (\bibinfo {year} {2018}), \href
 {https://agenda.infn.it/event/14377/contributions/24434/attachments/17481/19830/silvestriniLaThuile.pdf}
 {\emph {\bibinfo {title} {Flavour constraints on new physics}}}\BibitemShut {NoStop}%
\bibitem [{\citenamefont {Aad}\ \emph {et~al.}(2023{\natexlab{b}})\citenamefont
 {Aad} \emph {et~al.}}]{ATLAS:2023qzr}%
 \BibitemOpen
 \bibfield {author} {\bibinfo {author} {\bibfnamefont {G.}~\bibnamefont
 {Aad}} \emph {et~al.} (\bibinfo {collaboration} {ATLAS}),\ }\href {\doibase
 10.1103/PhysRevD.108.032019} {\bibfield {journal} {\bibinfo {journal}
 {Phys. Rev. D}\ }\textbf {\bibinfo {volume} {108}},\ \bibinfo {pages}
 {032019} (\bibinfo {year} {2023}{\natexlab{b}})},\ \Eprint
 {http://arxiv.org/abs/2301.11605} {arXiv:2301.11605 [hep-ex]}\BibitemShut
 {NoStop}%
 \end{thebibliography}

%

\end{document}